\documentclass[times, twocolumn, review, oneside]{elsarticle}
\pdfoutput=1
\usepackage{medima}
\usepackage{framed,multirow}
\usepackage{lineno,hyperref}
\usepackage{amssymb}
\usepackage{amsmath}
\usepackage{latexsym}
\usepackage{graphicx}
\usepackage{url}
\usepackage[table]{xcolor}
\usepackage{todonotes}
\usepackage{booktabs}
\usepackage[algo2e,vlined,linesnumbered]{algorithm2e}

\graphicspath{ {./figures/figures_revision/} }
\DeclareGraphicsExtensions{.png,.eps,.pdf}

\modulolinenumbers[5]
\definecolor{newcolor}{rgb}{.8,.349,.1}
\journal{Medical Image Analysis}
\verso{Oliver Maier \textit{et~al.}}
\biboptions{authoryear}
\newif\ifisresponse
\isresponsefalse  % Define FALSE for clean version
\usepackage{marginnote}
\usepackage{color}
\definecolor{changedcolorEditor}{rgb}{0.3,0.8,0.5}
\definecolor{changedcolorRone}{rgb}{1,0,0}
\definecolor{changedcolorRtwo}{rgb}{0,0.5,0}
\definecolor{changedcolorRthree}{rgb}{0,0,1}
\definecolor{changedcolorRfour}{rgb}{1,0.5,0}
\definecolor{darkolivegreen}{rgb}{0.33, 0.42, 0.18}

\ifisresponse
      \usepackage[draft]{changes}
			\newcommand{\responseEditor}[2]{{\color{changedcolorEditor}#2}\marginnote{\vspace*{-1.6ex}{\color{changedcolorEditor}#1}}}
			
			\newcommand{\responseRone}[2]{{\color{changedcolorRone}#2}\marginnote{\vspace*{-1.6ex}{\color{changedcolorRone}#1}}}
			
			\newcommand{\responseRtwo}[2]{{\color{changedcolorRtwo}#2}\marginnote{\vspace*{-1.6ex}{\color{changedcolorRtwo}#1}}}
			
			\newcommand{\responseRthree}[2]{{\color{changedcolorRthree}#2}\marginnote{\vspace*{-1.6ex}{\color{changedcolorRthree}#1}}}
			
			\newcommand{\responseRfour}[2]{{\color{changedcolorRfour}#2}\marginnote{\vspace*{-1.6ex}{\color{changedcolorRfour}#1}}}

		    \newcommand{\mdeleted}[2][None]{\deleted[]{#2}}       

\else

			\newcommand{\responseEditor}[2]{{\color{black}#2}\marginnote{\vspace*{-1.6ex}{\color{black}}}}
			\newcommand{\responseRone}[2]{{\color{black}#2}\marginnote{\vspace*{-1.6ex}{\color{black}}}}
			\newcommand{\responseRtwo}[2]{{\color{black}#2}\marginnote{\vspace*{-1.6ex}{\color{black}}}}
			\newcommand{\responseRthree}[2]{{\color{black}#2}\marginnote{\vspace*{-1.6ex}{\color{black}}}}
			\newcommand{\responseRfour}[2]{{\color{black}#2}\marginnote{\vspace*{-1.6ex}{\color{black}}}}	
			
			\newcommand{\mdeleted}[2][None]{\deleted[]{}}	        
			\usepackage[final]{changes}
\fi

\SetKw{KwBy}{by}
\begin{document}
%
%-----------------------------------------------------  REVIEW ----------------------------------------------------- 
\ifisresponse
\pagenumbering{roman}
\onecolumn
% --------------------------------------------------------------------
% Response to Referees
% --------------------------------------------------------------------

\fi

\clearpage
%
%\twocolumn
\setcounter{page}{1}

%-----------------------------------------------------  REVIEW ----------------------------------------------------- 

\begin{frontmatter}
\title{Non-linear fitting with joint spatial regularization in 
Arterial Spin Labeling} 
%\tnoteref{mytitlenote}}
%\tnotetext[mytitlenote]{possible note 
%\href{http://www.ctan.org/tex-archive/macros/latex/contrib/elsarticle}{CTAN}.}
%\author{Oliver Maier\fnref{myfootnote}}
%\address{Stremayrgasse 16/3, Graz, Austria}
\author[tug]{Oliver Maier}
\ead{oliver.maier@tugraz.at}
\author[tug]{Stefan M Spann}
\ead{stefan.spann@tugraz.at}
\author[mug]{Daniela Pinter}
\ead{daniela.pinter@medunigraz.at}
\author[mug,mug2]{Thomas Gattringer}
\ead{thomas.gattringer@medunigraz.at}
\author[mug2]{Nicole Hinteregger}
\ead{nicole.hinteregger@medunigraz.at}
\author[tug2,btm]{Gerhard G. Thallinger}
\ead{gerhard.thallinger@tugraz.at}
\author[mug,mug2]{Christian Enzinger}
\ead{chris.enzinger@medunigraz.at}
\author[siemens]{Josef Pfeuffer}
\ead{josef.pfeuffer@siemens-healthineers.com}
\author[kfu,btm]{Kristian Bredies}
\ead{kristian.bredies@uni-graz.at}
\author[tug,btm]{Rudolf Stollberger\corref{mycorrespondingauthor}}
\ead{rudolf.stollberger@tugraz.at}
\cortext[mycorrespondingauthor]{Corresponding author: \\
Tel.: +43-316-873-35401;  
fax: +43-316-873-1035400;}
\address[tug]{Institute of Medical Engineering, Graz University of Technology, 
Stremayrgasse 16/III, 8010 Graz, Austria}
\address[mug]{Department of Neurology, Division of General Neurology, Medical 
University of Graz, Auenbruggerplatz 22, 8036 Graz, Austria}
\address[mug2]{Division of Neuroradiology, Vascular and Interventional 
Radiology, Department of Radiology, Medical University of Graz, 
Auenbruggerplatz 
22, 8036 Graz, Austria}
\address[tug2]{Institute of Biomedical Informatics, Graz University of Technology, 
Stremayrgasse 16/I, 8010 Graz, Austria}
\address[btm]{BioTechMed-Graz, Mozartgasse 12/II, 8010 Graz, Austria}
\address[siemens]{Application Development, Siemens Healthcare, 
Henkestra\ss{}e 127, 91052 Erlangen, Germany}
\address[kfu]{Institute of Mathematics and Scientific Computing, University of 
Graz, Heinrichstraße 36, 8010 Graz, Austria}
\received{2020}
\finalform{xxxx}
\accepted{xxxx}
\availableonline{xxxx}
\communicated{NN}
\begin{abstract}
Multi-Delay single-shot arterial spin labeling (ASL) imaging provides accurate 
cerebral blood flow (CBF) and, in addition, arterial transit time (ATT) 
maps but the inherent low SNR can be challenging. Especially standard fitting 
using non-linear least squares often fails in regions with poor SNR, resulting 
in noisy estimates of the quantitative maps. State-of-the-art fitting 
techniques improve the SNR by incorporating prior knowledge in the estimation 
process which typically leads to spatial blurring. To this end, we propose a 
new estimation method with a joint spatial total generalized variation 
regularization on CBF and ATT. This joint regularization approach utilizes 
shared spatial features across maps to enhance sharpness and simultaneously 
improves noise suppression in the final estimates. The proposed method is 
\responseRone{R1.C7}{evaluated at three levels}, first on synthetic phantom data including 
pathologies, followed by in vivo acquisitions of healthy volunteers, and 
finally on patient data following an ischemic stroke. The quantitative estimates
 are compared to two reference methods, non-linear least squares fitting and a 
state-of-the-art ASL quantification algorithm based on Bayesian inference. The 
proposed joint regularization approach outperforms the reference 
implementations, substantially increasing the SNR in CBF and ATT while 
maintaining sharpness and quantitative accuracy in the estimates. 
\end{abstract}
\end{frontmatter}
\ifisresponse
\onecolumn
\fi
%\linenumbers
% --------------------------------------------------------------------
% Introduction
% --------------------------------------------------------------------
\section{Introduction}
\label{sec:Introduction}
Arterial spin labeling (ASL) is a non-invasive MRI technique for quantifying 
local tissue perfusion~\citep{Detre1992}.
\responseRtwo{R2.C6}{The method utilizes magnetically labeled blood water by inverting the blood water 
spins upstream the imaging region.} After waiting a specific period of time,
called the post labeling delay (PLD) which accounts for the time the 
magnetically labeled blood needs to flow into the region of interest, an image 
is acquired. This so called label image is subtracted from a second image, the 
control image, acquired without magnetization alterations of the inflowing 
blood. From this difference image, also known as perfusion weighted image (PWI),
 the cerebral blood flow (CBF) can be quantified using a general kinetic 
model~\citep{Buxton1998}. 
The recommended clinical ASL protocol\responseRtwo{R2.C7}{~\citep{Alsop2015,telischak2014}} consists of single-delay 
pseudo-continuous ASL (pCASL)~\citep{Dai2008} \responseRtwo{R2.C7}{\mdeleted{protocol}} combined 
with segmented 3D data acquisitions such as gradient and spin echo 
(GRASE)~\citep{Feinberg2009,Gunther2005} or turbo spin echo (TSE) stack of 
spirals (SoSP) \responseRtwo{R2.C7}{\mdeleted{\mbox{\citep{Ye2000,Vidorreta2013}}} \citep{Ye2000, Dai2008}} readout
due to efficient background suppression \responseRtwo{R2.C7}{\citep{Ye2000} \mdeleted{\mbox{\citep{Maleki2012}}}} and SNR 
gains~\citep{Alsop2015} of these methods. 
\responseRtwo{R2.C8}{\mdeleted{The main limitation of single-delay acquisition is that the CBF is 
underestimated in areas where the arterial transit time (ATT) of the blood is 
higher than the selected PLD.}
For single-delay acquisitions the signal in the PWI depends on both, the CBF and arterial transit time (ATT).
Hence, the accuracy of CBF estimation from single-PLD ASL data is dependent on both factors. 
Another limitation is that for cases with prolonged ATT (ATT $>$ PLD) some of the labeled blood 
may remain in larger vessels. This leads to bright spots and an overestimation in CBF in bigger 
vessels and an underestimation of CBF in brain areas. The bright spots are known as vascular artifacts and 
can complicate clinical diagnosis in patients with stroke, steno-occlusions, or moya-mayo disease~\citep{Zaharchuk2012}. 
A way to improve the clinical interpretation of single-delay ASL images is by applying additional coefficient of variation maps obtained from the multiple PWIs~\citep{Mutsaerts2017}. \mdeleted{The selection of the right PLD is difficult, 
because the ATT varies between healthy subjects and patients with vascular 
diseases, such as arteriosclerosis.}} Another way to reduce misquantification
is by using a longer PLD, ensuring that the blood has sufficient time to reach the 
tissue. However, this leads to longer acquisitions and additionally 
to a lower SNR due to the T1-relaxation of the labeled blood. Alternatively, 
multi-PLDs can be used to sample the inflowing blood at several time points, 
from a short PLD to long PLD. By fitting the acquired signal to a kinetic model, 
the potential bias in CBF due to unknown ATT can be reduced~\responseRtwo{R2.C8}{\citep{Buxton1998}}. In addition, this 
approach provides another important parameter, the ATT, which is helpful in 
characterization or detection of cerebrovascular diseases~\responseRtwo{R2.C8}{{\citep{Alsop2015, MacIntosh1892, MacIntosh2014, Zaharchuk2009, Zaharchuk2012}}}. 
However, the recommended segmented acquisitions have the drawback of a low 
temporal resolution with increased sensitivity to inter-segment motion. 
Therefore, only a limited number of PLDs can be acquired in a clinically 
acceptable time. Recently, accelerated single-shot 3D acquisition 
strategies~\citep{Dimo2017, Boland2018, Spann2019} were implemented to overcome 
this drawback, at the cost of reduced SNR. This makes 
the estimation of reasonable quantitative ATT and CBF maps from this low SNR 
perfusion weighted time series challenging. The standard voxel-wise non-linear 
least squares (NLLS) fitting approach leads to outliers in low-SNR voxels. 
To this end, a weighted delay approach~\citep{Dai2012} was proposed to reduce 
outliers in the quantitative maps. Further improvements could be 
achieved by inclusion of spatial priors on the CBF map~\citep{Groves2009} in a 
Bayesian inference model (BASIL~\citep{Chappell2010}). This stabilizes the 
fitting approach and reduces noise, ultimately leading to improved CBF estimates
 but introduces spatial blurring. Exploiting all available spatial information 
by means of joining the individual regularization of each unknown into a single,
 joint regularization functional can further improve reconstruction quality. 
Such an approach has been successfully applied in the context of 
relaxometry~\citep{Knoll2017a, Wang2017, Maier2019c}. Joint regularization 
utilizes information present in each map, such as tissue boundaries, by means of
 advanced spatial regularization functionals to avoid the loss of small features
 and promotes overall sharper parameter maps. In this study, we propose a new 
non-linear fitting algorithm with joint spatial constraints on the CBF and ATT 
map to stabilize the estimation procedure and hence enhance the image quality. 
To improve the motion robustness of the 3D acquisition, we combine the proposed 
method with a single shot CAIPIRINHA accelerated 3D GRASE readout. The method is
 \responseRone{R1.C7}{\mdeleted{validated}evaluated} on synthetic phantom datasets including simulated pathologies, on 
\responseRfour{R4}{\mdeleted{three}six} healthy subjects, as well as on \responseRfour{R4}{\mdeleted{three}seven} stroke patients and compared to \responseRtwo{R2.C9}{NLLS}
 and BASIL \responseRone{R1.C1}{without regularization on ATT (\textit{BASIL w/o}) and with regularization on CBF and ATT (\textit{BASIL w/}).}

% --------------------------------------------------------------------
% Theory
% --------------------------------------------------------------------
\section{Theory}
\label{sec:theory}
\subsection{Fixing notation}Throughout the work we fix the 
following notations. The image dimensions in 3D are denoted as $N_i$, $N_j$, 
$N_k$, defining the image space $U = \mathbb{R}^{N_i \times N_j \times N_k}$ 
with $x=(i,j,k)$ defining a point at location $(i,j,k) \in \mathbb{N}^3$. 
$u \in U^{N_u}$ expresses the space of unknown CBF- and ATT-maps with $N_u=2$ 
in this case. The measured data space is denoted as $D = 
\mathbb{C}^{ N_i \times N_j \times N_k}$ and consists of 
$N_d$ perfusion weighted images derived from Control/Label (C/L)-pairs, 
measured at time $t=(t_1, t_2, \hdots, t_{N_d})\in\mathbb{R}_{+}^{N_d}$. 

\subsection{Parameter fitting}
From a statistical point of view the problem of identifying the unknown 
parameters $u=(u_1,u_2,\dots,u_{N_u}) \in U^{N_u}$ given a series 
of noisy measurements $d = (d_1,d_2,\dots,d_{N_d}) \in D^{N_d}$ 
can be solved via maximum likelihood estimation. Assuming the measurements 
at time $t_n$ are generated by some function $A_{\phi,t_n}: u \mapsto d_n$ with 
fixed parameters $\phi$ the likelihood function of measuring $d$ is given by 
$p(d|u,A_{\phi,t_n})$. The realization of $p$ depends on the noise 
distribution in the measurements $d$. Under the assumption that 
additive independent and identically distributed zero-mean Gaussian noise with 
variance $\sigma^2$ (AWGN) corrupts the measurements $d$, the 
multivariate likelihood function turns into a product of single-variate 
functions. It is common to minimize the negative logarithm of the likelihood 
function, which is equivalent to maximizing the likelihood, as it turns the 
product into a sum and improves the numerical stability. Omitting constant terms
 with respect to $u$ yields
 
\begin{linenomath*}
\begin{align}
 u^* &\in \underset{u\in U^{N_u}}{\arg\min}\,\frac{1}{2\sigma^2} 
\sum_{n=1}^{N_d} {
\|A_{\phi,t_n}(u)-d_n\|_2^2}
\label{eq:stat}
\end{align}
\end{linenomath*}
which resembles the well known minimum least squares problem with 
$\|\cdot\|_2$ being the standard $L^2$-norm.

Typically, several measurements with varying sequence parameters are necessary 
to quantify tissue parameters. Especially in cases with a non-linear 
relationship between acquired signal and parameters, fitting is performed in an 
iterative fashion. 
\subsection{The ASL signal model} 
The quantification of CBF and ATT is based 
on the standard model for pseudo-continuous ASL (pCASL)~\citep{Buxton1998} 
which reads as

\begin{linenomath*}
\begin{equation}\label{eq:ASL}
{
 A(u)_{\phi,t_n} = 
   \begin{cases}
      0, & t_n < \Delta\\
      2 
M_{0\alpha}{f}T_{1}e^{-\frac{\Delta}{T_{1b}}}\left(1-e^\frac{
-t_n+\tau} {T_{1_{app}}} \right), &\Delta \leq t_n <
\Delta + \tau\\
      2 
M_{0\alpha}{f}T_{1}e^{-\frac{\Delta}{T_{1b}}-\frac{t_n-\tau-\Delta}
{ T_{1_{app}}}} \left(1-e^\frac {
-\tau} {T_{1_{app}}} \right), & \Delta + \tau \leq t_n
   \end{cases}}
\end{equation}
\end{linenomath*}
where $u=(f,\Delta)$ and $f$ amounts to CBF in \responseRone{R1.C11}{ml/g/s, but is normally quoted in ml/100g/min, and $\Delta$ to ATT 
in seconds.} The a priori known parameters of equation~\ref{eq:ASL} are  
combined into the variable   
$\phi = (M_{0\alpha},T_1,T_{1b},\tau)$. 
It is assumed that $T_1$, 
the apparent longitudinal relaxation decay constant of the tissue, amounts 
to 1.33 seconds at 3T. \responseRone{R1.C11}{\mdeleted{$M_0$ is the acquired 
proton density weighted image, and }$T_{1b}$ is} the longitudinal relaxation decay 
constant of blood, assumed to amount to 1.65 seconds at 3T~\citep{Lu2004}. $\tau$
 corresponds to the labeling duration, $\alpha$ is the labeling efficiency and 
set to \responseRtwo{R2.C14}{0.7}~\citep{Dai2008} and $t_n$ is the acquisition time point, i.e. 
the sum of post labeling delay and labeling duration, for the n$^{th}$ 
measurement. Further, the blood-brain partition coefficient $\lambda$ is 
assumed to be \responseRtwo{R2.C10}{0.9} ml/g~\citep{Herscovitch1985} thus 
$1/{T_{1_{app}}}(f) = {1/T_1+{f}/\lambda}$,  
and $M_{0\alpha}=\alpha M_{0}/\lambda$ \responseRone{R1.C11}{with $M_0$ being the acquired 
proton density weighted image.}
\subsection{Regularization}
As the acquired PWI images suffer from poor SNR the problem of quantifying CBF 
and ATT typically\responseRone{R1.C12}{\mdeleted{ill-posed} suffers from numerically instabilities}. A method to incorporate a priori
 knowledge of the parameters $u$ into the maximum 
likelihood estimation problem~\ref{eq:stat}  \responseRone{R1.C12}{\mdeleted{via Bayes' theorem }}
is known as maximum a posteriori estimation and leads to

\begin{linenomath*}
\begin{equation}\label{eq:l2minreg}
   \underset{u}{\min}\quad \frac{1}{2}\sum_{n=1}^{N_d}\left\|A_{\phi,t_n}(u)-
  d_n\right\|_2^2 + \gamma R(u),
\end{equation}
\end{linenomath*}
with $\gamma>0$ \responseRone{R1.C12}{\mdeleted{being used to balance}balancing} between the data fidelity term and the 
regularization $R$. $R(u)$ includes known information about $u$ such as 
its statistical distribution or spatial features, e.g. 
$u$ should consist of piece-wise constant areas. 
As the variance $\sigma^2$ is \responseRone{R1.C9}{\mdeleted{often} in general unknown, \mdeleted{it is 
included in the choice of $\gamma$} we will not consider $\sigma^2$ fixed but as something that can be chosen in the reconstruction process. Thus, we combine it with the regularization parameter $\gamma$}. The introduced prior can lead 
to a biased estimate of $u$ with reduced uncertainties~\citep{Brinkmann2017}. 
Thus a trade-off 
between faithfulness to acquired data and the prior needs to be determined 
according to the expected noise in the data. The most basic form consists of 
classical Tikhonov regularization which penalizes outliers in the parameter maps
 in an $L^2$-norm sense~\citep{tikhonov1977}. An extension to this basic form 
consists of penalizing the gradient of the maps which is known as $H^1$ 
regularization~\citep{tikhonov1977}, leading to a smoother appearance but comes 
at the cost of blurred edges. To preserve edges and to obtain a better visual
impression, a sparsity promoting functional
is usually preferred which can be realized by posing an $L^1$-norm based 
constraint on the sparse domain of the unknowns~\citep{Donoho2006, Lustig2007}. 
As $u$ is usually not sparse in its native domain, a sparsifying transform
such as a finite differences operation or a wavelet transformation is used.
 The total variation (TV) 
functional of Rudin-Osher-Fatemi (ROF)~\citep{Rudin1992} is based on an 
$L^1$-norm combined with a forward finite differences operator. This combination
 can be interpreted as a spatial piece-wise constant prior which is known to be 
prone to stair-casing artifacts in the final reconstruction 
results~\citep{Bredies2010}.
In order to avoid these stair-casing artifacts but leverage the edge-preserving 
feature of TV a generalization termed total generalized variation (TGV) 
functional was proposed by \citet{Bredies2010}. In the context of MRI, TGV$^2$, 
which enforces piece-wise linear solutions by balancing between a 
first order and approximated second order derivative, was shown to yield 
excellent reconstruction results, preserving fine details and edges while 
maintaining the denoising properties of TV~\citep{Knoll2011}. In the discretized
 form the TGV$^2$ regularization is realized via a minimization problem of the 
following form

\begin{linenomath*}
\begin{equation}
    \text{TGV}^2(u) := \underset{v}{\min}\,\beta_0\|\nabla u - 
v\|_{1,2} 
+ \beta_1\|\mathcal{E}v\|_{1,2}.
\end{equation}
\end{linenomath*}
The favorable properties of TGV$^2$ can be further improved by sharing common 
feature information between the unknown parameter maps by joining the TGV$^2$ 
functionals utilizing a Frobenius norm in parametric 
dimension~\citep{Bredies2014}. Recently, this combination was shown to yield 
improved reconstruction results compared to separate regularization on each map 
in the context of quantitative $T_1$ mapping~\citep{Maier2019c} and multi modal 
image reconstruction~\citep{Knoll2017a}. 
The combination by means of a Frobenius norm is justified by the assumption 
that quantitative maps share the same features at the same spatial positions. 
To incorporate the Frobenius norm the following adaptations to the TGV$^2$ 
semi-norm definitions are made

\begin{linenomath*}
\begin{equation}
\|v\|_{1,2,F} = \sum _{i,j,k} \sqrt{ \sum _{l=1}^{N_u} 
|v_{i,j,k}^{1,l}|^ 2 + 
|v_{i,j,k}^{2,l}|^ 2 + |v_{i,j,k}^{3,l}|^ 2}
\end{equation}
\end{linenomath*}
with $v = (v^{1,l},v^{2,l},v^{3,l})^{N_u}_{l=1}\in U^{3\times N_u}$ 
constituting the approximation of 3D spatial derivatives, and for the 
symmetrized gradient $\chi = 
(\chi^{1,l},\chi^{2,l},\chi^{3,l},\chi^{4,l},\chi^{5,l},\chi^{6,l})^{N_u}_{l=1} 
\in 
U^{6\times N_u}$

\begin{linenomath*}
\begin{equation}
\|\chi\|_{1,2,F} = \sum _{i,j,k} \sqrt{ \sum _{l=1}^{N_u} 
\begin{aligned}
& |\chi_{i,j,k}^{1,l}|^ 2 + |\chi_{i,j,k}^{2,l}|^ 2 + |\chi_{i,j,k}^{3,l}|^ 2 
 \\&+2|\chi_{i,j,k}^{4,l}|^ 2 + 2|\chi_{i,j,k} ^{5,l}|^ 2 + 2|\chi_{i,j,k}^ 
{6,l}|^ 
2\end{aligned}}.
\end{equation}
\end{linenomath*}

\subsection{The non-linear, non-smooth optimization problem}
\label{sec:opt}
The combination of TGV$^2$ with equation~\ref{eq:l2minreg} leads to

\begin{linenomath*}
\begin{align}
   \underset{u,v}{\min}\quad 
\frac{1}{2}\sum_{n=1}^{N_d}&\|A_{\phi,t_n}(u)-d_n\|_2^2 
+\nonumber\\ \gamma( \beta_0&\|\nabla u - v\|_{1,2,F} + 
\beta_1\|\mathcal{E}v\|_{1,2,F})
\end{align}
\end{linenomath*}
which is a non-linear problem in the unknowns $u$ and non-smooth due to 
the $L^1$-norms of the TGV$^2$ functional. Recall that for the ASL signal, 
the non-linear operator $A_{\phi,t_n}(u)$ is defined by equation~\ref{eq:ASL} 
and $u$ amounts to $u=(f,\Delta)$. A similar problem arises in model-based 
quantification of $T_1$ and $M_0$~\citep{Roeloffs2016, Wang2017, Maier2019c}. 
The problem is thus solved in analogy via a two-step procedure. First the data 
fidelity term is linearized in a Gauss-Newton (GN) fashion, second the 
linearized, non-smooth sub-problem is solved using a primal-dual splitting 
algorithm. The linearized sub-problem for each linearization step $k$ is 
given by 

\begin{linenomath*}
\begin{align} \label{eq:linearized}
   \underset{u,v}{\min}\quad 
\frac{1}{2}\sum_{n=1}^{N_d}&\|\mathrm{D}A_{\phi,t_n}\rvert_{u=u^{k}} u-\tilde{d_n}^k
\|_2^2 
+ \nonumber\\
\gamma_k(
\beta_0&\|\nabla u - v\|_{1,2,F} + \beta_1|\|\mathcal{E}v\|_{1,2,F}) +
\nonumber\\ \frac{\delta_k}{2}&\|u-u^k\|_{M_k}^2.
\end{align}
\end{linenomath*}
Constant terms stemming from the linearization at position $u^k$ are fused with 
the data by $\tilde{d}_n^k = d_n - {A}_{\phi,t_n}(u^{k}) + 
\mathrm{D}A_{\phi,t_n}u^{k}$ and the matrix $\mathrm{D}A_{\phi,t_n}
\rvert_{u=u^{k}} = \frac{\partial{A}_{\phi,t_n}}{\partial 
u}(u^k)$, i.e. the derivative of the signal with respect to each unknown,
 can be precomputed in each linearization step. The additional 
weighted $L^2$-norm penalty $\|u-u_k\|_{M_k}^2=\|M_k^{\frac{1}{2}}(u-u_k)\|_2^2$ 
improves convexity of the function and resembles a Levenberg-Marquadt update 
if the weight matrix $M$ is chosen as 
$M_k=diag(\mathrm{D}A_{\phi,t_n}\rvert_{u=u^{k}}^T
\mathrm{D}A_{\phi,t_n}\rvert_{u=u^{k}})$ or a 
simpler Levenberg update if $M_k$ is chosen as identity matrix. It was shown 
by~\citet{Salzo2012} that the GN approach converges with linear rate to a 
critical point for non-convex problems with non-differential penalty functions 
if the initialization is sufficiently close. By exploiting the Fenchel duality 
it is possible to transform the problem in equation~\ref{eq:linearized} into a 
saddle-point form

\begin{linenomath*}
\begin{equation}\label{eq:PD}
\underset{u}{\min}\,\underset{y}{\max}~ \left<\mathrm{K}u,y\right> + G(u) - 
F^*(y),
\end{equation}
\end{linenomath*}
which overcomes the non-differentiability issue 
of the $L^1$-terms. Problems of the form in~(\ref{eq:PD}) can be efficiently solved 
using a first order primal-dual splitting algorithm~\citep{Chambolle2011} in 
combination with a line search~\citep{Malitsky2018} to improve the 
convergence speed. The detailed derivation is given in the Appendix A. 
Pseudo-code for the implementation can be found in Appendix B.
\subsection{Reference Methods}
For comparison of the proposed algorithm we used the non-linear least squares 
(NLLS) as well as the Bayesian Inference for Arterial Spin Labeling MRI 
(BASIL) method~\citep{Chappell2009, Groves2009}. The NLLS method 
solves equation~\ref{eq:stat} without regularization by means of a 
trust-region reflective method implemented by \textit{lsqnonlin} in 
MATLAB (The Mathworks, Natick, MA, USA). This method uses additional box 
constraints on CBF and 
ATT to limit their values to a physiologically meaningful range of [0, 300] 
ml/100g/min for CBF
and [0, 6] seconds for ATT respectively.  
BASIL is included in FSL~\citep{Smith2004, Woolrich2009, Jenkinson2012} and uses
 Bayesian inference to estimate the unknown parameter maps. It incorporates 
fixed non-spatial priors as well as adaptive non-local spatial smoothing 
priors for the parameters. The spatial smoothing prior is used for CBF
and is directly based on evidence in the data. The smoothing strength is 
adjusted based on the local support in the specific area in the data. 
The arterial (macro-vascular) contribution flag was set to "OFF"
in BASIL to facilitate comparability to the proposed method which currently 
implements the pCASL model omitting the local arterial contribution. 
\responseRone{R1.C1}{In addition to this standard form of BASIL, termed \textit{BASIL w/o}, a simple duplication of
the line associated with spatial priors in the starting script enables priors on both, CBF and ATT, as described in~\citep{Chappell2009, Groves2009}. 
This modification serves as second BASIL reference and is termed \textit{BASIL w/}.}
% --------------------------------------------------------------------
% Methods
% --------------------------------------------------------------------
\section{Methods}
\label{sec:methods}
\subsection{Synthetic ASL data}
\subsubsection{Phantom generation}
To \responseRone{R1.C7}{\mdeleted{validate}evaluate} the proposed method, synthetic ASL data was generated from brain 
$T_1$ and PD maps supplied by MRiLab~\citep{Liu2017} for MATLAB with a 
matrix size of $216\times 180\times 180$ and 
1 mm$^3$ isotropic resolution. Gray (GM) and white matter (WM) CBF values of 65 
ml/100g/min and 20 ml/100g/min as well as ATT values of 0.8 s and 1.5 s, 
reported for the healthy human brain were assigned to the tissue maps. 
In a subsequent step the quantitative maps were down sampled onto a 
$72\times 60 \times 60$ grid, matching 
typical matrix sizes and resolution (3 mm$^3$ isotropic) of 3D ASL 
acquisitions. \responseRtwo{R2.C2}{A small linear phase variation of -0.5 to 0.5 rad going from anterior to posterior was added to the $M_0$ image to obtain
a complex valued image}. The signal equation~(\ref{eq:ASL}) was used to generate a series 
of PWI images. The control images $C\in D^{N_d}$ were assumed to correspond to 
\responseRtwo{R2.C11}{$C = (1-\alpha)M_0$\mdeleted{$C = (1-\lambda)M_0$}}, which models the background suppression applied to $M_0$. 
The label images $L\in D^{N_d}$ were simply given by $L = C - \text{PWI}$. \responseRtwo{R2.C2}{3-D coil sensitivity profiles
 were computed using Biot-Savart's law. The 20 coils were placed equally spaced on a spherical surface. The coil sensitivity maps were multiplied with the complex control- and label-images. The resulting weighted images were then transformed into k-space via Fourier transformation
\responseRfour{R4.C4}{ and zero mean complex Gaussian noise was added to each k-space.} 
\responseRfour{R4.C7}{The standard deviation of the added Gaussian noise was
0.65 which results approximately in an SNR of 4 in WM and 6 in GM (assuming a LD of 1.8s and a PLD of 1.75s).}
Noisy complex control and label images were then computed using a phase sensitive
reconstruction based on~\citet{Roemer1990}, using the complex conjugate coil images.\mdeleted{Complex Gaussian noise was added to each control and label image 
separately to simulate the MR imaging acquisition.}} The noisy 
control and label pairs constituted the final PWI sequence. \responseRtwo{R2.C2}{For our approach
the complex valued PWIs\mdeleted{which} were used as 
input for the fitting process. For BASIL, the PWIs were calculated by subtracting the absolute valued
label image form the control image as BASIL does not support complex valued images.} In total $N_d=32$ time points in two series, 
each containing 16 time points, 
were simulated with the parameters of the ASL model amounting to  
$\lambda=0.9$, $T_{1b}=1650$ ms, $\tau = (1050, 1300, 1550, 1800, \hdots, 
1800)$ ms, $t = (1050:250:4800)$ ms, \responseRthree{R2.C14}{$\alpha = 0.7$}. $M_0$ and $T_1$ 
amounted to the down sampled values supplied by the MRiLab phantom.
In addition to the healthy brain phantoms of MRiLab we simulated perfusion 
changes in GM and WM. In total, \responseRone{R2.C2}{\mdeleted{three}six} cases, denoted by Case 1 to \responseRone{R2.C2}{\mdeleted{C3}Case 6, were generated. It should be noted that all pathologies were drawn in the high-resolution space to get smoother transitions in the ASL space. \mdeleted{, and
realizing t} For Case 1 to Case 3 the following pathologies in a region in frontal WM and in a part of the putamen (GM) were \mdeleted{generated}simulated as illustrated in Figure~\ref{fig:fig1}}:
\begin{itemize}
 \item [Case 1:] No changes to values, i.e. healthy.
 \item [Case 2:] Hyperperfusion \responseRfour{R4.C5}{of 113.75 ml/100g/min in GM and 50 ml/100g/min in WM} without changes in corresponding areas in ATT.
 \item [Case 3:] Hyperperfusion \responseRfour{R4.C5}{of 113.75 ml/100g/min in GM and 40 ml/100g/min in WM} with corresponding reduced arterial transit time \responseRfour{R4.C5}{of 0.4 s in GM and 0.75 s in WM}.
\end{itemize}

\responseRone{R2.C2}{In addition we simulated cases with normal CBF and increased ATT by a partly-blockage of the arteria cerebri media and a small region in the frontal lobe illustrated in Figure~\ref{fig:fig1_2}:}

\begin{itemize}
\responseRone{R2.C2}{ 
 \item [Case 4:] Normal perfusion with increased ATT of 2.75 s.
 \item [Case 5:] Normal perfusion with increased ATT of 3 s.
 \item [Case 6:] Normal perfusion with increased ATT of 3.25 s.}
\end{itemize}

\responseEditor{E.C1}{To assess the influence of joint regularization the proposed method
was run without regularization on ATT (\textit{Proposed w/o}) by simply setting the finite differences based image gradient of ATT to zero in all iterations. The joint regularization was denoted as \textit{Proposed w/}.
The CBF and ATT estimates of the individual fitting algorithms were compared
to each other based on visual inspection, pixel-wise absolute difference plots, and evaluation of
median and inter quartile range (IQR) in the simulated ROIs, i.e. median and IQR of GM, WM, and
each simulated lesion. In addition, 2D histograms of CBF and ATT for each method were generated by 
combining all simulated cases into one distribution of CBF and ATT, respectively.}

\responseEditor{E.C2}{
\subsubsection{Error propagation and stability}
To asses the error propagation and stability due to the non-linear fitting 
procedure we performed a pseudo replica analysis for all three methods. To this 
end, 100 different noise realizations with \responseRfour{R4.C7}{\mdeleted{the same standard 
deviation}a standard deviation of 0.65} were simulated for Case 3. Due to the non-linear fitting 
process a Gaussian noise assumption in the parameter maps could be violated, 
thus the median and inter-quartile range between the 25$^{th}$ and 
75$^{th}$ quartile were used for evaluation. \responseRone{R1.C2}{\mdeleted{In addition we performed a 
box-plot based analysis of possible biases in the medians towards the ground truth values of GM and WM.} 
We assessed potential biases using the medians of differences to the ground truth of the 100 realisations and} \responseRone{R1.C2}{compared the differences in the IQRs between the methods.
For the synthetic dataset GM and WM binary masks, generated on the ground 
truth phantom, are employed.\responseRone{R1.C6}{\mdeleted{Tissue is assumed to belong to WM if reference CBF 
values are within [15, 30] ml/100g/min and assumed to belong to GM if CBF is 
 within [55, 65] ml/100g/min.} Based on the down sampled GM and WM mask, low resolution mask were generated by
thresholding the corresponding GM/WM masks with 0.7. Mask for the simulated lesions were generated in analogy.} 
\responseRone{R1.C6}{\responseRthree{R3.C20}{\mdeleted{For the in vivo data the GM and WM binary mask were used for evaluation\mdeleted{segmented and 
co-registered GM and WM mask were thresholded by 0.9 for the evaluation.}}}

Additionally, we compared the estimated CBF and ATT maps of the proposed method 
with the results of BASIL and NLLS by means of a relative difference to the 
numerical ground truth parameter maps. }}

\responseRone{R1.C2}{To assess if differences in median or IQR are statistically significant, we applied Mann-Whitney-U tests, as medians and IQR distributions may deviate from a normal distribution. 
The medians and IQRs were computed in ROIs for the 100 reported noise realizations. Medians were then compared to the median of the noise free reference. 
For the IQRs the proposed method was tested against the other methods to show
if statistically significant differences were observable. p-values were adjusted for multiple comparisons using the method of~\citet{BONFERRONI1936}.
Each method and tissue was considered as parallel test case. Median and IQR of CBF and ATT were considered as separate cases.
 Results were considered statistically significant for p-values less than 0.05 (p$<$0.05).}}

\subsection{In vivo measurements}
All measurements were performed on a 3T MAGNETOM Prisma (Siemens Healthcare, 
Erlangen, Germany) system using a 20-channel head coil. 
Written informed consent was obtained by all healthy volunteers as 
well as by all patients following the local ethics committee's regulations. 
In total, \responseRfour{R4}{\mdeleted{three}six} healthy volunteers\responseRfour{R4.C6}{, consisting of five male and one female subject with an age of 29.5$\pm$2.6 years were analyzed.}\responseRfour{R4}{\mdeleted{ and three}Additionally, seven} patients with ischemic stroke due 
to middle cerebral artery occlusion who received successful re-canalization 
therapy (i.e.  intravenous thrombolysis followed by mechanical thrombectomy)\responseRfour{R4.C6}{, consisting of six male and one female subject with an age of 57.1$\pm$13 years,} 
were considered. Patient data was acquired 24 hours after recanalization therapy. 
ASL images were acquired using a prototype 3D pCASL sequence with a 2D CAIPIRINHA
 accelerated single-shot 3D GRASE 
readout \responseRtwo{R2.C12}{\citep{Dimo2017}} and two background suppression pulses \responseRtwo{R2.C12}{~\mdeleted{\mbox{\citep{Dimo2017}}}\citep{Vidorreta2013}}. Labeling 
efficiency for this sequence \responseRtwo{R2.C12}{\mdeleted{amounts to} was experimentally determined in~\citet{Vidorreta2013} and assumed as} $\alpha=0.7$.
The following image parameters were used: FOV = 192 x 192 x 114 mm$^3$, matrix 
= 
64x64x38 resulting in 3 mm$^3$ isotropic resolution, 10\% phase and 17.5\% 
slice oversampling, TR = 5260 ms, TE = 14.44 ms, 2x2$^{(1)}$ CAIPIRINHA scheme,
phase-partial Fourier 6/8, refocusing FA = 180$^\circ$, EPI-factor = 25, 
turbo-factor (TF) = 22, resulting in one segment. 16 time points were acquired 
with a labeling duration of 
$\tau = (1050, 1300, 1550, 1800, \hdots, 1800)$ ms and a PLD of $(0, 0, 
0, 0:250:3000)$ ms. \responseRthree{R3.C19}{The PLDs were acquired in sequential order and the sequence of PLDs 
was repeated four times. \mdeleted{Four averages per PLD and}Additionally,} one proton density weighted image
 was acquired for each healthy subject, resulting in an acquisition time of 
11 min 29 s.  Due to time restrictions 
only two averages per PLD were acquired for the patients (5 min 53 s). The 
ASL labeling plane was placed according to a time-of-flight angiography in the neck 
area above the bifurcation of the carotid artery. 

Additionally, for each healthy subject a $T_1$ weighted image was acquired using
 a 3D-MPRAGE sequence with the following imaging parameters:
1 mm$^3$ isotropic resolution, 176 slices, TR = 1900 ms, TE = 2.7 ms, TI = 900 
ms, flip angle = 9$^\circ$, acquisition time = 5 min 58 s.

\responseEditor{E.C2}{\mdeleted{\textit{Data Processing and correction}}}

\subsubsection{ASL Data Processing}

The accelerated ASL images were reconstructed directly on the scanner console 
by means of a prototype reconstruction pipeline provided by the vendor.
 The reconstructed ASL images were 
motion corrected using Statistical Parameter Mapping 
(SPM)12\footnote{\url{https://www.fil.ion.ucl.ac.uk/spm/software/spm12/}} 
(Wellcome Trust Centre for Neuroimaging, University College London, UK) 
\citep{Friston2007} and the ASL-Toolbox~\citep{Wang2008,Wang2012}. \responseRone{R1.C13}{This rigid-body based motion correction process involved three sub-steps as described in~\citet{Wang2012}.} After reconstruction and motion correction the perfusion weighted time series were calculated. From this perfusion weighted 
time-series the CBF and ATT maps were estimated using the proposed method as 
well as the two reference methods. The fixed parameters $\phi$ amount to the
same values as in the synthetic data set except for \responseRtwo{R2.C14}{\mdeleted{$\alpha=0.7$ and}} $T_1$=1330 
ms, the approximate tissue $T_1$ relaxation constant.

\subsubsection{Anatomical Image Processing}

\responseRone{R1.C6}{
For each healthy subject brain masks and PV estimates for GM and WM were computed using FSL (FMRIB Software Library, Oxford, UK~\citep{Jenkinson2012}) and BASIL. In a first step, non-brain tissue was removed from the high resolution structural ($T_1$ weighted) image using the FSL tool \textit{BET}~\citep{smith_2002}. In a second step, PV estimates for GM and WM were obtained from the T1w image using the FSL tool \textit{FAST}~\citep{Zhang2001}. Third, the structural image and brain mask were registered to the mean ASL image using the FSL tool \textit{FLIRT} with 6 degrees of freedom~\citep{JENKINSON2001143,JENKINSON2002825}. The obtained transformation matrix served as initial guess for the next registration refinement step, implemented in BASIL. This step used the \textit{epi\_reg} tool of FSL for boundary based registration of the perfusion image with the segmented white matter mask~\citep{GREVE200963}. In the last step, the PV estimates for GM and WM were transformed to the ASL-space by a process that integrates over the volume of the low resolution voxels as described by~\citet{Chappell2011} and implemented in the FSL tool \textit{applywrap}. Finally GM and WM binary masks in ASL space were computed by thresholding the PV estimates at 70\% in WM and GM respectively. For the patient data brain masks were generated from the $M_0$ image using the FSL tool BET due to the missing $T_1$ weighted image. 
\mdeleted{the high resolution $T_1$ weighted images were 
segmented into GM and WM using SPM12 and CAT12 
 toolbox \mbox{footnote{\url{http://www.neuro.uni-jena.de/cat/}}} (C. 
Gaser, Structural Brain Mapping Group, Jena University Hospital, Jena, 
Germany). 
The high resolution $T_1$ image and the segmented GM- and WM-maps were 
co-registered to the mean PWI as suggested by. From the 
co-registered GM- and WM-maps a brain mask was generated for each subject. 
This was achieved by summing up the corresponding GM- and WM-maps followed 
by a 3D dilation with a cubic kernel element of size 3x3x3.}}

\responseEditor{E.C2}{
\subsubsection{Method Comparison}
Healthy subjects were compared based on visual inspection of ATT and CBF for 1 and 4 acquired averages.
In addition, WM and GM masks were used to compute median and IQR which were visualized with box-plots. 
Statistically significant differences in median and IQR between methods were assessed using Mann-Whitney-U tests, similar to the ones in the numerical simulation. p-values were adjusted for multiple comparisons~\citep{BONFERRONI1936}. Results were considered statistically significant for p-values less than 0.05 (p$<$0.05).

Stroke patients are compared based on visual inspection only.}
\subsection{Parameter optimization}
To identify a good set of model and regularization parameters a grid search was 
performed on the synthetic dataset \responseEditor{E.C0}{and in vivo healthy subjects}. The resulting regularization parameters 
amounted to \responseEditor{E.C0}{\mdeleted{$\gamma_{init}=10^{-4}$}$\gamma_{init}=10^{-3}$} and \responseEditor{E.C0}{\mdeleted{$\delta_{init}=10^{-2}$}$\delta_{init}=1$} which were 
reduced by 0.5 and 0.1 respectively after each Gauss-Newton step down to 
\responseEditor{E.C0}{\mdeleted{$\gamma_{final}=2\cdot 10^{-6}$}$\gamma_{final}=6.5\cdot 10^{-6}$ and \mdeleted{$\delta_{final}=10^{-8}$}$\delta_{final}=10^{-2}$}. A reduction of 
regularization parameters was observed to be beneficial for overall 
convergence in IRGN methods~\citep{Bakushinsky2004, Barbara2008, Blaschke1997}. 
Relative tolerance for convergence was set to $10^{-8}$ between consecutive 
evaluations of function value. Regarding the inner iterations, 50 were used in 
the initial Gauss-Newton step and the number was increased by a factor of two \responseRtwo{R2.C15}{until the maximum allowed number of 1000 iterations is reached \mdeleted{ up
 to 1000}}, i.e. $iter^k=\min\{50*2^k, 1000\}$\mdeleted{{50, 100, 200, ..., 1000}}. A total of ten Gauss-Newton steps were
 performed. The ratio of the model parameters $\beta_0/\beta_1=1/2$ of TGV$^2$
 was chosen according to~\citet{Knoll2011}.
\responseEditor{E.C0}{\mdeleted{These parameter settings were used for all parameter map estimations from 
synthetic and in vivo data throughout this work.} For in vivo data, the TGV$^2$ related weight $\gamma_{final}=3\cdot 10^{-5}$ was used.}

\responseEditor{E.C2}{\mdeleted{
\textit{Error propagation and stability}\\
To asses the error propagation and stability due to the non-linear fitting 
procedure we performed a pseudo replica analysis for all three methods. To this 
end, 100 different noise realizations with the same standard 
deviation were simulated for C3. Due to the non-linear fitting 
process a Gaussian noise assumption in the parameter maps could be violated, 
thus the median and inter-quartile range between the 25$^{th}$ and 
75$^{th}$ quartile were used for evaluation. In addition we performed a 
box-plot 
based analysis of possible biases in the medians towards the ground truth values of GM and WM. 
For the synthetic dataset GM and WM binary masks generated on the ground 
truth phantom are employed. Tissue is assumed to belong to WM if reference CBF 
values are within [15, 30] ml/100g/min and assumed to belong to GM if CBF is 
 within [55, 65] ml/100g/min.
For the in vivo data the segmented and 
co-registered GM and WM mask were thresholded by 0.9 for the evaluation.
Additionally, we compared the estimated CBF and ATT maps of the proposed method 
with the results of BASIL and NLLS by means of a relative difference to the 
numerical ground truth parameter maps.}}

\subsection{Implementation}
The proposed method is implemented in Python 3.7 (Python Software Foundation, 
\url{https://www.python.org/}) with 
OpenCL~\citep{Stone2010, Klockner2012a} based on a recently proposed 
quantitative 
MRI framework~\citep{Maier2019c, Maier2019d, Maier2020} which utilizes GPU acceleration. 
Evaluation was done using Python 3.7 
with NumPy 1.17.4 and SciPy 1.3.2. Computations were performed on a GPU 
server running Ubuntu 18.04, equipped with four Nvidia Titan XP cards (Nvidia 
Corporation, Santa Clara, CA, USA) and an Intel Xeon Gold 6136 CPU (Intel 
Corporation, Santa Clara, CA, USA) running at 3 GHz with 503 GB of RAM. \\
The fitting code is available at \\
\url{https://github.com/IMTtugraz/PyQMRI} \\
and exemplary data as well as the used 
configuration files for the optimization can be downloaded from \\
\url{https://doi.org/10.5281/zenodo.4493854}. \\
Scripts to generate the figures of this paper are available at \\
\url{https://doi.org/10.5281/zenodo.4494236}.
% --------------------------------------------------------------------
% Results
% --------------------------------------------------------------------
\section{Results}
\label{sec:results}
\begin{figure*}[!htbp]
\centering
 \includegraphics[width=0.95\textwidth, height=0.85\textheight, keepaspectratio]{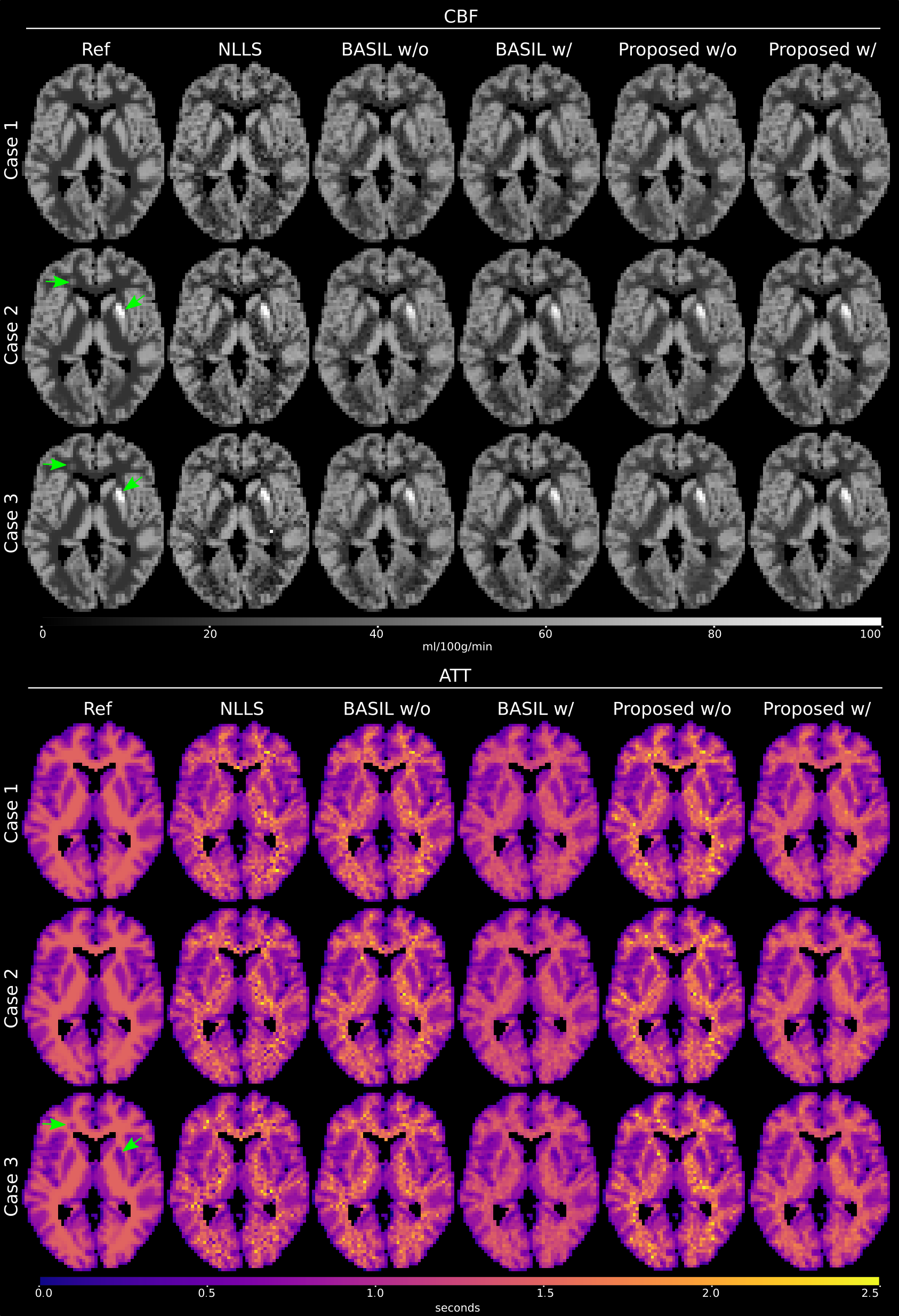}
  \responseRtwo{R2.C2}{\caption{CBF and ATT maps of synthetic phantoms for three different cases, one in each row 
(Case 1 - Case 3, top down). 
Case 1 has no simulated pathologies, Case 2 shows hyperperfusion in CBF only and 
Case 3 hyperperfusion in CBF and increased ATT in the corresponding area. 
In the first column the numerical ground truth is shown and in the following 
columns the estimated CBF- and ATT-maps from NLLS, BASIL without regularization 
on ATT (\textit{BASIL w/o}), and with regularization on ATT (\textit{BASIL w/}), and the proposed 
method without and with regularization on ATT, respectively. 
The proposed method with regularization on both unknown maps shows improved noise removal in CBF 
and ATT compared to the other methods due to joint spatial constraints.
Median and 25\% - 75\% IQR for selected ROIs is given in table~\ref{tab:tab1}\label{fig:fig1}}}
\end{figure*}
\begin{figure*}[!htbp]
\centering
 \includegraphics[width=0.95\textwidth, height=0.85\textheight, keepaspectratio]{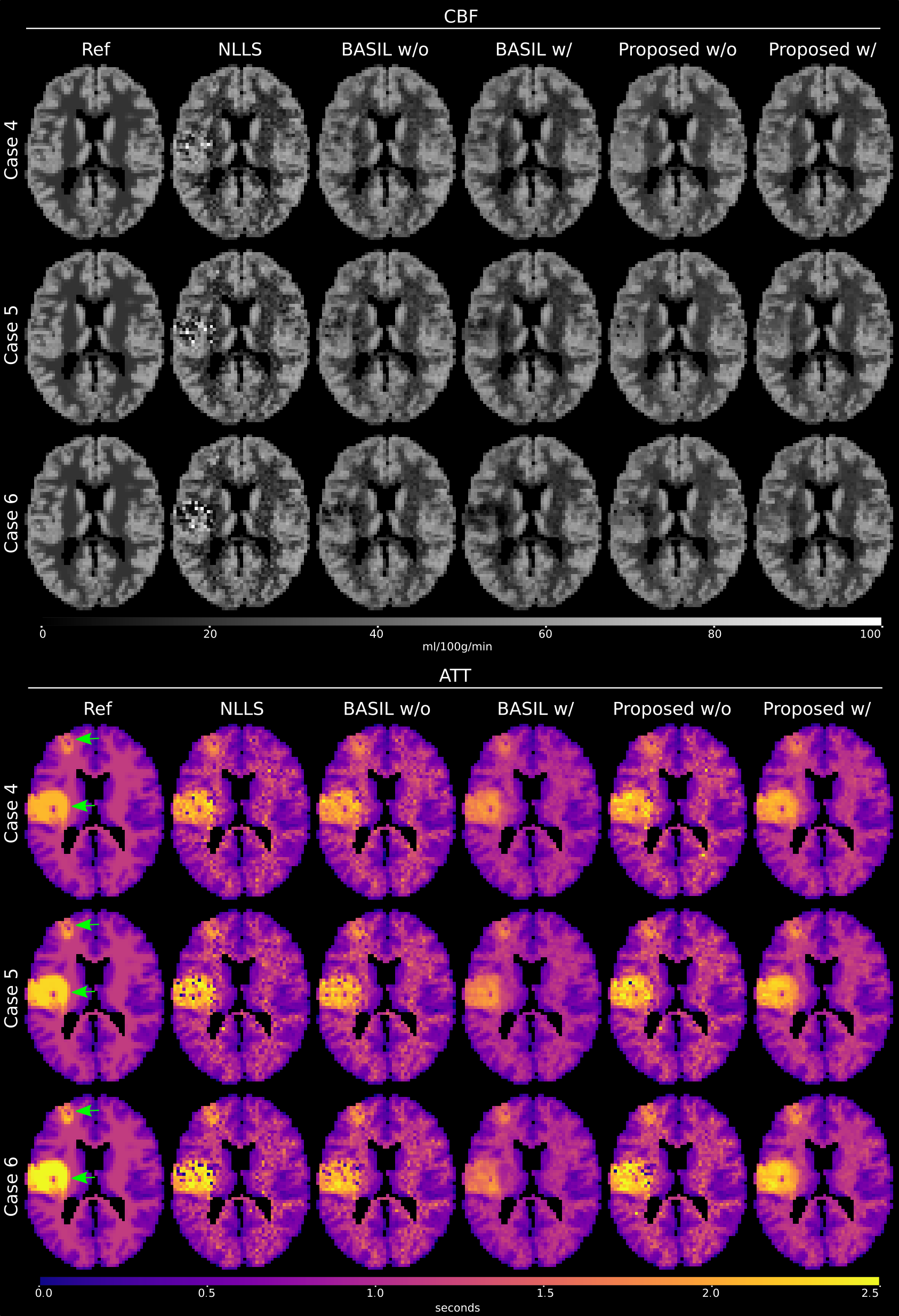}
 \responseRtwo{R2.C2}{\caption{CBF and ATT maps of synthetic phantoms
of three cases with pathologies are shown. Each case represents a large partly occlusion of the arteria media combined with a small partly occlusion in frontal gray matter. No variation in CBF is simulated but each case shows an increase in ATT, which gets more severe from Case 4 to Case 6.  
The order of reference and displayed reconstruction algorithms is the same as in figure~\ref{fig:fig1}.
The proposed method shows the least influence of the highly increased ATT on the CBF estimates
and is able to recover higher ATT values in the affected areas than the other methods,
as can be seen in table~\ref{tab:tab1}.\label{fig:fig1_2}}}
\end{figure*}

\begin{figure*}[!htbp]
\centering
 \includegraphics[width=0.95\textwidth, keepaspectratio]{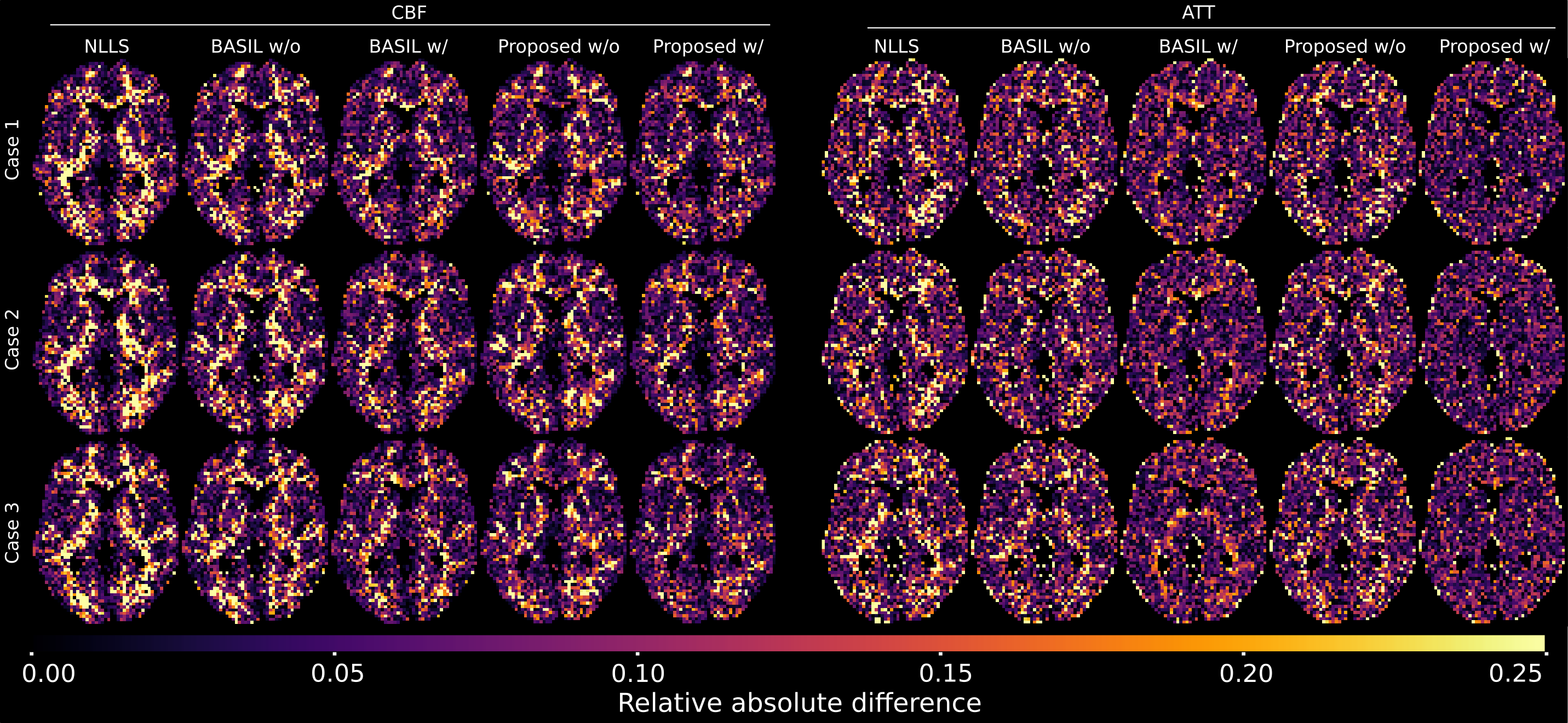}
 \responseRthree{R3.C23}{\caption{Pixel-wise relative absolute difference between the ground truth numerical 
reference and the quantitative maps, estimated with the algorithms of 
figure~\ref{fig:fig1}. The NLLS shows the greatest deviation in ATT and CBF. 
\textit{BASIL w/o} reduces the relative difference in the CBF due to the spatial prior and
\textit{BASIL w/} is able to reduce deviations even further. The \textit{proposed w/o} method shows 
similar results on CBF and ATT as \textit{BASIL w/o}. The least relative 
difference is achieved with the proposed method due to joint spatial constraints on CBF and ATT 
simultaneously.\label{fig:fig2}}}
\end{figure*}

\begin{figure*}[!htbp]
\centering
 \includegraphics[width=0.95\textwidth, keepaspectratio]{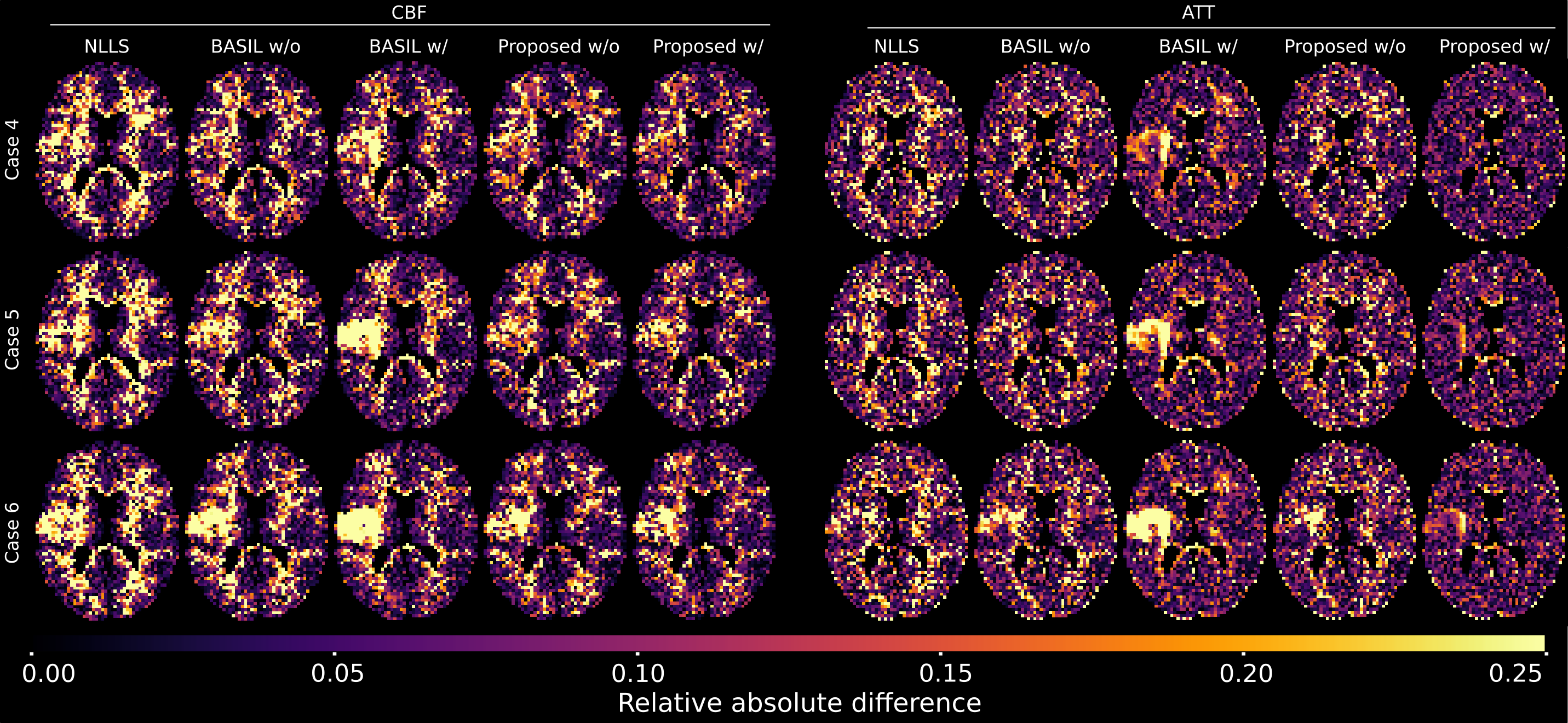}
 \responseRthree{R3.C23}{\caption{Pixel-wise relative absolute difference between the ground truth numerical 
reference and the quantitative maps, estimated with the algorithms of 
figure~\ref{fig:fig1_2}. NLLS shows large deviation across the brain and in the
simulated pathologies. \textit{BASIL w/o} and \textit{BASIL w/} can reduce this variation in normal
appearing white and gray matter but variations in the large effected area remain. 
These variations are especially pronounced for \textit{BASIL w/}. The proposed method
is able to reduce variations in normal tissue and in addition shows smaller
deviations in the pathological area, especially visible in ATT. Again, the proposed
algorithm with regularization on CBF and ATT shows the least deviations overall. \label{fig:fig2_2}}}
\end{figure*}
\begin{figure}[!htbp]
\centering
 \includegraphics[width=0.95\columnwidth, height=0.85\textheight, keepaspectratio]{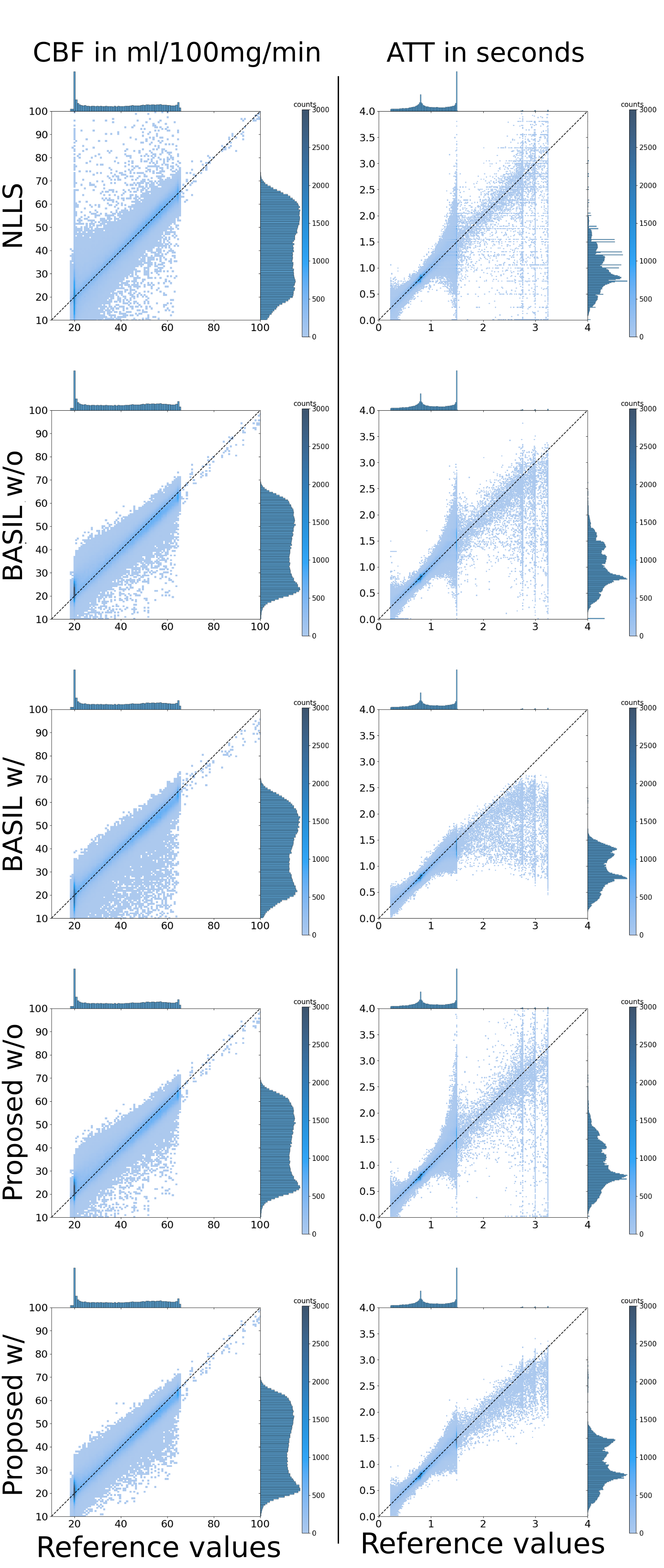}
 \responseRtwo{R2.C3}{\caption{
2D histograms between reference values and fitted values for all methods. The dashed line represents identity. Points below correspond to underestimation and points above to overestimation compared
to the reference value. Pixel values from all six cases are combined into one 2D histogram for each method.}
 \label{fig:fig2_3}}
\end{figure}
\responseRone{R1.C8}{\mdeleted{Throughout this chapter we will show the advantage of posing a joint spatial 
TGV$^2$ regularization strategy on the CBF- and ATT-maps, 
directly embedded in the fitting process, compared to two commonly used 
quantification strategies for multi-delay ASL data.}
We compare the fitting quality of our proposed joint spatial
TGV$^2$ regularization strategy to established quantification on the CBF- and ATT-maps at three levels: }
First, synthetic phantom 
simulations demonstrate the accuracy and precision of NLLS, BASIL, and the 
proposed fit by means of a comparison to the ground truth values. This is 
further supported by a pseudo replica method to evaluate noise propagation into 
the final results. \responseRone{R1.C8}{Violin-plots show the median and IQR distributions of estimated quantitative values and statistical tests assess the significance of differences between the methods.} Second, the fitting 
algorithms are applied to in vivo ASL data acquired from six healthy 
volunteers. Results are compared visually and quantitatively by means of 
box-plots. Third, quantification of CBF and ATT is performed in patients 
following a stroke. \responseEditor{E.C0}{\mdeleted{All results are produced with a fixed set of parameters for 
the fitting algorithms to facilitate a fair comparison.}}

\subsection{Synthetic phantom simulations}

\begin{table*}[!t]
\centering
\responseRthree{R3.C4}{\caption{Median and (25\%-75\%) IQR 
for the simulated phantom data. Each ROI is based on the ground truth mask which was used for generating the lesion. w/o and w/ refers to regularization on CBF only and regularization on CBF and ATT, respectively. For cases 4-6, evaluation is performed only in the affected areas as normal appearing white and gray matter will be similar to cases 1-3. CBF values are given in ml/100g/min and ATT values in seconds.\label{tab:tab1}}}
\rowcolors{1}{gray!15}{white}
\resizebox{\textwidth}{!}{
\begin{tabular}{llrlrlrlrlrlrl}
\rowcolor{gray!35}
& Method &       \multicolumn{2}{c}{Ref} &      \multicolumn{2}{c}{NLLS} &     \multicolumn{2}{c}{BASIL w/o} &      \multicolumn{2}{c}{BASIL w/} &        \multicolumn{2}{c}{Proposed w/o} &  \multicolumn{2}{c}{Proposed w/}\\
\rowcolor{gray!35}
& Tissue & & & & & & & & & & & &\\
\midrule
\cellcolor{white}& WM CBF      & 21.79& (14.91-28.68) &  23.72& (12.95-34.48) &  25.74& (16.31-35.17) &  23.29& (14.29-32.28) &  25.68& (17.91-33.45) &  24.16& (16.55-31.76) \\
\cellcolor{white}& WM ATT      &  1.47& (1.36-1.58) &     1.39& (1.06-1.72) &     1.44& (1.14-1.74) &     1.32& (1.17-1.48) &     1.47& (1.16-1.78) &     1.42& (1.26-1.58) \\
\cellcolor{white}& GM CBF      & 57.12& (48.9-65.34) &  57.24& (48.62-65.86) &  55.81& (47.64-63.97) &   55.31& (47.3-63.32) &  55.74& (47.55-63.94) &  55.83& (47.64-64.02) \\
\multirow{-4}{*}{\cellcolor{white}Case 1} & GM ATT      &    0.8& (0.66-0.94) &     0.79& (0.64-0.95) &     0.78& (0.63-0.93) &     0.78& (0.63-0.93) &     0.78& (0.63-0.93) &     0.78& (0.63-0.93) \\
\midrule
\cellcolor{white} & WM lesion CBF&  45.51& (42.89-48.13) &    45.27& (36.04-54.49) &   43.02& (37.76-48.28) &   40.39& (36.6-44.19) &     39.53& (34.06-45.0) &    39.96& (36.42-43.5) \\
\cellcolor{white} & WM lesion ATT&  1.5& (1.5-1.5) &       1.53& (1.46-1.61) &       1.5& (1.39-1.61) &      1.47& (1.4-1.53) &       1.46& (1.35-1.57) &      1.47& (1.41-1.54) \\
\cellcolor{white} & GM lesion CBF&  105.81& (97.02-114.59) &  104.41& (92.69-116.13) &  97.51& (88.41-106.62) &  96.44& (86.8-106.08) &  101.03& (89.51-112.55) &  101.01& (89.5-112.52) \\
\multirow{-4}{*}{\cellcolor{white}Case 2} & GM lesion ATT &  0.8& (0.8-0.8) &       0.81& (0.77-0.85) &      0.77& (0.73-0.81) &      0.77& (0.73-0.8) &        0.8& (0.76-0.83) &       0.8& (0.76-0.83) \\
\midrule
\cellcolor{white} & WM lesion CBF &    37.01& (35.26-38.75) &    38.52& (35.95-41.08) &   38.27& (36.39-40.14) &  37.48& (34.34-40.61) &    35.59& (33.95-37.24) &     36.6& (35.26-37.94) \\
\cellcolor{white} & WM lesion ATT &    0.86& (0.8-0.93) &       0.94& (0.77-1.11) &      0.94& (0.74-1.14) &     0.93& (0.77-1.08) &       0.82& (0.67-0.97) &         0.85& (0.7-1.0) \\
\cellcolor{white} & GM lesion CBF &    105.81& (97.02-114.59) &  105.16& (93.12-117.21) &  100.04& (89.28-110.8) &  99.4& (88.66-110.15) &  102.85& (92.81-112.89) &  102.86& (92.88-112.85) \\
\multirow{-4}{*}{\cellcolor{white}Case 3} & GM lesion ATT &   0.47& (0.39-0.54) &        0.46& (0.4-0.52) &      0.45& (0.37-0.52) &     0.44& (0.38-0.51) &       0.45& (0.39-0.52) &       0.45& (0.38-0.52) \\
\midrule
\cellcolor{white}&Large Stroke WM CBF &  20.12& (16.17-24.06) &   14.0& (-6.07-34.07) &    17.0& (3.08-30.92) &    9.44& (-2.62-21.5) &   20.42& (7.42-33.42) &  22.23& (12.92-31.54) \\
\cellcolor{white}&Large Stroke WM ATT &    2.75& (2.73-2.77) &        2.3& (1.0-3.6) &     2.14& (1.31-2.96) &     1.64& (1.03-2.26) &     2.49& (1.64-3.35) &     2.53& (2.24-2.83) \\
\cellcolor{white}&Large Stroke GM CBF &  55.33& (49.22-61.45) &   54.18& (39.67-68.7) &  47.26& (40.29-54.22) &  42.01& (34.99-49.03) &  47.67& (42.18-53.15) &  47.73& (41.41-54.06) \\
\cellcolor{white}&Large Stroke GM ATT &    2.55& (2.18-2.92) &     2.49& (2.05-2.92) &     2.42& (2.07-2.76) &     2.26& (1.98-2.54) &     2.43& (2.07-2.79) &     2.43& (2.14-2.72) \\
\cellcolor{white}&Small stroke GM CBF &  55.34& (48.07-62.6) &  53.68& (42.47-64.89) &   48.74& (41.27-56.2) &   44.32& (39.14-49.5) &  49.09& (42.37-55.81) &  48.74& (42.63-54.85) \\
\multirow{-6}{*}{\cellcolor{white}Case 4}& Small stroke GM ATT &     2.36& (2.05-2.68) &      2.3& (1.97-2.64) &      2.23& (1.96-2.5) &     2.14& (1.93-2.35) &      2.2& (1.95-2.45) &      2.19& (1.99-2.4) \\
\midrule
\cellcolor{white}&Large Stroke WM CBF &   20.12& (16.17-24.06) &   9.81& (-8.44-28.07) &  11.68& (-1.73-25.09) &     5.9& (-3.41-15.2) &  15.33& (-0.65-31.32) &   20.52& (9.32-31.72) \\
\cellcolor{white}&Large Stroke WM ATT &       3.0& (2.97-3.03) &      2.3& (0.58-4.02) &      1.95& (0.99-2.9) &      1.44& (0.7-2.18) &     2.42& (1.14-3.71) &     2.63& (2.24-3.01)  \\
\cellcolor{white}&Large Stroke GM CBF &  55.33& (49.22-61.45) &  54.23& (36.09-72.38) &  44.28& (36.49-52.07) &  36.41& (27.34-45.48) &  45.33& (39.62-51.04) &  45.75& (38.77-52.73) \\
\cellcolor{white}&Large Stroke GM ATT &      2.78& (2.37-3.19) &      2.7& (2.22-3.18) &     2.59& (2.23-2.95) &      2.33& (2.05-2.6) &     2.62& (2.26-2.98) &     2.62& (2.32-2.91) \\
\cellcolor{white}&Small stroke GM CBF &  55.34& (48.07-62.6) &  52.17& (37.33-67.02) &  47.94& (43.06-52.82) &  41.07& (36.81-45.32) &  46.48& (41.25-51.71) &  46.74& (42.03-51.45) \\
\multirow{-4}{*}{\cellcolor{white}Case 5}& Small stroke GM ATT &      2.56& (2.21-2.92) &     2.48& (2.08-2.87) &     2.37& (1.98-2.77) &     2.22& (2.02-2.43) &      2.39& (2.0-2.78) &     2.35& (2.03-2.67)  \\
\midrule
\cellcolor{white}&Large Stroke WM CBF &  20.12& (16.17-24.06) &   6.88& (-5.98-19.75) &    6.91& (-2.88-16.7) &    2.92& (-3.16-8.99) &   8.37& (-2.31-19.05) &    15.68& (3.77-27.6) \\
\cellcolor{white}&Large Stroke WM ATT &    3.25& (3.22-3.28) &     2.25& (0.23-4.27) &     1.68& (0.79-2.56) &     1.17& (0.55-1.78) &     2.17& (0.29-4.05) &     2.57& (2.11-3.03) \\
\cellcolor{white}&Large Stroke GM CBF &  55.33& (49.22-61.45) &  50.61& (25.83-75.39) &  37.62& (27.57-47.67) &  26.08& (14.71-37.45) &  41.21& (33.75-48.67) &  40.37& (32.33-48.42) \\
\cellcolor{white}&Large Stroke GM ATT &    3.01& (2.56-3.46) &      2.81& (2.31-3.3) &     2.69& (2.35-3.03) &      2.22& (1.9-2.53) &      2.72& (2.35-3.1) &     2.75& (2.47-3.02) \\
\cellcolor{white}&Small stroke GM CBF &  55.34& (48.07-62.6) &  51.99& (31.45-72.53) &  45.59& (39.66-51.52) &  38.03& (29.86-46.21) &   42.97& (37.43-48.5) &  43.27& (36.83-49.71) \\
\multirow{-6}{*}{\cellcolor{white}Case 6} & Small stroke GM ATT &    2.76& (2.38-3.14) &      2.5& (2.05-2.95) &     2.51& (2.23-2.79) &     2.27& (2.07-2.47) &     2.46& (2.18-2.74) &     2.42& (2.23-2.62)\\
\bottomrule
\end{tabular}}
\end{table*}
All synthetic phantoms were simulated with the ground truth values of CBF and 
ATT given in the leftmost column in figures~\ref{fig:fig1}\responseRtwo{R2.C2}{~and~\ref{fig:fig1_2}}. Each row corresponds 
to a different realization of pathologies as previously described. Fits in 
the second column, representing NLLS, show the highest noise in both CBF and 
ATT. The spatial prior on CBF incorporated in BASIL is able to reduce outliers, 
especially in low signal areas such as WM tracts, as shown in the 
third \responseRone{R1.C1}{and fourth }row. The most efficient noise reduction in both, CBF and ATT, is achieved 
by the proposed joint regularization method \responseRone{R1.C5}{\textit{Proposed w/}}, which delivers comparable 
quantitative maps to the numerical ground truth. This visual impression is 
confirmed by the relative pixel wise differences to the ground truth in 
figures~\ref{fig:fig2} \responseRtwo{R2.C2}{and~\ref{fig:fig2_2}}. The lack of spatial information in the NLLS 
approach leads to strong variations of values in low signal areas, evident in 
the first column of figure~\ref{fig:fig2}. \responseRone{R1.C1}{\textit{BASIL w/o}} is able to reduce these 
outliers, showing overall smoother appearance of errors but \responseRone{R1.C1}{\mdeleted{comes at the cost 
of overestimating CBF values in WM tracts} also the highest deviation in median WM CBF (table~\ref{tab:tab1}).} However, for the estimated ATT-map 
both methods show similar relative difference. \responseRone{R1.C1}{Using spatial priors on CBF and ATT (\textit{BASIL w/}) reduces this variations but also seems to introduce a slightly lower value in ATT (table~\ref{tab:tab1}).}
\responseRone{R1.C5}{\mdeleted{The proposed method
 shows the lowest relative difference for both maps.} The proposed method without joint regularization (\textit{Proposed w/o}) shows reduced noise in CBF but similar results in ATT as \textit{BASIL w/o}. 
The highest reduction in noise in both maps could be achieved with joint regularization on CBF and ATT (\textit{Proposed w/}). 
This visual impression is also supported by the median and IQR values, evaluated for each tissue and presented in table~\ref{tab:tab1}. 
All methods show a slight overestimation of median CBF in WM. Approaches using spatial prior show a slight underestimation of median GM CBF. 
Similar, median WM ATT is slightly underestimated by all methods and \textit{BASIL w/} shows the strongest deviation. This deviation is also visually noticeable in figure~\ref{fig:fig1}. \textit{BASIL w/} and \textit{Proposed w/} are showing
the highest reduction of IQR in CBF and ATT compared to NLLS.
For small lesions (Case~2~and~3), regularized methods show slight deviations in the median but again lower IQR than NLLS. 
Of all regularized methods, the proposed joint regularization performs best in GM lesions for both cases and WM lesions in CBF and ATT (Case 3). 
WM lesions in CBF only (Case 2) show the least deviations with \textit{BASIL w/o}.}
\responseRone{R1.C5}{
For lesions with a severe increase in ATT but no variation of CBF (Case 4-6), the proposed method shows overall
the least deviations to median CBF and ATT in WM but starts to degenerate for Case 6. Other methods are not
able to recover WM CBF at all (Case 4-6).
In large GM lesions median CBF estimates of NLLS are closest to the reference
followed by the proposed method, but NLLS show an up to three fold increase in IQRs. 
Small GM lesions show similar results with \textit{BASIL w/o} being second closest to the reference, 
followed by the proposed method. 
ATT values are underestimated by all methods. \textit{Proposed w/} showing the least deviations to
WM ATT in the large effected area and second closest for GM with NLLS being closer to the true median in GM. 
Small stroke GM ATT shows the least deviations using NLLS, followed by methods that do not regularize
ATT.} 
\responseRtwo{R2.C3}{The differences to the reference method are visualized by means of 2D histograms in figure~\ref{fig:fig2_3}.
These histograms combine pixels of all six cases into one CBF and ATT plot per method. As seen in the pixel-wise difference maps in figure~\ref{fig:fig2} and figure~\ref{fig:fig2_2}, all regularized methods show a slight tendency to overestimate low CBF values and underestimate high CBF values but NLLS shows the highest deviations. \textit{BASIL w/o} is able to reduce these variance in CBF but shows minor underestimation of high CBF values and more severe underestimation of high ATT values. \textit{BASIL w/} shows even more underestimation of CBF and a pronounced underestimation of high ATT values. The proposed method w/o shows reduced variations in CBF and variations similar to NLLS in ATT. The least variations in CBF and ATT can be achieved with \textit{Proposed w/}, showing a minor underestimation of high CBF and ATT values.}
\begin{figure}[!htbp]
\centering
 \includegraphics[width=0.95\columnwidth, 
height=0.6\textheight, keepaspectratio]{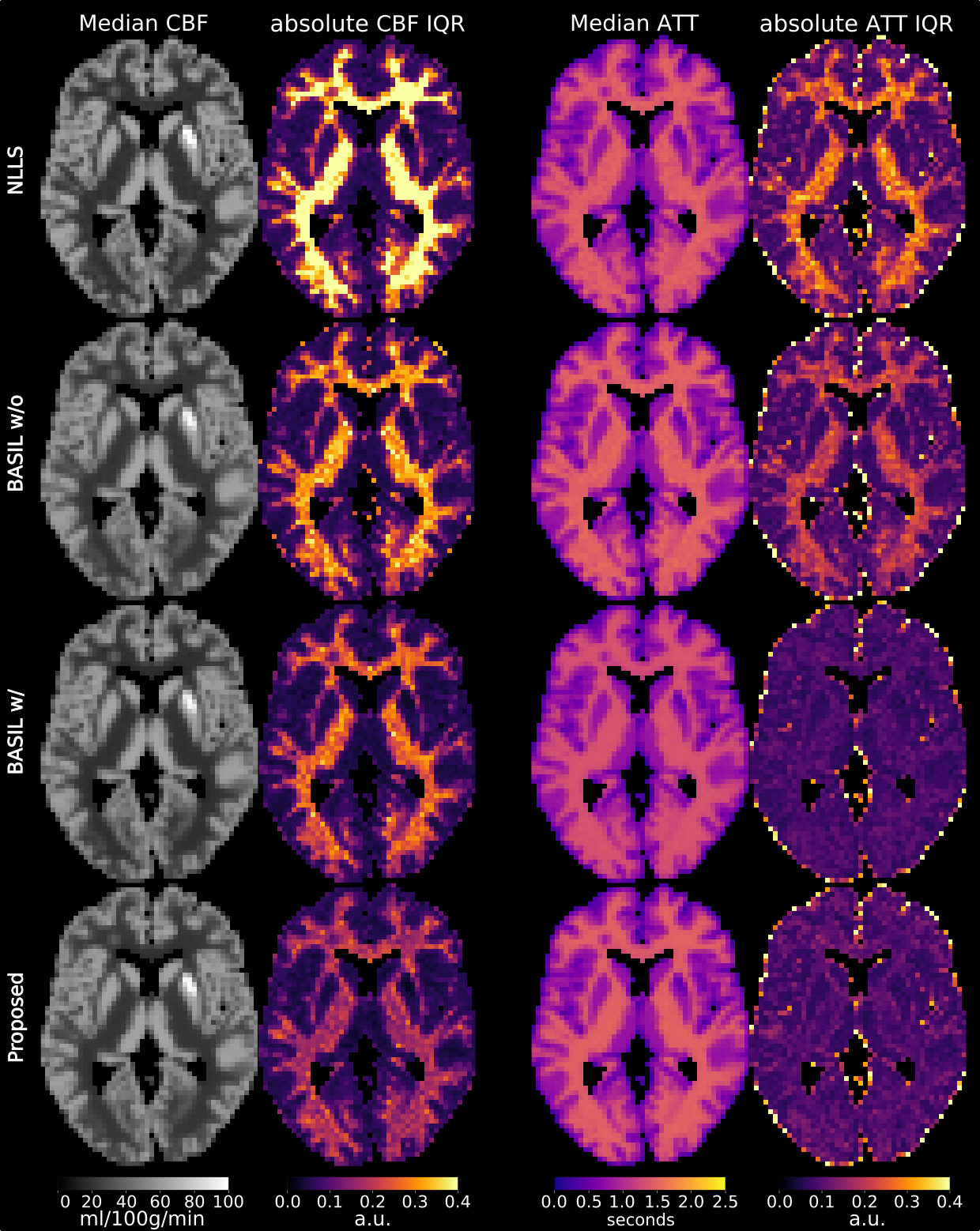}
 \responseRone{R1.C1}{\caption{Fitting results for the pseudo replica method of case 3, 
showing the median and 25\%-75\% IQR for the NLLS, \textit{BASIL w/o}, \textit{BASIL w/} and \textit{Proposed w/} algorithm, respectively. 100 different noise realizations were used. For all four methods the median 
CBF- and ATT-maps are visually close to the noise free maps~(figure~\ref{fig:fig1}).
 The IQR in CBF and ATT is lowest for the proposed method, showing the least fluctuations between individual runs. \label{fig:fig3}}}
\end{figure}
\begin{table*}[!htbp]
\centering
\responseRone{R1.C2}{\caption{Medians of 100 noise realizations of Case 3 for differences to reference and IQRs for both CBF and ATT based on four different tissue types: WM, GM, and lesions in WM and GM, respectively. Mann-Whitney-U-tests were performed to assess whether differences between approaches are statistically significant. p-values are corrected for multiple testing using the Bonferroni method. For each run of the 100 noise realizations the median is computed in a selected ROI and these 100 medians are compared to the reference median by means of a Mann-Whitney-U test. Similar, the IQR between 25 \% and 75 \% is computed in the same ROI and
the different methods are compared to the proposed reconstruction.\label{tab:tab2}}}
\rowcolors{1}{gray!15}{white}
\resizebox{\textwidth}{!}{
\begin{tabular}{llrrrrrrrr}
\toprule
& Tissue &        \multicolumn{2}{c}{WM} &        \multicolumn{2}{c}{GM} & \multicolumn{2}{c}{WM lesion} & \multicolumn{2}{c}{GM lesion} \\
Value &   Comparison & median & adj. p-value & median & adj. p-value &   median & adj. p-value &   median & adj. p-value \\
\midrule
 \cellcolor{white}& NLLS - Ref & 1.88e+00 & 4.51e-38 &  9.43e-02 & 4.51e-38 &  7.96e-01 & 2.35e-04 & -2.01e-01 & 2.05e-03 \\
\cellcolor{white} & BASIL w/o - Ref& 3.93e+00 & 4.51e-38 & -1.32e+00 & 4.51e-38 & -2.51e-01 & 5.41e-01 & -5.30e+00 & 4.51e-38 \\
 \cellcolor{white}& BASIL w/ - Ref& 1.46e+00 & 4.51e-38 & -1.83e+00 & 4.51e-38 &  1.95e-01 & 1.00e+00 & -6.04e+00 & 4.51e-38 \\
\multirow{-4}{*}{\cellcolor{white}\shortstack[2]{Median diff. \\to Ref CBF in \\ ml/100g/min}} & Prop - Ref& 2.46e+00 & 4.51e-38 & -1.33e+00 & 4.51e-38 & -7.89e-01 & 5.59e-06 & -3.26e+00 & 4.51e-38 \\
\midrule
 \cellcolor{white}& NLLS - Ref& -7.91e-02 & 4.51e-38 & -5.79e-03 & 4.51e-38 & -2.16e-02 & 2.35e-04 &  3.19e-03 & 2.94e-01 \\
\cellcolor{white} & BASIL w/o - Ref& -2.37e-02 & 4.51e-38 & -1.61e-02 & 4.51e-38 & -2.25e-02 & 2.93e-09 & -1.79e-02 & 2.31e-28 \\
\cellcolor{white} & BASIL w/ - Ref& -1.43e-01 & 4.51e-38 & -2.10e-02 & 4.51e-38 &  1.80e-02 & 2.35e-04 & -2.09e-02 & 9.81e-34 \\
\multirow{-4}{*}{\cellcolor{white}\shortstack[2]{Median diff. \\to Ref ATT in \\ seconds}} & Prop - Ref& -3.86e-02 & 4.51e-38 & -2.05e-02 & 4.51e-38 & -9.07e-03 & 2.05e-03 & -1.13e-02 & 4.95e-19 \\
 \midrule
\cellcolor{white} & Prop &7.83e+00&& 8.15e+00&& 4.58e+00&& 7.71e+00&\\
\cellcolor{white} & Prop vs. NLLS & 1.10e+01 & 3.07e-33 & 8.61e+00 & 3.07e-33 &  5.96e+00 & 1.38e-06 &  7.54e+00 & 1.00e+00 \\
\cellcolor{white} &Prop vs. BASIL w/o & 9.75e+00 & 3.07e-33 & 8.12e+00 & 4.01e-04 &  4.43e+00 & 1.00e+00 &  9.37e+00 & 2.48e-08 \\
\multirow{-4}{*}{\cellcolor{white}\shortstack[l]{IQR CBF in \\ ml/100g/min}} &Prop vs. BASIL w/ & 9.21e+00 & 3.07e-33 & 8.03e+00 & 5.36e-28 &  4.00e+00 & 1.00e+00 &  9.26e+00 & 6.80e-08 \\
 \midrule
\cellcolor{white} & Prop & 1.79e-01&& 1.48e-01&& 1.26e-01&& 6.69e-02&\\
\cellcolor{white}& Prop vs. NLLS & 3.33e-01 & 3.07e-33 & 1.53e-01 & 3.07e-33 &  1.58e-01 & 1.13e-04 &  6.72e-02 & 1.00e+00 \\
\cellcolor{white} &Prop vs. BASIL w/o  & 2.99e-01 & 3.07e-33 & 1.50e-01 & 3.04e-30 &  1.42e-01 & 8.36e-01 &  6.26e-02 & 4.69e-01 \\
\multirow{-4}{*}{\cellcolor{white}\shortstack[l]{IQR ATT in \\ seconds}}  &Prop vs. BASIL w/ &  1.58e-01 & 3.07e-33 & 1.49e-01 & 5.62e-21 &  1.16e-01 & 1.00e+00 &  6.41e-02 & 9.30e-01 \\
\bottomrule
\end{tabular}}
\end{table*}
\responseRone{R1.C2}{\mdeleted{The voxel-wise stability of the fitting methods over 100 different noise 
realizations is shown in figure~\ref{fig:fig3}. The evaluation is performed based on case 3 of the phantom simulations. To account for possible 
deviations of a Gaussian distribution in the reconstructed maps, median and 
inter-quartile range (IQR) are used for evaluation.} 

The voxel-wise stability of the fitting methods was assessed based on 100 different noise realizations of Case 3 of the phantom simulations. For the evaluation, medians and IQRs of the 100 runs were visualized and differences statistically analyzed.} A qualitative comparison 
of the median values for CBF- and ATT-maps shows a good agreement to the noise 
free numerical maps for all \responseRone{R1.C1}{\mdeleted{three} four} methods\responseRone{R1.C1}{(figure~\ref{fig:fig3})}.\responseRone{R1.C2}{\mdeleted{However, high variations in WM CBF 
(median 30.9 \%) and ATT (median 20.5 \%), reflected by the relative IQR, 
are visible in the NLLS fits. BASIL is able to reduce these variations in 
both maps but still suffers from deviations with a median of 24.1 percent in WM 
CBF maps. The proposed method reduces these variations throughout both maps down
 to 10.1 \% for WM CBF and 8.0 \% for WM ATT, respectively. 
The box-plots in figure~\ref{fig:fig4} support the visual impression of reduced 
variations in the quantitative maps. BASIL shows a bias in CBF of 12.1 \%/-2.3 
\% in WM/GM compared to the reference and the proposed method 11.9 \%/-2.9 \%, 
respectively. NLLS values are bias free 
but have a much higher IQR, especially in low signal WM areas, amounting to 
approximately a factor of 1.3 and 3.1 in WM 
compared to BASIL and the proposed method. 
In contrast, in the ATT maps the bias is lowest for the proposed method which 
underestimates the ATT in WM by 0.7 \% compare to NLLS with 6.8 \% and BASIL 
with 2.7 \%. Furthermore the variations are reduced by a factor of 2.6 compared 
to NLLS and 2.1 compared to BASIL. Results for ATT in GM are similar for all 
methods with a slight reduction of variance by factor of 1.1 using the proposed 
regularization method.} However, median WM and GM differences to the ground truth of the four approaches are statistically different at a significance level of 0.05 (table~\ref{tab:tab2}). With the exception of both BASIL methods, deviations in the CBF WM lesion are also statistically significant. ATT deviations in
the WM lesion are statistically significant for all methods. The CBF GM lesion also shows statistically significant differences to the reference in all methods.
ATT deviations in the GM lesion are not statistically significant for NLLS. These differences to the median reference value are visualized by a violin plot in the left half of figure~\ref{fig:fig4}.

Variations of the reconstruction algorithms over the 100 runs are reflected by the IQR in each ROI.
The median of these variations is plotted in the right half of figure~\ref{fig:fig4}.
The IQRs are highest for NLLS, except for the CBF GM lesion where both BASIL approaches show higher fluctuations. 
The proposed method shows a lower median IQR in WM CBF which is also statistically significant lower compared to all other approaches (table~\ref{tab:tab2}).
For WM ATT, similar results are obtained with the exception of \textit{BASIL w/}, which shows lower IQR than the proposed method.
In GM CBF, the proposed method shows statistically significantly lower IQR than NLLS but both BASIL approaches are able to
further reduce IQR than the proposed method. IQR of GM ATT shows the least deviations in the proposed method.
In the WM lesion the proposed method reduces IQR compared to NLLS in CBF and ATT statistically significant but no statistically significant difference
to both BASIL methods is observed. The CBF GM lesion shows statistically significant reduction of IQR 
using the proposed method over BASIL but no difference to NLLS. In the corresponding ATT,
no statistically significant differences of IQR are observable. All p-values and median IQRs are reported in table~\ref{tab:tab2}.
}
\begin{figure}[!htbp]
\centering
 \includegraphics[width=0.95\columnwidth, 
height=0.9\textheight, keepaspectratio]{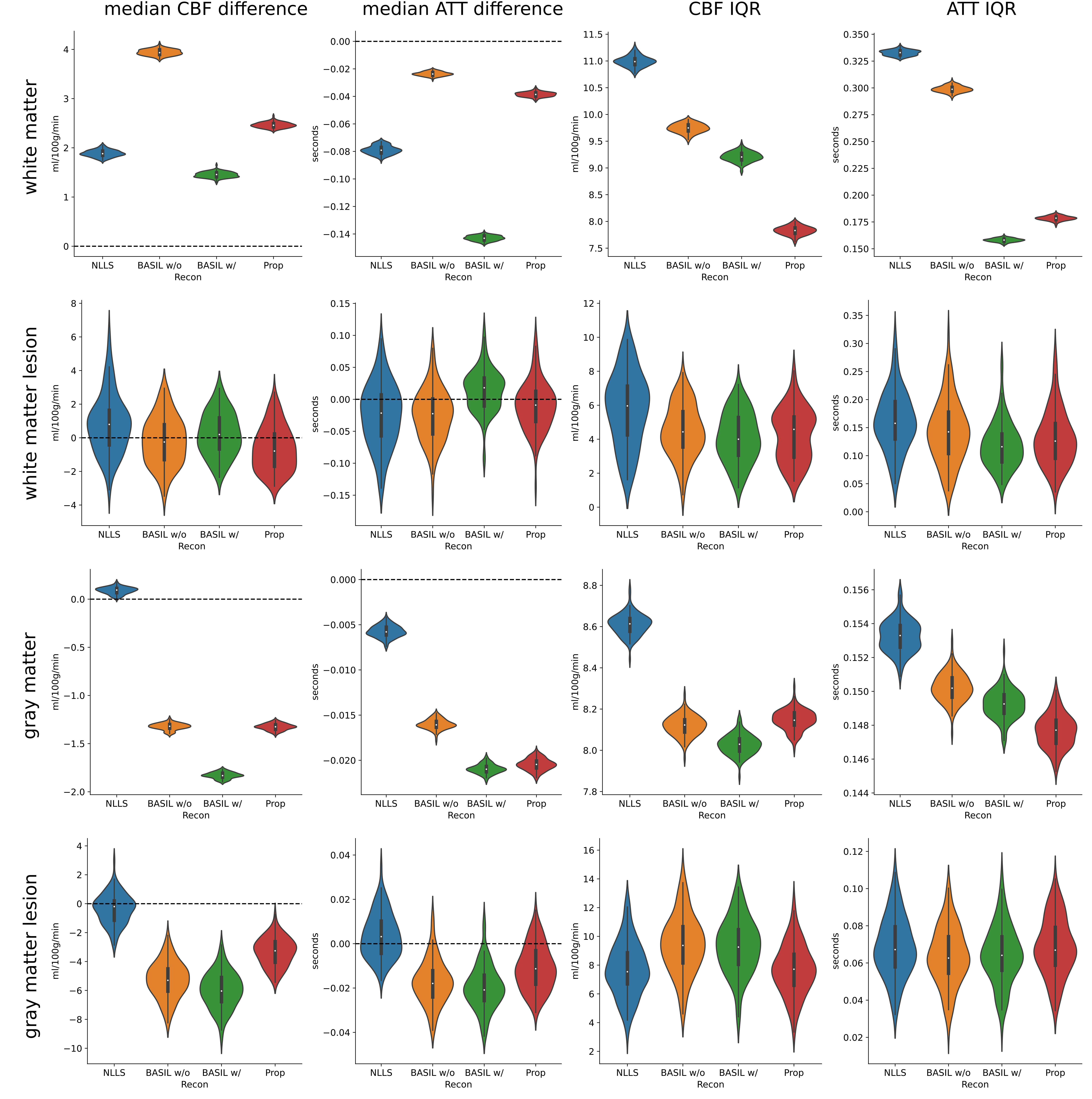}
 \responseRone{R1.C2}{\caption{Violin-plots for median differences in the median and IQR of the simulated phantom Case 3. The difference to the reference for the 100 median values for each
of the evaluated ROIs is given in the left half of the figure. The dotted lines represent
zero deviation to the reference median. The right half of the figure shows the 25\%-75\% 
IQR in each of the evaluated ROIs for the 100 runs. Mann-Whitney-U tests are used to assess
if the differences are statistically significant. The corresponding adjusted p-values
 are given in table~\ref{tab:tab2}. The scale on the y-axis is non-uniform due to the
large differences in observed values. \label{fig:fig4}}}
\end{figure}

\subsection{Healthy volunteers}
\begin{table}[!t]
\centering
\responseRone{R1.C2}{\caption{Statistical evaluation of median and IQR for the investigated methods.
Displayed p-values are corrected for multiple testing using the Bonferroni method.
For each subject the median is computed in a white and gray matter ROI and compared to the median of NLLS with 4 averages (NLLS-4) by means of a double-sided Mann-Whitney-U test, respectively. 
Similar, the IQR between 25\% and 75\% is computed in the same ROI and
the different methods are compared to the proposed method.\label{tab:tab3}}}
\rowcolors{1}{gray!15}{white}
\resizebox{\columnwidth}{!}{
\begin{tabular}{llrrrr}
\toprule
& Tissue &        \multicolumn{2}{c}{WM} &        \multicolumn{2}{c}{GM} \\
Value &   Comparison & median & adj. p-value & median & adj. p-value \\
\midrule
\cellcolor{white} & NLLS-4 & 3.03e+01 & & 6.93e+01 & \\
\cellcolor{white} & NLLS-4 vs. BASIL w/o& 3.12e+01 & 1 & 6.91e+01 & 1 \\
 \cellcolor{white}& NLLS-4 vs. BASIL w/&  2.71e+01 & 6.49e-01 & 6.85e+01 & 1 \\
\multirow{-4}{*}{\cellcolor{white}\shortstack[2]{Median CBF \\4 aver}} & NLLS-4 vs. Prop& 2.92e+01 & 1 & 6.74e+01 & 1 \\
\midrule
\cellcolor{white} & NLLS-4 vs. NLLS& 3.00e+01 & 1 & 6.81e+01 & 1 \\
\cellcolor{white} & NLLS-4 vs. BASIL w/o& 3.41e+01 & 1 & 6.72e+01 & 1 \\
 \cellcolor{white}& NLLS-4 vs. BASIL w/&  2.59e+01 & 5.77e-02 & 6.47e+01 & 1 \\
\multirow{-4}{*}{\cellcolor{white}\shortstack[2]{Median CBF \\1 aver}} & NLLS-4 vs. Prop& 2.96e+01 & 1 & 6.60e+01 & 1 \\
\midrule
\cellcolor{white} & NLLS-4 & 1.18e+00 & & 1.15e+00 & \\
\cellcolor{white} & NLLS-4 vs. BASIL w/o & 1.27e+00 & 1 & 1.17e+00 & 1 \\
\cellcolor{white} & NLLS-4 vs. BASIL w/ & 1.07e+00 & 1 & 1.13e+00 & 1 \\
\multirow{-4}{*}{\cellcolor{white}\shortstack[2]{Median ATT \\4 aver}} & NLLS-4 vs. Prop& 1.19e+00 & 1 & 1.14e+00 & 1 \\
\midrule
\cellcolor{white} & NLLS-4 vs. NLLS& 1.07e+00 & 1 & 1.14e+00 & 1 \\
\cellcolor{white} & NLLS-4 vs. BASIL w/o & 1.27e+00 & 1 & 1.17e+00 & 1 \\
\cellcolor{white} & NLLS-4 vs. BASIL w/ & 8.31e-01 & 3.55e-02 & 1.06e+00 & 1 \\
\multirow{-4}{*}{\cellcolor{white}\shortstack[2]{Median ATT \\1 aver}} & NLLS-4 vs. Prop& 1.14e+00 & 1 & 1.14e+00 & 1 \\
 \midrule
\cellcolor{white} & Prop & 2.27e+01 &  & 2.41e+01 & \\
\cellcolor{white} & Prop vs. NLLS-1 & 2.25e+01 & 1 & 2.58e+01 & 1 \\
\cellcolor{white} & Prop vs. BASIL w/o & 2.19e+01 & 1 & 2.46e+01 & 1 \\
\multirow{-4}{*}{\cellcolor{white}\shortstack[2]{IQR CBF\\ 4 aver}} & Prop vs. BASIL w/ & 2.41e+01 & 1 & 2.60e+01 & 1 \\
 \midrule
\cellcolor{white} & Prop & 2.49e+01 &  & 2.55e+01 & \\
\cellcolor{white} & Prop vs. NLLS-1 & 2.74e+01 & 5.44e-01 & 2.84e+01 & 7.87e-01 \\
\cellcolor{white} & Prop vs. BASIL w/o & 2.47e+01 & 1 & 2.31e+01 & 7.87e-01 \\
\multirow{-4}{*}{\cellcolor{white}\shortstack[2]{IQR CBF\\ 1 aver}} & Prop vs. BASIL w/ & 2.60e+01 & 1 & 2.58e+01 & 1 \\
 \midrule
\cellcolor{white} & Prop & 7.36e-01 & & 5.41e-01 & \\
\cellcolor{white}& Prop vs. NLLS-1 & 9.94e-01 & 6.09e-02 & 5.69e-01 & 1 \\
\cellcolor{white} & Prop vs. BASIL w/o  & 8.52e-01 & 1.57e-01 & 5.49e-01 & 1 \\
\multirow{-4}{*}{\cellcolor{white}\shortstack[2]{IQR ATT\\ 4 aver}}  & Prop vs. BASIL w/ &  5.84e-01 & 5.44e-01 & 5.39e-01 & 1 \\
 \midrule
\cellcolor{white} & Prop & 7.52e-01 & & 5.76e-01 & \\
\cellcolor{white}& Prop vs. NLLS-1 & 1.12e+00 & 1.57e-01 & 6.34e-01 & 1 \\
\cellcolor{white} & Prop vs. BASIL w/o  & 7.87e-01 & 1 & 5.65e-01 & 1 \\
\multirow{-4}{*}{\cellcolor{white}\shortstack[2]{IQR ATT\\ 1 aver}}  & Prop vs. BASIL w/ &  5.16e-01 & 3.68e-01 & 5.32e-01 & 1 \\
\bottomrule
\end{tabular}}
\end{table}

\begin{figure*}[!htbp]
\centering
 \includegraphics[width=0.95\textwidth, 
height=0.85\textheight, keepaspectratio]{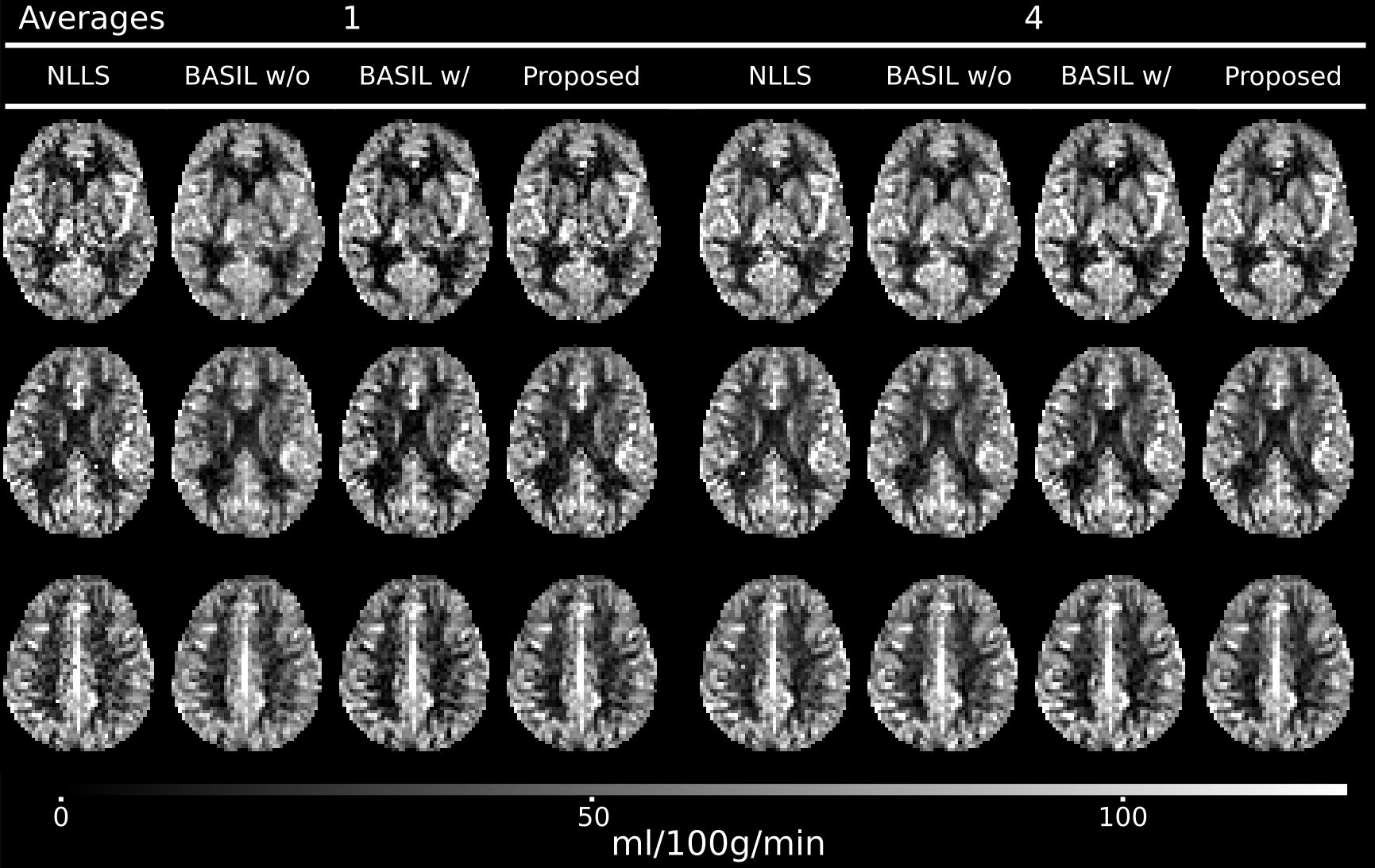}
 \responseRthree{R3.C4}{\caption{Performance comparison of the four fitting algorithms in a healthy 
subject for a different number of averages. The CBF-maps estimated with NLLS 
become noisier and show more outliers as the number of averages is
decreased. 
\textit{BASIL w/o}, on the other hand, counteracts the worse SNR by smoothing, 
leading to a loss of detail in the CBF-maps. Both, \textit{BASIL w/} and the proposed method, are able to 
suppress noise while maintaining high fidelity CBF-maps, independent of the 
number of averages.\label{fig:fig5}}}
\end{figure*}

Figure~\ref{fig:fig5} and figure~\ref{fig:fig6} illustrate three 
different exemplary slices of the estimated CBF- and ATT-maps for subject 3 in 
dependence of the numbers of averages.\responseRone{R1.C3}{\mdeleted{Subject 1 and 2} The remaining five 
healthy subjects are available as 
Supplementary~Material~Figures~\ref{fig:figS1_1}-~\ref{fig:figS1_5}.\mdeleted{ in the supplementary material.}} 
For \responseRthree{R3.C5}{\mdeleted{the highest number of} four} averages each 
method  produces meaningful CBF- and ATT-maps. The influence of regularization 
becomes more pronounced \responseRthree{R3.C5}{\mdeleted{with reduced number of averages} if no averaging is performed}, showing increased 
number of outliers in the NLLS fits.\responseRone{R1.C1}{\mdeleted{and strong spatial blurring using BASIL} Both BASIL methods show a reduction in outliers in CBF and ATT. \textit{BASIL w/o} shows slight blurring in CBF
and only minor reduction of outliers in ATT. \textit{BASIL w/} shows a similar reduction of outliers in CBF but much stronger smoothing of ATT.} The proposed joint approach is able to maintain similar visual quality \responseRthree{R3.C5}{\mdeleted{over all averages, 
even}} if just one average per PLD is used\responseRone{R1.C1}{, reducing outliers in CBF and ATT without introducing blurring.}\responseRone{R1.C1}{\mdeleted{Visually this is most evident in the 
ATT maps, as the other methods employ no spatial prior information to estimate 
ATT}} \responseRone{R1.C3}{Median and IQR of each subject are computed in WM and GM ROIs and visualized using a box-plot in figure~\ref{fig:fig7}. 
} Due to a lack of ground truth values, the quantitative accuracy \responseRone{R1.C3}{in the median} is compared
 to NLLS with 4 averages \responseRone{R1.C3}{(NLLS-4). 
The IQRs are
compared to the proposed method. None of the methods shows a statistically significant difference in CBF medians after p-value correction (table~\ref{tab:tab3}). However, median ATT using one average 
shows a statistically significant differences to the NLLS reference for \textit{BASIL w/} in WM. No statistically significant difference in IQR could be observed.}

\responseRone{R1.C3}{\mdeleted{
Box-plot evaluations in figure~\ref{fig:fig7} show a 
quantitative comparison of the four methods for different numbers of averages 
per PLD. For the highest SNR case with 4 averages, the median values of 
BASIL and the proposed method are close to NLLS. 
The least deviations (IQR) are 
achieved with the proposed method utilizing joint regularization.}} 

\responseRtwo{R2.C5}{\mdeleted{Optimization 
time for the full volume amounted to 9 min 16 s.} The run time for all algorithms
was evaluated on a server with an Intel Core Xenon Gold CPU and an NVIDIA Titan XP GPU. 
The NLLSQ took 8 min 40 s independent of the number of averages used. BASIL with one
spatial prior needs 2 min 36 s using one average as input and 6 min 20 s using four averages. 
The run time for BASIL with two spatial priors was slightly higher with 2 min 41 s and 
6 min 26 s for one and four averages respectively. Our algorithm took 4 min 54 s for one average and
7 min for four averages, respectively, running on a single GPU of the server at a given time. 
We have also evaluated the run time of
 our algorithm on a workstation equipped with an NVIDIA Geforce GTX 1080Ti which resulted in
4 min 20 s for 4 averages and 2 min 17s for a single average.}
\responseRthree{R3.C21}{Fitting on the GPU takes roughly 1 GB of memory. 
Other than sufficient memory, no hardware restrictions are present. However, 
the algorithm is currently only tested on NVIDIA hardware.}

\responseRone{R1.C4}{\mdeleted{
Reducing the number of averages, BASIL increasingly overestimates the CBF and 
ATT in WM and underestimates the CBF and ATT in GM. 
This may be in part due to the increase of spatial blurring seen in 
figure~\ref{fig:fig5}. In contrast the proposed method shows sharper 
quantitative maps with a lower bias.}} 

\subsection{Stroke patients}
Exemplary quantitative maps from the first patient are given in 
figure~\ref{fig:fig8}. \responseRthree{R3.C5}{\mdeleted{Six}Three} slices of the central brain region clearly show 
reactive hyperperfusion in areas affected by the stroke or close to 
the stroke after successful re-canalization therapy. The corresponding arrival 
time is reduced, as can be seen in ATT. The area of ischemic infarction can be 
delineated in all fitting strategies. However, the NLLS fits are noisy and show 
several outliers compared to the other methods. The CBF-map estimated with 
BASIL shows no outliers but admits over-smoothing\responseRthree{R3.C5}{\mdeleted{, especially in deep brain 
areas}, blurring tissue boundaries compared to NLLS and the proposed method. These differences are highlighted 
with arrows in CBF and ATT}. \responseRthree{R3.C5}{An exemplary slice for three additional 
stroke patients is given in figure~\ref{fig:fig9}. Differences between the methods are again 
highlighted by arrows. Over smoothing of CBF in both BASIL methods leads to a loss of small structures
in the stroke area. Similar loss of detail can be observed in ATT, especially severe for \textit{BASIL w/}.}
The proposed method achieves the highest contrast between GM and WM 
tissue without blurring \responseRthree{R3.C5}{or loss} of structure while suppressing noise in the 
quantitative maps. \responseRone{R1.C1}{\mdeleted{This effect is again most visible in ATT as the other 
methods are not employing any prior information for this map.}} The same behavior 
is observable for \responseRfour{R4}{\mdeleted{patient 2 and 3 in Figure~S3 and S4} the remaining patients, shown} in Supplementary~Material~Figure~\ref{fig:figS3}.
% --------------------------------------------------------------------
% Discussion
% --------------------------------------------------------------------
\section{Discussion}
\label{sec:discussion}

In this study we present a novel joint spatial regularization technique for 
quantitative ASL imaging, combining non-linear fitting with a TGV$^2$ functional. The 
proposed method poses a joint spatial TGV$^2$ prior on both, CBF and ATT, to improve the robustness of 
the fitting procedure. Synthetic ASL datasets with different pathologies as 
well as in vivo data from healthy and stroke patients with different SNR levels were considered. 

\responseRtwo{R2.C3}{\responseRone{R1.C1}{CBF and ATT pathologies in the simulated cases could be identified in all methods. 
The proposed method showed an improvement in noise reduction in CBF and ATT in all simulated phantom cases.
\textit{BASIL w/} showed similar reduction of noise but leads to underestimation of high values in CBF and ATT, reflected by the increased number of points below the identity line in figure~\ref{fig:fig2_3}. This effect is less severe for \textit{BASIL w/o} and thus might be related to the additional prior on ATT. 
With the exception of NLLS, all methods showed a tendency to overestimate low CBF values and underestimate high CBF values. 
CBF estimates of the proposed method without and with regularization on ATT produce similar results. However, joint regularization on both maps could greatly reduce noise in ATT while not showing a substantial increase in bias. \responseRone{R1.C5}{
This improvements of joint regularization over single regularization
are in concordance with previous studies~\citep{Knoll2017a, Huber2019, Maier2019c}. Thus,
regularization on only CBF for the proposed method is excluded in further evaluations.}}}
\responseRtwo{R2.C2}{\responseRone{R1.C4,5}{
The joint regularization approach most accurately recovered ATT and CBF in simulated cases of partly occlusion of the arteria cerebri media, Cases 4-6 in figure~\ref{fig:fig1_2}. The visual impression is supported by the least deviations in the pixel-wise difference (figure~\ref{fig:fig2_2}) and the median and IQR values of the large stroke in table~\ref{tab:tab1}. Even though GM values of NLLS are closer in the median, the large IQR, especially in WM, makes interpretation of the image difficult. This effect becomes more severe for increased delays in ATT (figure~\ref{fig:fig1_2}, Case 6). \responseRthree{R3.C7}{Both BASIL approaches tend to underestimate CBF in the large stroke with increasing ATT. This might be due to the
combination of spatial and non-spatial priors, assuming certain values of ATT in WM and GM which are further smoothed by the spatial prior. As the high ATT in the simulated strokes might be well out of this range, BASIL is not able to recover correct ATT maps. This in turn could be the reason for the incorrect CBF estimates. The proposed approach makes no assumption of underlying CBF and ATT values and thus does not suffer as much if one parameter is strongly varied. Even though the joint regularization approach combines gradient information of both maps, no artificially introduced structures in CBF could be observed. The combined information of both structures helps in stabilizing fitting, especially in low signal areas. This joint regularization approach is advantageous compared to separate regularization (\textit{BASIL w/}) or regularizing on one quantitative map (\textit{BASIL w/o}, \textit{Proposed w/o}). In Case 6 the proposed method starts to degenerate, showing an underestimation of CBF values due to the strong increase in ATT. In combination with the used LD and PLD, this strongly delayed ATT results in only few non-zero data points. Such cases require an adjustment of simulation/sequence parameters as suggested in~\citet{Woods2019}.}}} \responseRthree{R3.C16}{Another simplification in our simulated phantoms affects the employed background suppression. In the simulated cases we assume instant and constant background suppression after labeling, which is not possible in practice for our used LD and PLD combinations.}

\responseRone{R1.C4}{\mdeleted{
All datasets show improved noise suppression, for low as well as high SNR 
regime, while 
maintaining high image fidelity and preserving fine details compared to the two
reference methods. 
This is especially visible in the small simulated pathologies, which could be 
easily identified 
even though residual noise could be nearly completely 
eliminated using the proposed method (figure~\ref{fig:fig1}). 
The joint regularization approach allows 
for higher regularization as edges in both structures correlated well with each 
other and thus are easily separable from noise. Especially in low signal areas 
like white matter such an regularization approach is advantageous compared to 
separate regularization or only regularizing one quantitative map.}}

\begin{figure*}[!htbp]
\centering
 \includegraphics[width=0.95\textwidth, 
height=0.85\textheight, keepaspectratio]{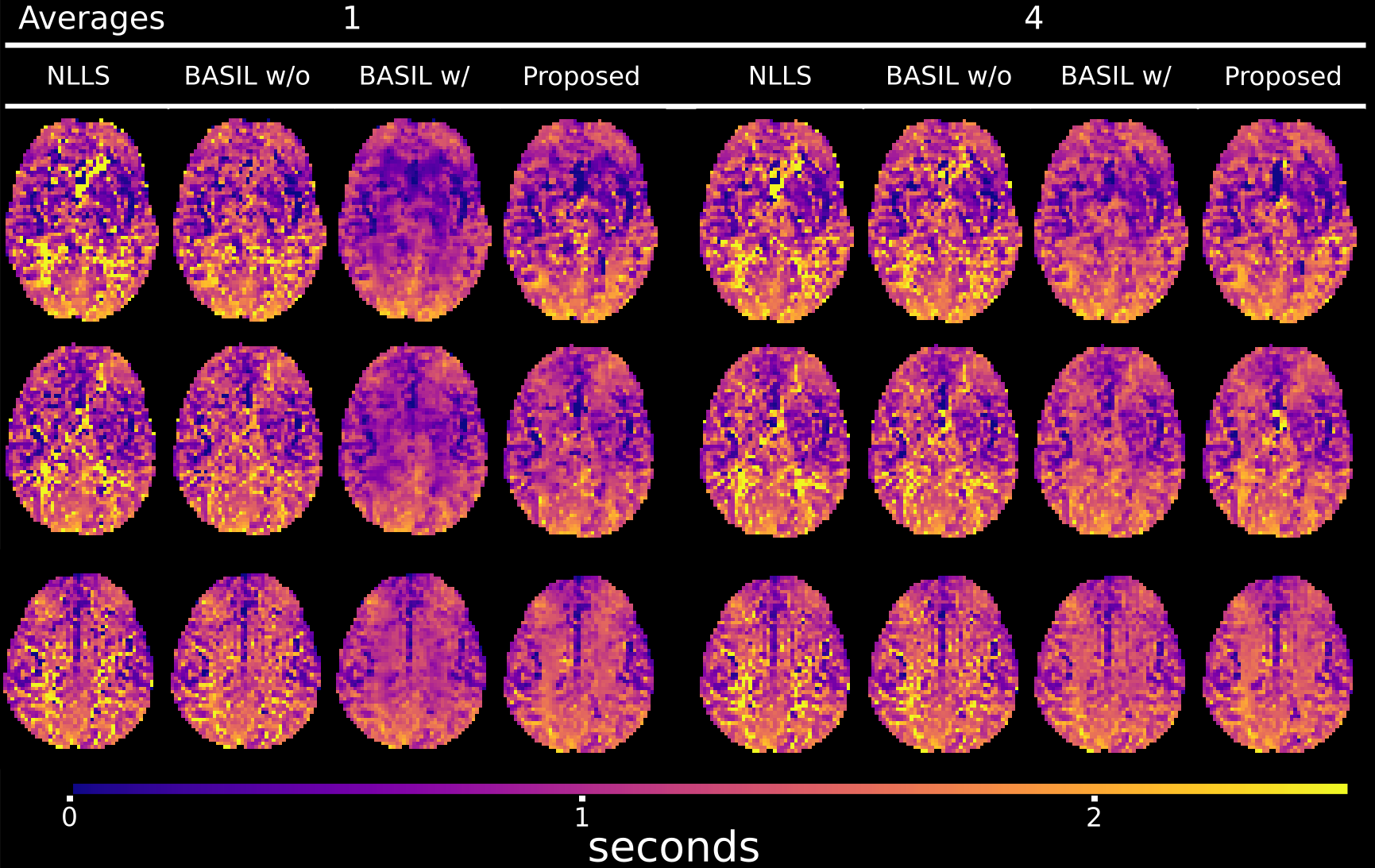}
 \responseRthree{R3.C4}{\caption{ATT maps corresponding to the CBF values in figure~\ref{fig:fig5}. 
Similar to the CBF-maps, more outliers are visible in the NLLS fit if no averaging is performed. The same trends are observable for \textit{BASIL w/o}. \textit{BASIL w/} shows a reduction of noise in ATT but comes at the cost of slightly lower values in 
WM tissue. In contrast, the 
estimated ATT maps with the proposed method show reduced outliers and noise 
with only minimal loss of detail and no visually observable difference in WM values.\label{fig:fig6}}}
\end{figure*}

As TGV$^2$ assumes piece-wise linear structures, flat areas perfectly fit this 
model and variations can be nearly eliminated\responseRone{R1.C2}{\mdeleted{with only 3 percent of 
variations}, leading to the smallest IQRs in WM CBF and WM/GM ATTs over different noise realizations (figure~\ref{fig:fig3}), which is statistically significantly lower than for all other approaches (table~\ref{tab:tab2}). Both BASIL approaches are also able to reduce IQRs. The reductions are even statistically significantly lower than proposed method in the GM lesion, and in WM ATT for \textit{BASIL w/}. For the GM lesion, BASIL shows a statistically significant increase in IQR over NLLS and the proposed method. Even though not visible in the median images itself, this might indicate increased blurring of CBF structures, as later seen in stroke patients in figures~\ref{fig:fig8}~and~\ref{fig:fig9}.} \responseRthree{R3.C2}{\responseRtwo{R2.C1}{All methods have in common that they show a statistically significant difference to the reference median in most cases. Deviations seem to be more severe for tissue with low SNR. As our phantom generation process mimics the MRI acquisition pipeline, noise transformation due to the multiple receive coils could be the reason for this deviations in the simulations. Such effects are more pronounced in tissue with low SNR which matches the observation of increased deviations in WM structures. Even though the proposed method takes complex noise into account, the multiple coil setting was not included. The extension to multi-coil reconstruction can be modelled by a simple multiplication of the forward model with the a priori determined receive coils and could be included in a future study.}} 
\responseRone{R1.C2}{\mdeleted{
However, at tissue boundaries variations up to 10 percent are still visible. 
The box-plots in figure~\ref{fig:fig4} support the visual impression of reduced 
variations in the quantitative maps. The inclusion of spatial prior information 
in the fitting procedure of both BASIL and the proposed method leads to lower 
variations and IQRs. However, this property comes at the cost of a slight 
under- or overestimation of quantitative values in areas where optimization 
heavily relies on the prior information.}} 

\responseRtwo{R2.C2}{The improved noise suppression of the proposed method in the simulated phantom may be in part 
due to the way the phantom has been simulated. Flat areas in CBF and ATT favour the TGV$^2$ assumptions. To mitigate
this effect, phantoms were generated with fine details in high resolution (1x1x1 mm$^3$) followed by down sampling
to ASL resolution. This gives smoother transitions at tissue boundaries and reduces the amount of completely flat areas, which is reflected by the variation present in the reference itself (table~\ref{tab:tab1}). The influence is higher in gray matter as it consists of fewer voxels, compared to white matter.}

\begin{figure*}[!t]
\centering
 \includegraphics[width=0.95\textwidth, 
height=0.95\textheight, keepaspectratio]{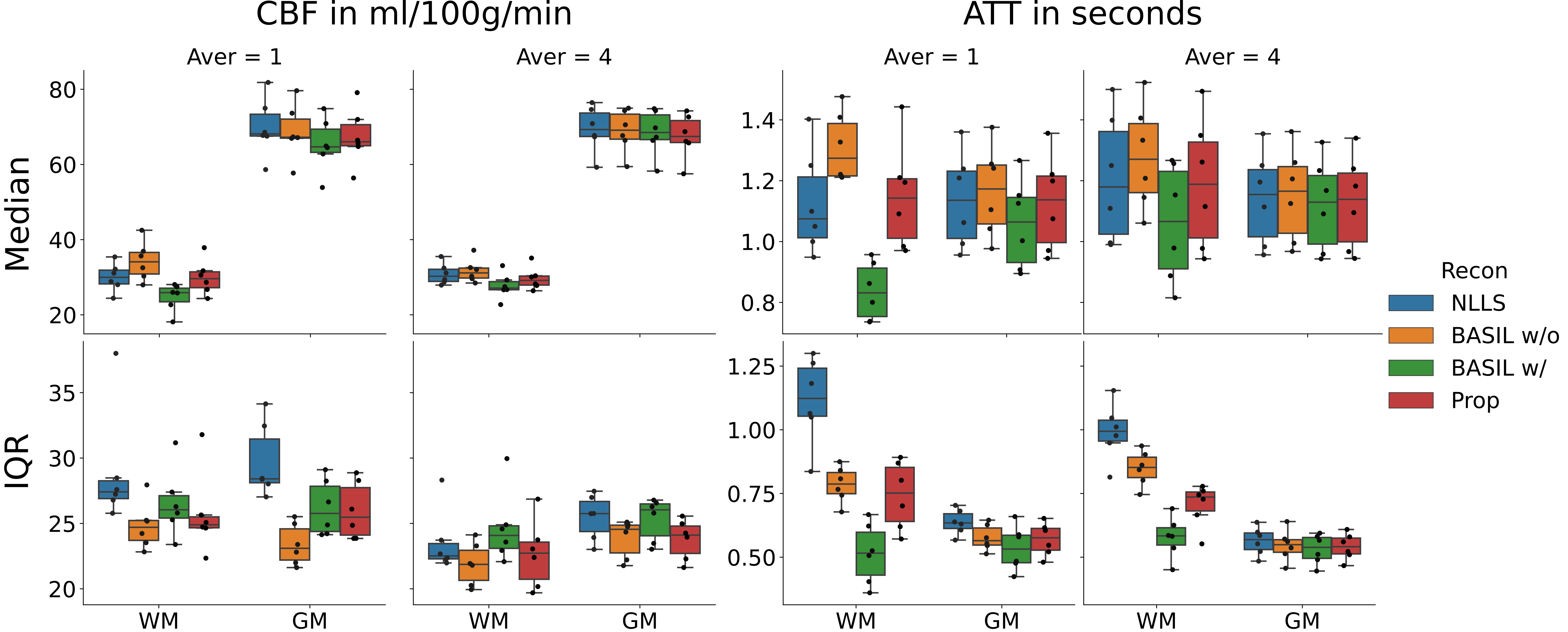}
 \responseRone{R1.C3}{\caption{Box-plot for median and IQR in
white and gray matter CBF and ATT, respectively. Black dots correspond to mean or IQR of the six individual healthy subjects within the specified ROI. ROIs are defined by GM and WM mask, 
generated from co-registered $T_1$ images. Median and IQR are tested for statistically significant differences using a Mann-Whitney-U test. 
The corresponding p-values are given in table~\ref{tab:tab3}. \label{fig:fig7}}}
\end{figure*}

\begin{figure}[!htbp]
\centering
 \includegraphics[width=0.95\columnwidth, 
height=0.9\textheight, keepaspectratio]{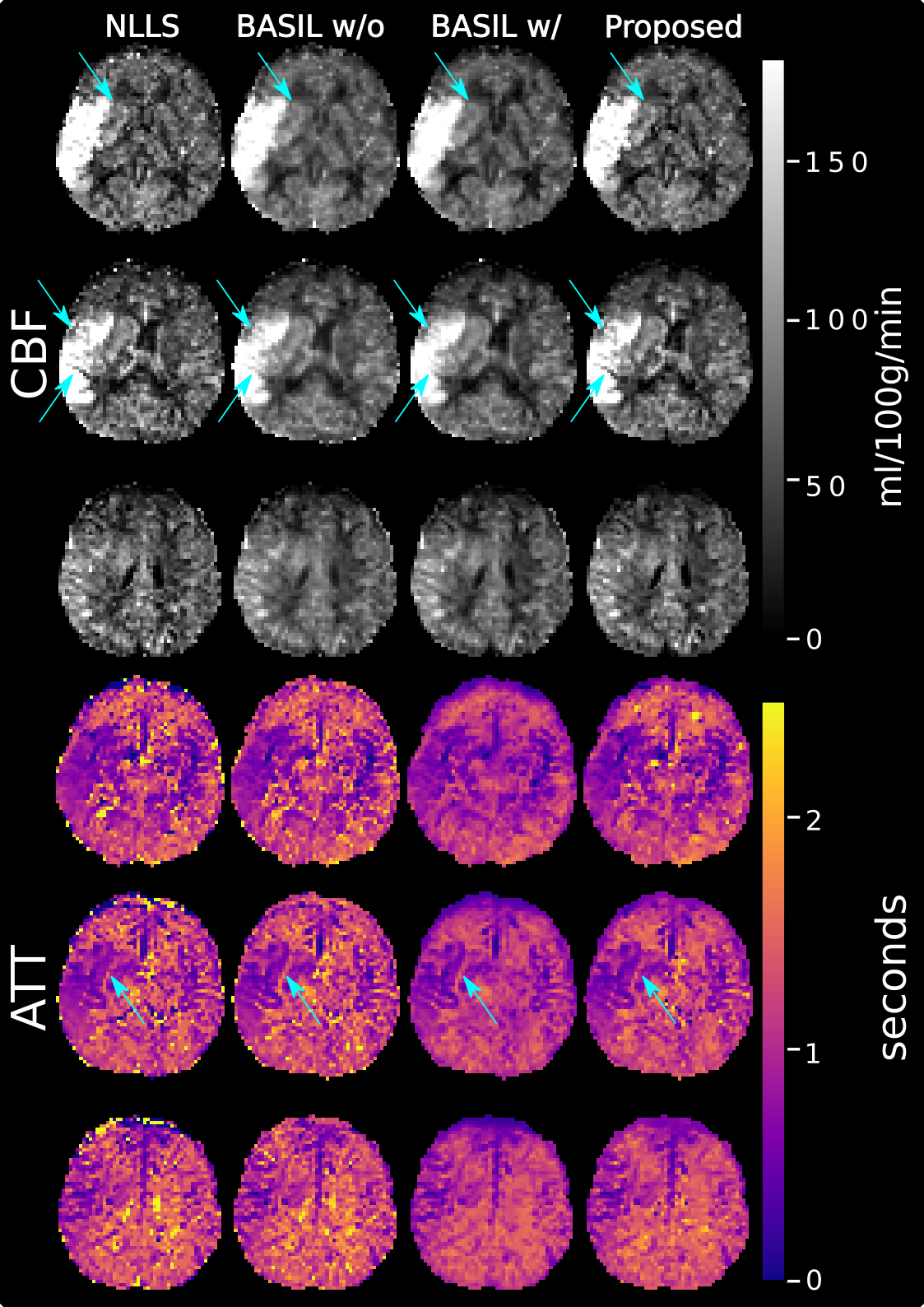}
 \caption{CBF and ATT maps of a patient 24 hours post ischemic stroke. 
The reactive hyperperfusion areas after successful recanalization are 
clearly visible in all fits. BASIL CBF maps appear to be oversmoothed 
compared to NLLS and the proposed method. Especially deep brain regions 
show worse contrast compared to the proposed regularization strategy. \label{fig:fig8}}
\end{figure}

\begin{figure}[!htbp]
\centering
 \includegraphics[width=0.95\columnwidth, 
height=0.9\textheight, keepaspectratio]{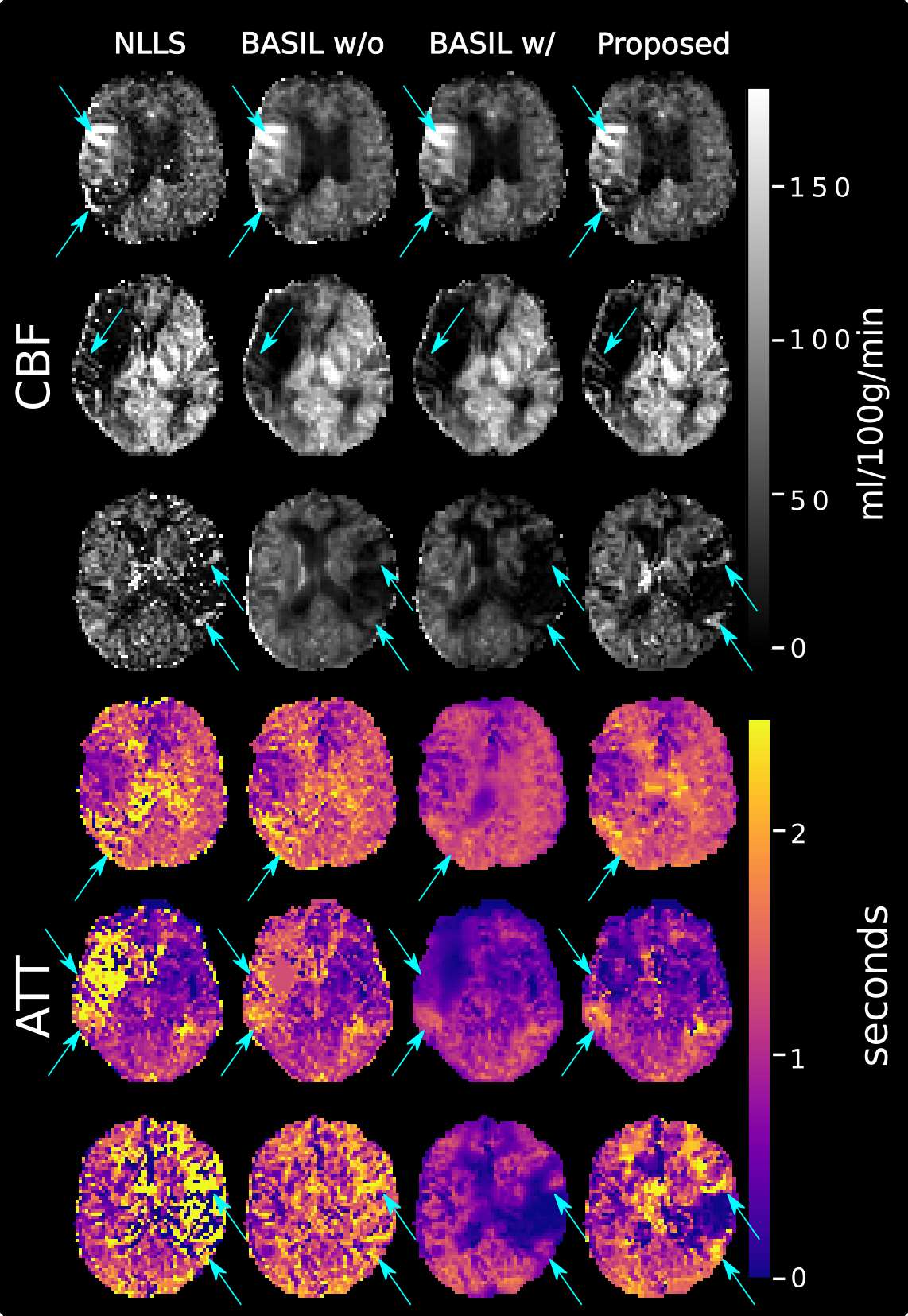}
 \responseRfour{R4}{\caption{An exemplary slice of CBF and ATT for three selected stroke patients. 
Patients are shown in rows, different reconstruction methods in columns. 
Difference between methods are highlighted by arrows. \label{fig:fig9}}}
\end{figure}

\responseRone{R1.C9}{Imposing prior knowledge on the unknown parameter maps leads to the fundamental problem of bias/variance trade-off, as the solution will depend on the used prior information. This is also true for the used MAP-based approach shown in this work.} The amount 
of bias, however, can be controlled by the used prior and the weight 
between data and prior information in optimization, 
respectively~\citep{Brinkmann2017}. To this end, the NLLS approach without any 
regularization can be considered bias free in the mean value under the 
assumption of Gaussian noise. \responseRone{R1.C3}{As we have seen in the phantom study, violation of this noise assumption can lead to a statistically significant bias for all methods. However, due to the lack of ground truth values for in vivo measurements and the wide spread use of NLLS
we consider NLLS values for the case of four averages as our reference values for GM and WM. 
Using NLLS as reference is further strengthened by the overall least deviations to identity in our 2D histogram of all simulation cases (figure~\ref{fig:fig2_3}).

CBF estimates of all methods match closely for four averages (figure~\ref{fig:fig5}), showing only minor noise in NLLS. 
Using a single average, NLLS shows increased noise while \textit{BASIL w/o} shows slight blurring of CBF. No visual differences between
\textit{BASIL w/} and the proposed method are observable. Both show clear structures with good noise suppression. None of the methods showed a statistically significant difference to the median of NLLS in our six subjects (table~\ref{tab:tab3}). A slight reduction of median CBF in WM and GM of \textit{BASIL w/} can be seen
as well as an increase in WM CBF in \textit{BASIL w/o}, which both are not statistically significant.
ATT in figure~\ref{fig:fig6} shows similar results. \textit{BASIL w/} and the proposed method can reduce noise in ATT but the proposed method is simultaneously maintaining the correct median value in WM ATT. The proposed method shows no tendency to increased over- or underestimation in dependence on the number of used averages (figure~\ref{fig:fig7}). In contrast to the simulation study, no statistically significant reduction in IQR could be observed with the proposed method. The small number of healthy subjects (six) for this evaluation limits the statistical power as physiological variance between subjects~\citep{Henriksen2012, Heijtel2014} may dominate effects introduced by the different fitting methods. We expect that differences become statistically significant if the number of subjects is increased as certain trends can be seen in figure~\ref{fig:fig7}.

\mdeleted{
The proposed method was able to closely match 
NLLS mean values in CBF and ATT even for just one average, suggesting that a 
good balance between data and regularization could be found. Regularization 
weights are scaled by the estimated SNR of the data by taking a Fourier 
transform and computing the ratio between the maximum of the central 5\% of 
k-space and the standard deviation of the outer 5\% of k-space. Even though this
 is a very crude way to estimate the SNR, it seems to capture the SNR loss over 
a reduction of averages well. The fact that no parameter tuning was performed 
between synthetic and in vivo fits also supports the effectiveness of the chosen
 SNR estimate based scaling of regularization parameters.}} 

\responseRtwo{R2.C2}{\mdeleted{
The synthetic phantoms were down sampled to typical ASL resolution, leading
to more realistic tissue boundaries due to interpolation. This 
results in not perfectly single valued reference tissue, as can be seen by the 
existence of an IQR for the reference in
figure~\ref{fig:fig4}. The influence is higher in gray matter as it consists 
of fewer voxels, compared to white matter.}} 

In general, the deviations of the fits are higher in WM compared to GM due to 
three times lower
signal and additionally longer arterial transit time resulting in a lower SNR. 
For high SNR regimes (4 averages as well as in GM) the performance of the 
methods is close to each other. 
However, the smallest deviations are achieved with the proposed method\responseRone{R1.C4}{, even though we could not show 
statistical significance}. The 
advantage of posing a joint regularization becomes especially 
clear in low SNR areas such as WM and in the ATT maps. Leveraging all available 
spatial information
drastically improves noise suppression and outlier elimination even if a very 
limited amount of data is used, \responseRone{R1.C4}{as in stroke patients in figure~\ref{fig:fig8} and ~\ref{fig:fig9}}.
 This leads to a clear delineation of GM and WM
compared to the reference methods. \responseRone{R1.C4}{\mdeleted{Fits of CBF and ATT in 
the stroke patient reveal similar information in regions of the stroke for 
BASIL and the proposed method but a loss of information in ATT using NLLS. In 
addition, TGV$^2$ regularization is able to preserve sharp edges in the CBF 
maps which are lost in the BASIL fit. However, this improvement in fitting 
stability comes at the cost of a 
slight underestimation of high values in CBF and ATT due to the nature of the 
TGV$^2$ regularization~\mbox{\citep{Brinkmann2017, Deledalle2017a}}. 
Nevertheless the deviations are small and within 
physiologically inter subject variation~\mbox{\citep{Henriksen2012, Heijtel2014}.}}

Compared to NLLS, CBF and ATT maps of the proposed method show a reduction in noise and outliers in all stroke patients (figure~\ref{fig:fig8}). As highlighted by arrows, the proposed method is able to recover fine details which are lost
in the other regularized approaches but visible in NLLS. This loss of details may hamper detection, or even lead to a complete removal of structures in areas affected by the stroke. While all regularized approaches can reduce noise in CBF and ATT, the proposed method recover such fine structures best, as highlighted by arrows in figure~\ref{fig:fig9}. These differences are most severe in the patient in the third line, leading to a complete miss of certain structures close to the affected area.} 

The utilized CAIPIRINHA accelerated 3D single-shot acquisition is especially 
important for patients where subject movement can lead to uninterpretable CBF 
and ATT maps. Due to the single shot acquisition it is possible to account for
subject motion prior to fitting. \responseRtwo{R2.C13}{Careful selection of
imaging parameters is necessary to avoid excessive blurring in slice direction. 
For the employed sequence, minor blurring can be seen in sagittal views (Supplementary~Material~Figure~\ref{fig:figS4}). Additionally, the employed sequence} provides a more flexible approach for multi-PLD data, 
allowing to sample a broader range of the inflowing blood, which could be 
especially beneficial for patients where the transit time varies over a broad 
range. However, the improved temporal resolution using 4-fold acceleration
 comes at the cost of a reduced SNR which either requires a dedicated denoising 
step prior to fitting 
or makes robust parameter quantification necessary. In contrast to denoising a
direct fitting approach offers the advantage of the inclusion of the signal 
equation which serves as additional a priori knowledge, 
further stabilizing the fit.

Compared to BASIL, the current method only implements the simple ASL model 
given in equation~\ref{eq:ASL}, whereas BASIL allows for simultaneous fitting 
of CBF, ATT, and the arterial inflow contribution in vessels. To facilitate a 
fair comparison this functionality of BASIL was not used in the shown fits. 
However, an extension to the complex model with the proposed method could be 
easily obtained by an adaptation of the signal equation used for fitting, which
will be done in a future step. 
\responseRfour{R4.C2}{
As the proposed approach has been implemented into a Python toolbox~\citep{Maier2020}, addition of new models
can be achieved in a straight forward manner.
Extension to other ASL models, e.g. pulsed ASL (PASL), is simply done by replacing the forward model in equation~\ref{eq:ASL} with the appropriate one. Simple models, not consisting of composed functions, can be included using a plain text file. Complex models
need to be implemented in Python by the user but templates exists to help in the implementation process.
A detailed description of the employed software and how to include new models can be found in~\citet{Maier2020}. An exemplary PASL fit for phantom Case 0 is given in Supplementary~Material~Figure~\ref{fig:figS5}.}

In this study we employed an advanced ASL method with a flexible number of 
combinations of averages and PLDs. Sixteen equally spaced PLDs were used, as 
the range of the arterial transit time was not known in advance. Additionally, 
a different range of ATT is expected for healthy subjects and ischemic stroke 
patients where mostly a prolonged ATT 
is observed. Further improvements in ATT and CBF maps for all methods could be 
expected by optimizing the imaging protocol for the healthy and patient cohort 
separately using the general framework proposed by~\citet{Woods2019}.

Joint regularization could potentially lead to a feature creep from one map 
into another. \responseRone{R1.C5}{While theoretically possible, previous work using joint regularization strategies based on the Frobenius norm have shown that such adverse effects are virtually never observed in practice~\citep{Knoll2017a, Huber2019, Maier2019c}. 
Feature creep usually occurs if reconstruction relies heavily on regularization or if
the chosen regularization weight is too high for the given data and stronger coupling norms, like the Nuclear norm, are used.}  
To investigate if feature creep occurs, we introduced Case 2 in our phantom series. The pathology 
in the frontal left area is clearly visible in the fitted CBF but no adverse 
affect can be observed in the corresponding area of ATT, neither in the 
quantitative map itself~(figure~\ref{fig:fig1}) nor in the pixel wise 
difference~(figure~\ref{fig:fig2}). \responseRone{R1.C5}{In addition, Ccases 4-6 in figure~\ref{fig:fig1_2} and figure~\ref{fig:fig2_2} also show no adverse effect of the imposed ATT pathology on the CBF map. The contrary can be observed, the joint regularization produces the most stable CBF estimates of all methods.} A totally wrong choice of the 
regularization weight compared to the supporting data could nevertheless 
introduce such errors. However, such a strong weight for regularization would 
also lead to a severely hampered visual impression and such fits would likely 
be discarded. 

The assumption of AWGN is typically valid in the real and imaginary channel of 
single channel complex MRI data but violated for commonly used magnitude 
images, especially if array coils are used~\citep{Aja-Fernandez2015}. 
\responseRtwo{R2.C1}{Even though the AWGN assumption might be violated in the 
PWI series, convergence of the algorithm itself is not hampered. However, results obtained
from data that violates the AWGN assumption might show some bias compared to the true value. Such bias could be 
avoided by adapting the data norm to account for the altered noise distribution, e.g. $L^1$ data norm for salt-and-pepper noise.
In general, the resulting data measure need not be convex and optimization could get stuck in 
critical points, e.g. local minima or points of inflection, resulting in even worse fitting than the generic $L^2$ data norm. 
In addition, the $L^2$-norm possesses several desirable properties for 
optimization (differentiability, convexity) which usually outweigh the drawback of a slight bias and is thus widely used, also in our reference methods.\mdeleted{
However, under the assumption of sufficient SNR the noise in the magnitude 
images of control and label can be assumed to be Gaussian distributed. The 
subtraction operations in generating the PWIs is linear, thus preserves the 
Gaussian distributions. To this end the assumption of Gaussian noise in the 
PWIs holds true for typically used image protocols in ASL.}} Nevertheless it 
should be noted that the optimal case for fitting complex MRI data to a given 
signal model would be to use the raw complex k-space signal from each receive 
coil separately. 

A natural extension of the proposed fitting approach would be to incorporate 
the whole MRI signal acquisition pipeline, i.e. coil sensitivity profiles and 
Fourier sampling, directly fitting k-space data. This would leverage the 
natural Gaussian distribution present in the raw k-space signal, fully 
validating the choice of the $L^2$-norm for fitting. \responseRthree{R3.C8}{Contrary to recent works on
improved ASL-perfusion image reconstruction such as MOCHA~\citep{Mehranian2020} or
ASL-TGV~\citep{SPANN2020116337}, which reconstructs the perfusion weighted images from raw k-space data, no subsequent
fitting step would be required to estimate CBF and ATT. Further, noise characteristics in such improved PWI images might severely deviate from the
assumption of AWGN due to the non-linear reconstruction process and might lead to a bias in the estimation process.}
The model based fitting approach potentially allows for higher acceleration compared to 
the two separate steps, image reconstruction followed by fitting\responseRthree{R3.C8}{, due to the reduction of unknowns and combination of information of the different PWIs by means of the signal equation}. \responseRtwo{R2.C4}{Further extensions of the proposed method could include partial volume correction based on two compartment models, which are already included in BASIL.}

ASL imaging is very sensitive to \responseRthree{R3.C2}{\mdeleted{motion thus, motion correction is vital.}
signal variations from motion or changes in blood velocity due to the cardiac cycle~\citep{verbree2017}. 
Currently, the proposed method does not directly account for these variations. 
Motion related variations are corrected for in a preprocessing step but no correction for blood velocity changes is applied. As it is planned to extend the method to use raw k-space data, motion could be included in the
forward model, e.g. based on determined motion fields prior to fitting, as it is done in MOCHA.}
As estimation of motion directly from highly undersampled k\nobreakdash-space data can be 
challenging, a robust estimation and correction needs to be found. Another 
possibility would be to include a motion term into the \responseRthree{R3.C2}{\mdeleted{forward model}fitting process} but this poses a mathematically challenging problem, especially for forming forward and 
adjoint operation pairs.

\responseRthree{R3.C4}{If ASL data is evaluated over large ROIs, NLLS seems to be
the favourable method as it shows the least bias in our simulations~(figure~\ref{fig:fig4}) for
most investigated ROIs, especially in GM. However, this might only be advantageous if a large population of
subjects is used. Single reconstruction can be subject of significant noise, as 
reflected by the highest IQR of all methods. The proposed approach is preferable in
single observation scenarios as it shows comparable bias to the other methods
but generally the least deviations (IQR). Even though \textit{BASIL w/} can reduce IQR to a 
similar range as the proposed method, it shows the highest bias of all investigated
algorithms in most cases, especially in ATT.}

\responseRfour{R4.C1}{In addition to the 3D acquisition used in this work, ASL is often performed in a 2D slice-by-slice fashion.
Such data can also be fitted with the presented reconstruction framework, either, by applying regularization in 2D only or by adapting the third
gradient direction to account for non-isotropic voxels. This adaptation amounts to a simple scaling of the gradient with respect to the 
ratio between in-plane and the acquired slice resolution, taking into account slice thickness and inter slice gap. Setting this scaling to zero equals 2D regularization. However, 3D regularization outperforms 2D, as has been shown in previous work~\citep{Huber2019, Maier2019c}.}

\responseRthree{R3.C21}{Reconstruction on the GPU required roughly 1 GB of memory which should be available on any recent GPU. Computation speed varies with hardware and further reduction could be expected with the recent increase in GPU performance. If memory requirements supersede the available
GPU memory, a double buffering strategy is available, as introduced in~\citet{Maier2020}. New scanner consoles already often include GPUs, thus the proposed method could be directly integrated into the scanner reconstruction process.}
% --------------------------------------------------------------------
% Conclusion
% --------------------------------------------------------------------
\section{Conclusion}
\label{sec:conclusion}

The proposed non-linear fitting approach with joint spatial priors on CBF and 
ATT provides high-quality quantitative maps of the whole brain from a 
single-shot 3D acquisition. The combination of single-shot 3D acquisition and 
robust parameter quantification addresses important clinical demands in terms of
 motion robustness, scan-time reduction and a much better sampling of the 
kinetic curve. This makes this approach promising for challenging image acquisition 
conditions of patients with cerebrovascular disease.

\section*{Acknowledgments}
The authors would like to acknowledge Dr. Marta Vidorreta Diaz De Cerio, 
Siemens Healthcare, Spain for providing the updated pCASL module with 
improved background suppression.

We would like to thank Lukas Pirpamer from the Medical University Graz for assistance in data export.

\section*{Funding}
Oliver Maier is a recipient of a DOC Fellowship (24966) of the Austrian Academy 
of Sciences at the Institute of Medical Engineering at TU Graz. The authors 
would like to acknowledge the NVIDIA Corporation Hardware grant support.

\clearpage
\appendix
\section{Mathematical derivations}
\label{sec:appA}
As described in section~\ref{sec:opt}
it is our goal to solve the following optimization task within each GN iteration

\begin{linenomath*}
\begin{align} \label{eq:linapp}
   \underset{u,v}{\min}\quad 
\frac{1}{2}\sum_{n=1}^{N_d}&\|\mathrm{D}A_{\phi,t_n}\rvert_{u=u^{k}} u-\tilde{d_n}^k
\|_2^2 
+ \\ 
\gamma_k(
\beta_0&\|\nabla u - v\|_{1,2,F} + \beta_1|\|\mathcal{E}v\|_{1,2,F}) +
\nonumber\\ \frac{\delta_k}{2}&\|u-u^{k}\|_{M_k}^2,\nonumber
\end{align}
\end{linenomath*}
where $\nabla: U^{N_u} \to U^{3\times N_u} $ and $\mathcal{E}: U^{3\times 
N_u} \to U^{6\times N_u}$ are defined as

\begin{linenomath*}
\[
\nabla u = \Big( \delta_{i+} u^l, \delta_{j+} u^l, 
\delta_{k+} u^l\Big)_{l=1}^{N_u}
\]
\end{linenomath*}
and

\begin{linenomath*}
\begin{align*}
\mathcal{E}v =  \Big( 
&\delta_{i-}v^{1,l},\delta_{j-}v^{2,l},\delta_{k-}v^{
3
,l}, \\
&\frac{\delta_{j-}v^{1,l} + \delta _{i-}v^{2,l}}{2}, 
\frac{\delta_{k-}v^{1,l} + \delta _{i-}v^{3,l}}{2},
\frac{\delta_{k-}v^{2,l} + \delta _{j-}v^{3,l}}{2} 
\Big)_{l=1}^{N_u}.
\end{align*}
\end{linenomath*}
The operators 
$\delta_{i+},\,\delta_{j+},\,\delta_{k+}$ and 
$\delta_{i-},\,\delta_{j-},\,\delta_{k-}$ define 
forward and backward finite difference operators, 
respectively, with respect to the $i$, $j$ and $k$ 
coordinate. The image is symmetrically extended outside the domain.

The required saddle point formulation of the form,

\begin{linenomath*}
\begin{equation}\label{eq:PD_saddle_ap}
\underset{x}{\min}\,\underset{y}{\max}~ \left<\mathrm{K}x,y\right> + G(x) - 
F^*(y),
\end{equation}
\end{linenomath*}
equivalent to Eq. \ref{eq:linearized} can be obtained using the convex 
conjugate as follows:

\begin{linenomath*}
\begin{alignat*}{3}
&  &&\min_{x=(u,v)} ~
&&\frac{1}{2}\sum_{n=1}^{N_d}\|\mathrm{D}A_{\phi,t_n}|_{u=u^{k}} u-\tilde{d_n}^k
\|_2^2 + \\
& && &&\gamma_k(\beta_0\|\nabla u - v\|_{1,2,F} + 
\beta_1|\|\mathcal{E}v\|_{1,2,F}) + \\ 
& && &&\frac{\delta_k}{2}\|u-u^{k}\|_{M_k}^2\\
&{\Leftrightarrow}
 &&\min_{x} \max_{y=(z_0,z_1,r)}
 &&\sum_{n=1}^{N_d}\left\{\left<\mathrm{D}A_{\phi,t_n}|_{u=u^{k}} u,r_n\right> -
\left<\tilde{d_n}^k,r_n\right> - \frac{1}{2} \| r_n \|_2^2 \right\}+\\
& && &&\left<K_1x,z\right> 
-\mathcal{I}_{\{\|\cdot \|_{\infty \leq \beta_0\gamma_k}\}}(z_0)
-\mathcal{I}_{\{\|\cdot \|_{\infty \leq \beta_1\gamma_k}\}}(z_1) \\
& && && +\frac{\delta_k}{2}\|u-u^{k}\|_{M_k}^2\\
 &{\Leftrightarrow}
  &&\min_{x} \max_{y} ~ &&\left<\mathrm{K}x,y\right> + G(x) - F^*(y). 
\end{alignat*}
\end{linenomath*}
with

\begin{linenomath*}
\begin{align*}
 K = \left( \begin{matrix}
	DA_{\phi} & 0	\\
 \nabla & -id	\\
 0 & \mathcal{E}	\\
\end{matrix}\right), \quad  K_1 = \left( \begin{matrix}
 \nabla & -id	\\
 0 & \mathcal{E}	\\
\end{matrix}\right), z = (z_0, z_1).
\end{align*}
\end{linenomath*}
$id$ amounts to the identity matrix. 
$F^*(y) = F^*(z_0,z_1,r) = 
\sum_{n=1}^{N_d}\left\{\left<\tilde{d_n}^k,r_n\right>+ \frac{1}{2} \| r_n 
\|_2^2\right\} +\mathcal{I}_{\{\|\cdot \|_{\infty \leq \beta_0\gamma_k}\}}(z_0)
+\mathcal{I}_{\{\|\cdot \|_{\infty \leq \beta_1\gamma_k}\}}(z_1) $, and \\
$G(x)=G(u)=\frac{\delta_k}{2}\| u - u^{k} \|_{M_k}^2$. 
$\mathcal{I}_{\{\|\cdot \|_{\infty \leq \alpha_p\gamma_k}\}}(z_p)$
 amounts to the convex conjugate of the $L^1$-norm which is defined as the 
indicator function of the unit ball of the $L^\infty$-norm
scaled with the corresponding regularization parameter $\beta_p\gamma$ ($p={0,1}$)

\begin{linenomath*}
\begin{align*}
\mathcal{I}_{\{\|\cdot \|_{\infty \leq \beta_p\gamma_k}\}}(z_p) = \begin{cases}
0 & \|z_p\|_\infty \leq \beta_p\gamma_k \\
\infty & else
\end{cases}
\end{align*}
\end{linenomath*}
$\mathrm{D}A_{\phi,t_n}|_{u=u^{k}}$ is the Jacobian matrix evaluated at 
$u=u^{k}$ of the non-linear ASL signal equation for all scans $_nt$:

\begin{linenomath*}
\begin{center}
\begin{equation}
  \mathrm{D}A_{\phi}:  u = (u_l)_{l=1}^{N_u}
                \mapsto  \left( \sum\limits_{l=1}^{N_u}  \left[ \left. 
\frac{
                	\partial A_{\phi,t_n}(u)}{\partial u_l} \right|_{u = u^{k}}  
                u_l \right] \right)_{n=1}^{N_d}       = (\eta_n)_{n=1}^{N_d}.
\end{equation}
\end{center}
\end{linenomath*}
To compute the update steps of the PD algorithm as

\begin{linenomath*}
\begin{center}
\begin{equation}
\begin{aligned} 
  y^{n+1} &= (id+\sigma \partial F^*)^{-1}(y^n+\sigma K \overline{x}^n)\\
  x^{n+1} &= (id+\tau \partial G)^{-1}(x^n-\tau K^H y^{n+1})\\
  \overline{x}^{n+1} &= x^{n+1} + \theta(x^{n+1} - x^n)
\end{aligned}
\end{equation}
\end{center}
\end{linenomath*}
with $\theta\in[0,1]$, 
the following operations need to be defined.

The adjoint operations of $K$, $K^H$ are 

\begin{linenomath*}
\begin{center}
\begin{equation}
\begin{aligned}
  K^H = &\begin{pmatrix}    
            DA_{\phi}^H & - \text{div}^1 & 0\\
            0 & -id &-\text{div}^2 
            \end{pmatrix},    
\end{aligned}
\end{equation}
\end{center}
\end{linenomath*}
where the divergence operators $\text{div}^1$ and $\text{div}^2$ are the 
negative adjoints of $\nabla $ and $\mathcal{E}$, respectively. 
The adjoint of $\mathrm{D}A_{\phi}$ reads as

\begin{linenomath*}
\begin{center}
\begin{equation*}
  DA_{\phi}^H: \eta = (\eta_n)_{n=1}^{N_d} \mapsto 
            \left(  \sum\limits_{n=1}^{N_d}\left. \overline{
              \frac{\partial A_{\phi,t_n}(u)}
              {\partial u_l}} \right|_{u = u^{k}} \eta_n 
                    \right)_{l=1}^{N_u} = (u_l)_{l=1}^{N_u} = u.
\end{equation*}
\end{center}
\end{linenomath*}
The operators $P$ corresponding to the proximal mapping of $F^*$, i.e. 
the convex conjugate of $F$, and $G$ in the algorithm are given by

\begin{linenomath*}
\begin{align*}
(id+\sigma \partial F^*)^{-1}(\xi) &:=\left\{\begin{matrix}
&P_{\beta_0}(\xi_0)  = \frac{\xi_0}{\max \left( 1,  \frac{ 
\vert \xi_0 \vert }{\beta_0\gamma_k} \right)} \\ 
&P_{\beta_1}(\xi_1)  = \frac{\xi_1}{\max \left( 1,  \frac{ 
\vert \xi_1 \vert }{\beta_1\gamma_k} \right)} \\ 
&P_{\sigma L^2}(\xi_2) = \frac{\xi_2 - \sigma \tilde{d}^k}{1+{\sigma}}
\end{matrix}\right.\\
(id+\tau \partial G)^{-1}(\xi) &:= P_{\tau G}(\xi)={(id+{\tau\delta_k M_k})^{-1}}{(\tau\delta_k M_k u^{k}+\xi)}
\end{align*}
\end{linenomath*}
where the operations in $P_{\beta_0}(\xi_1)$, $P_{\beta_1}(\xi_2)$, 
and $P_{\sigma L^2}(\xi_3)$ are performed point wise.
The multiplication with $M_k$ can be easily computed as $M_k$ is a
 diagonal matrix. The inversion of $(id+{\tau\delta_k M_k})$ is simply
a inversion of each element on the diagonal. Thus $P_{\tau G}(\xi)$ can be
computed easily in a point-wise fashion.
\newpage
\section{Pseudo Code}
\begin{algorithm2e}
    \DontPrintSemicolon

    \textbf{Initialize:}  $(u^0,v^0)$, 
$(\overline{u^0},\overline{v^0}),$ 
$(z_0^0,z_1^0,r^0)$, $\tau^0 > 0 $, $\kappa = 1, \theta^0 = 1$, $\mu = 0.5$, $\delta=0.99$\\

    \For{m$\gets$0 \KwTo $maxit$}{
    \vspace*{1em}
    \textbf{Primal Update: }\\
    \vspace*{0.2em}
    \hspace*{1em} $u^{m+1} \leftarrow P_{\tau^m G}\left( u^m - \tau^m \left( 
-\text{div}^1 z_0^{m}
+ \mathrm{D}A_{\phi}^H r^{m}  \right)  \right)$ \\
    \hspace*{1em}$v^{m+1} \leftarrow v - \tau^m \left(- \text{div}^2 z_1^{m} - 
z_0^{m}
\right) $ \\
    \textbf{Update $\tau$:} \\
    \hspace*{1em}$\tau^{m+1} \leftarrow \tau^m\sqrt{(1+\theta^m)}$\\
    \vspace*{1em}
    \textbf{Start Linesearch:} \\
   \textbf{	Update $\theta$:} \\
    \hspace*{1em}$\theta^{m+1} \leftarrow \frac{\tau^{m+1}}{\tau^m}$\\
    \vspace*{0.2em}    
    \textbf{Extrapolation:} \\
    \vspace*{0.2em}
    \hspace*{1em}$(\overline{u}^{m+1},\overline{v}^{m+1}) 
    \leftarrow (u^{m+1},v^{m+1}) + \theta^{m+1}
((u^{m+1},v^{m+1})-(u^m,v^m)) $    \\
    \vspace*{0.2em}    
    \textbf{Dual Update:} \\
    \vspace*{0.2em}
    \hspace*{1em}$z_0^{m+1}\leftarrow P_{\beta_0} \left( z_0^m + 
\kappa\tau^{m+1} (\nabla
\overline{u}^{m+1} - \overline{v}^{m+1}) \right)$ \\
    \hspace*{1em}$z_1^{m+1} \leftarrow P_{\beta_1} \left( z_1^m + 
\kappa\tau^{m+1} 
(\mathcal{E}\overline{v}^{m+1}) \right)$ \\
    \hspace*{1em}$r^{m+1} \leftarrow P_{\kappa\tau^{m+1}L^2} \left( r^m + 
\kappa\tau^{m+1}(\mathrm{D}A_{\phi}\,\overline{u}^{m+1})\right)$ \\
    \vspace*{0.2em}
    \textbf{break Linesearch if:} \\
    \vspace*{0.2em}
    \hspace*{1em}$\sqrt{\kappa}\tau^{m+1}\|K^H\,y^{m+1}-K^H\,y^m\|_2 \leq \delta\|y^{m+1}-y^m\|_2$\\
    \vspace*{0.2em}
    \textbf{else:} \\
    \vspace*{0.2em}    
    \hspace*{1em}$\tau^{m+1} \leftarrow \tau^{m+1}\mu $\\
    \vspace*{1em} 
    }\caption{Primal-dual algorithm for solving the TGV$^2$ 
regularized ASL parameter quantification task in every Gauss-Newton step. Note 
that linearity of involved operations can be used to decrease computational 
load.} 
    \label{alg:tgvsolve}
\end{algorithm2e} 

\clearpage

\bibliography{ASL}

%\ifisresponse
\onecolumn
\clearpage
\pagenumbering{Roman}
\setcounter{page}{1}
\section*{Supplementary Material}
\label{sec:submissions}

\renewcommand{\figurename}{Supplementary Material Figure }
\setcounter{figure}{0} \renewcommand{\thefigure}{S\arabic{figure}}

\renewcommand{\tablename}{Supplementary Material Table }
\setcounter{table}{0} \renewcommand{\thetable}{S\arabic{table}}
\begin{figure*}[!htbp]
\centering
 \includegraphics[width=0.95\textwidth, height=0.85\textheight, keepaspectratio]{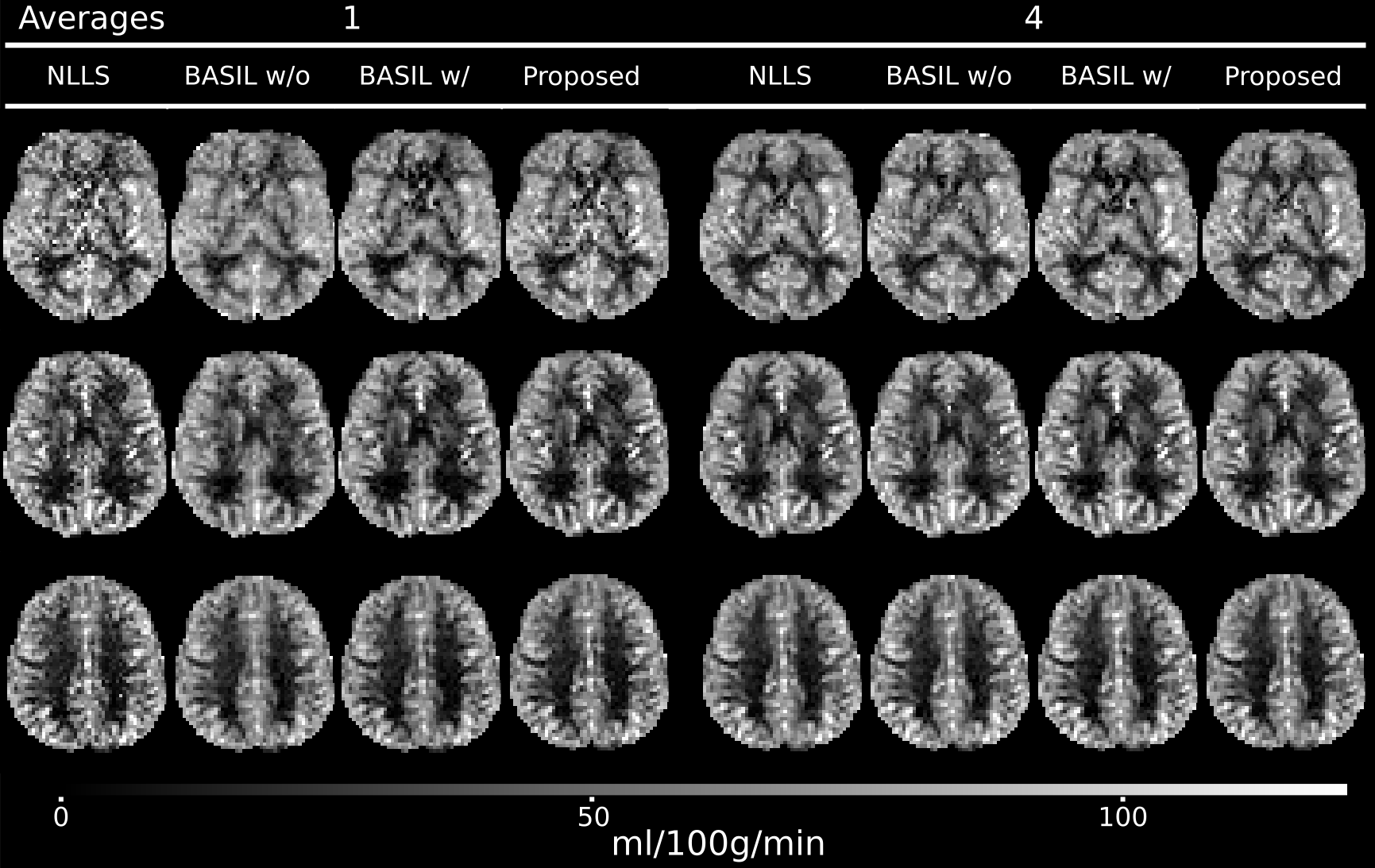}\\
 \includegraphics[width=0.95\textwidth, height=0.85\textheight, keepaspectratio]{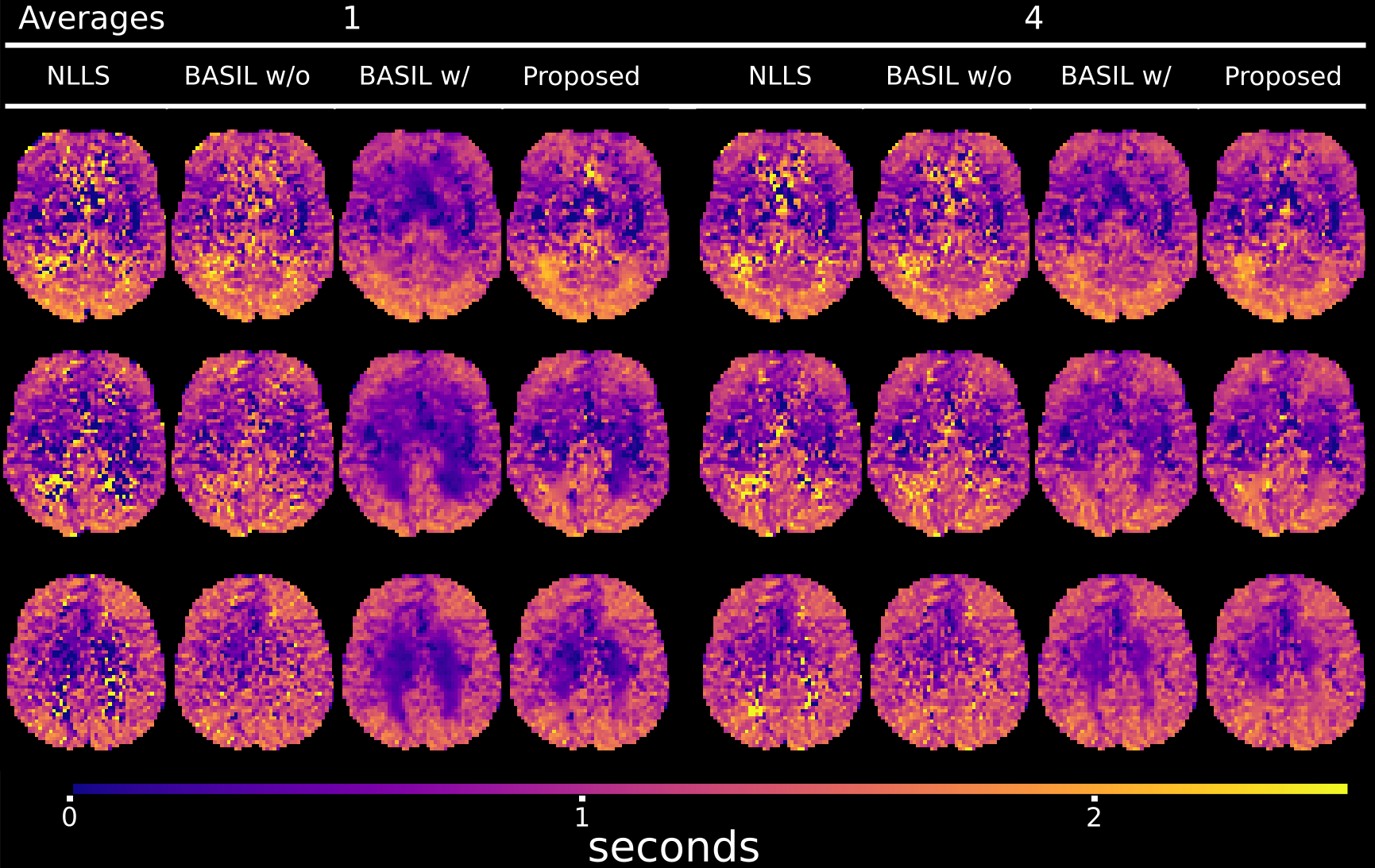}
  \caption{Three representative slices of the CBF and ATT maps of subject 1}
 \label{fig:figS1_1}
\end{figure*}

\begin{figure*}[!htbp]
\centering
 \includegraphics[width=0.95\textwidth, height=0.85\textheight, keepaspectratio]{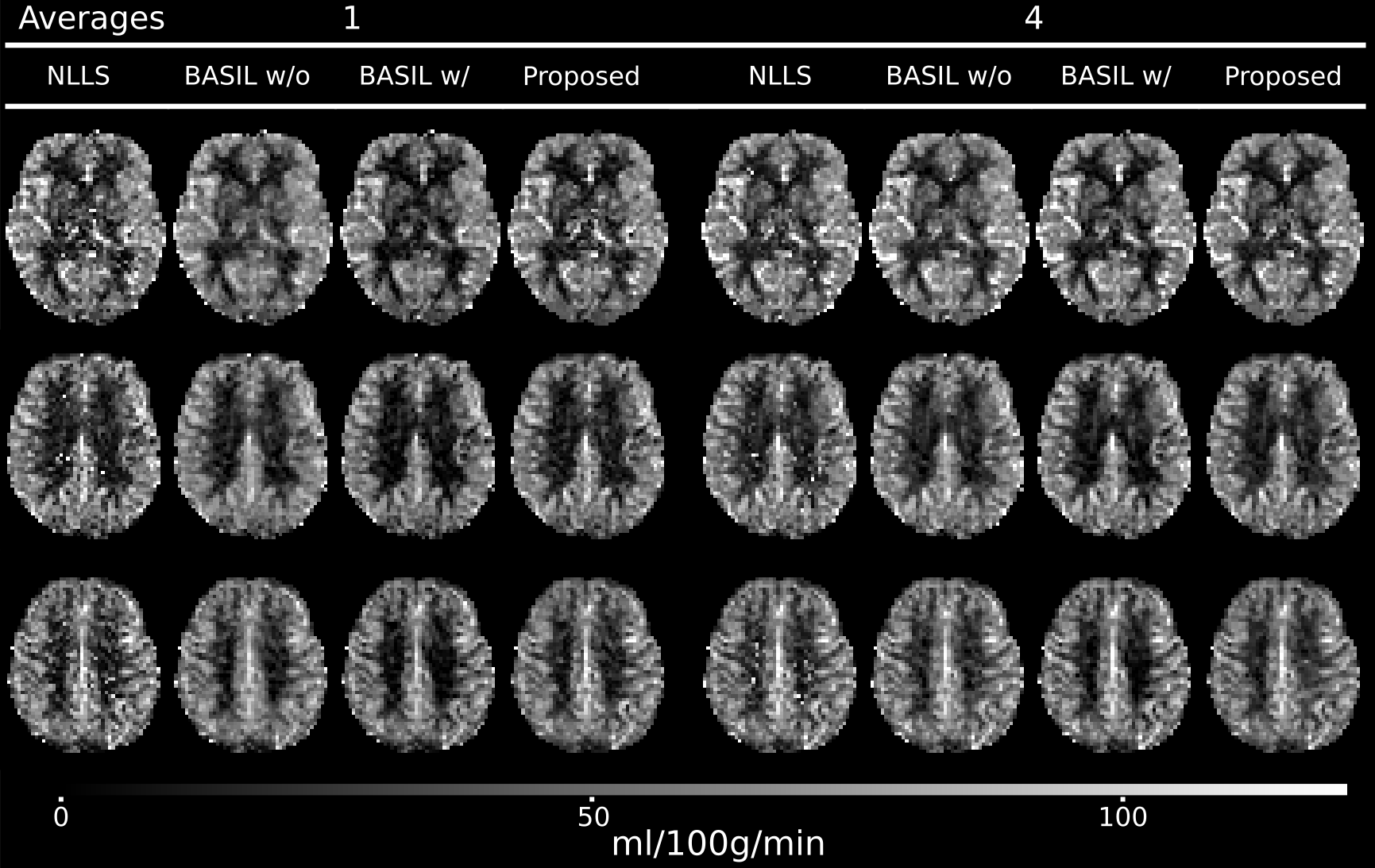}\\
 \includegraphics[width=0.95\textwidth, height=0.85\textheight, keepaspectratio]{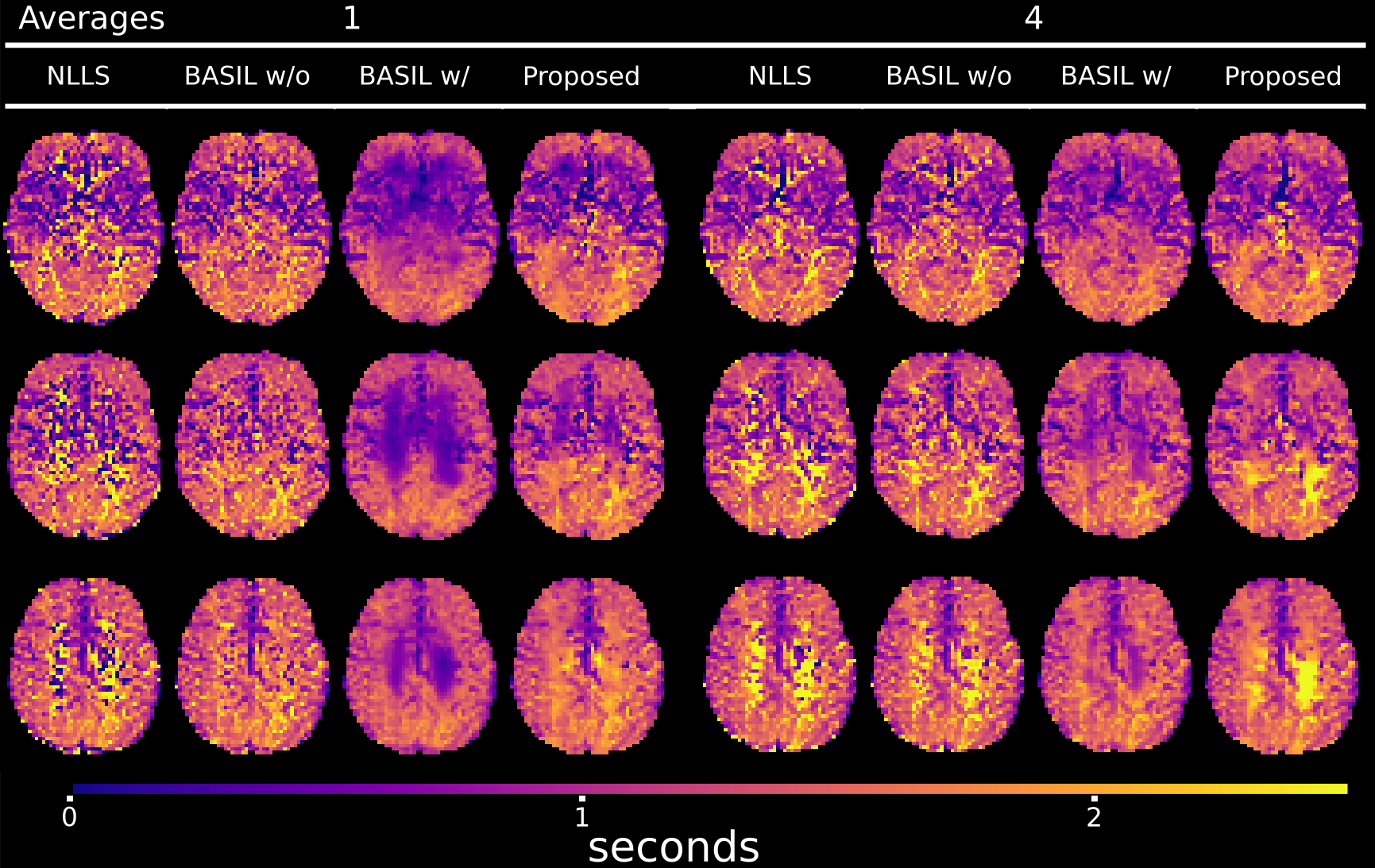}
  \caption{Three representative slices of the CBF and ATT maps of subject 2}
 \label{fig:figS1_2}
\end{figure*}

\begin{figure*}[!htbp]
\centering
 \includegraphics[width=0.95\textwidth, height=0.85\textheight, keepaspectratio]{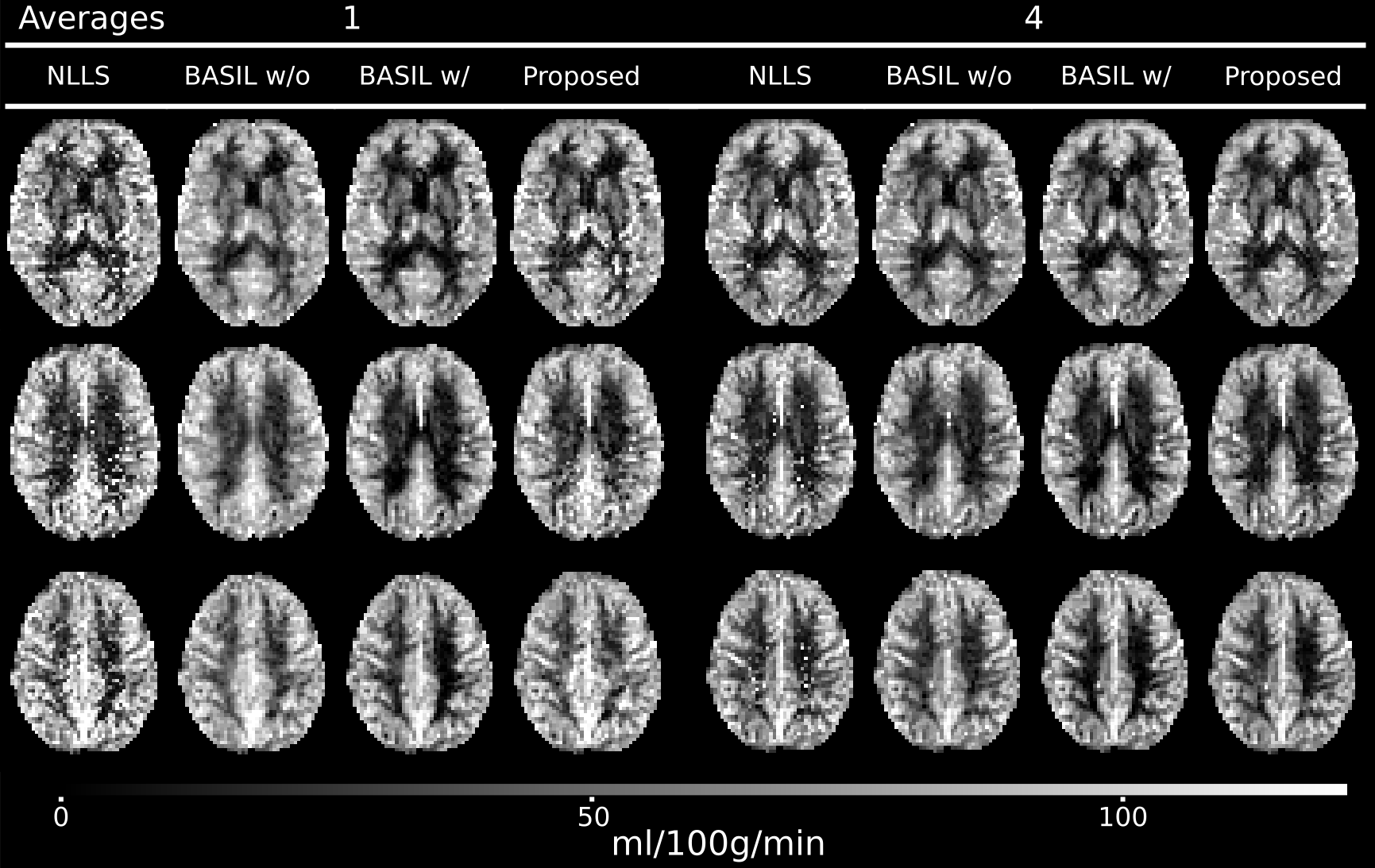}\\
 \includegraphics[width=0.95\textwidth, height=0.85\textheight, keepaspectratio]{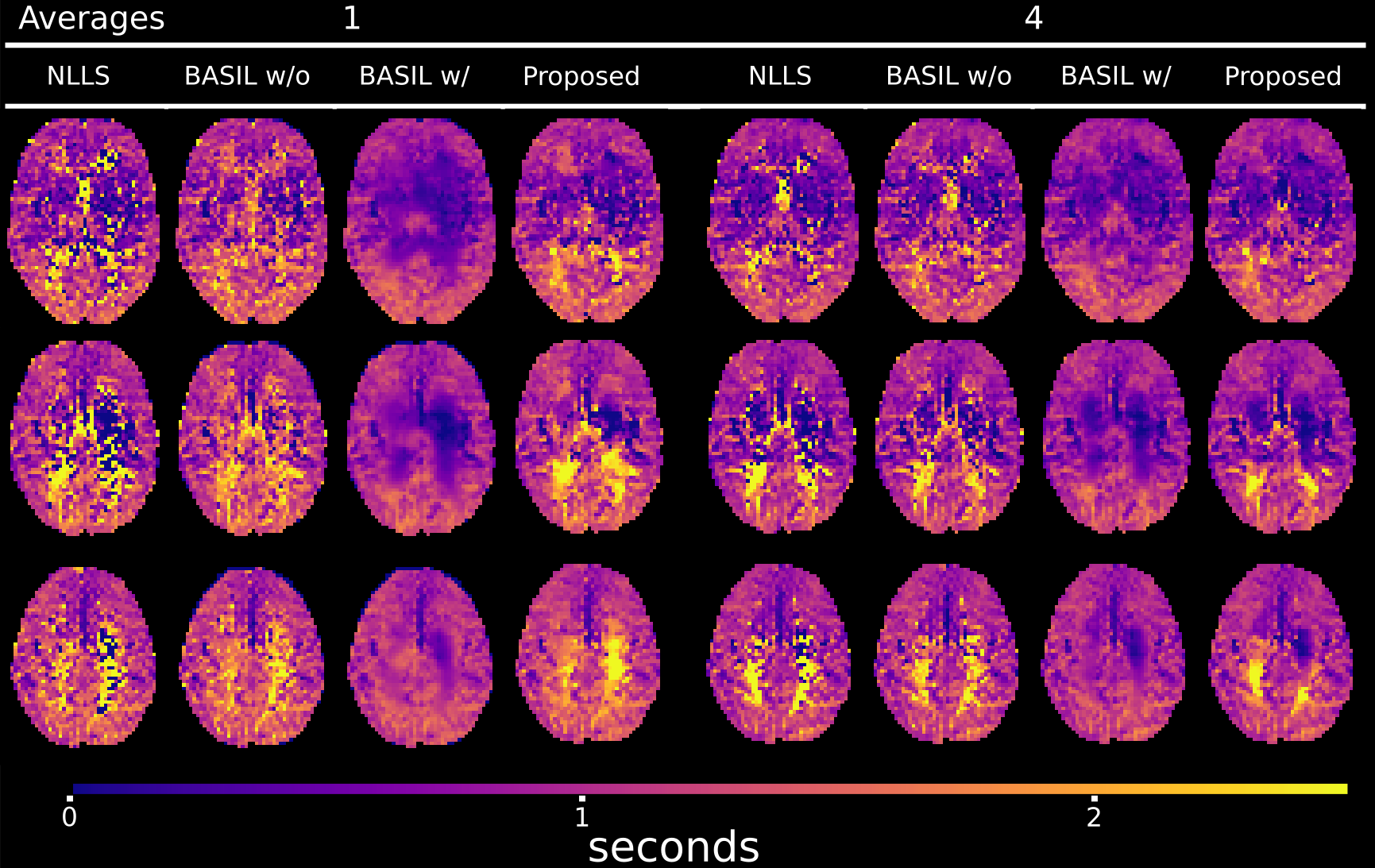}
  \caption{Three representative slices of the CBF and ATT maps of subject 4}
 \label{fig:figS1_3}
\end{figure*}

\begin{figure*}[!htbp]
\centering
 \includegraphics[width=0.95\textwidth, height=0.85\textheight, keepaspectratio]{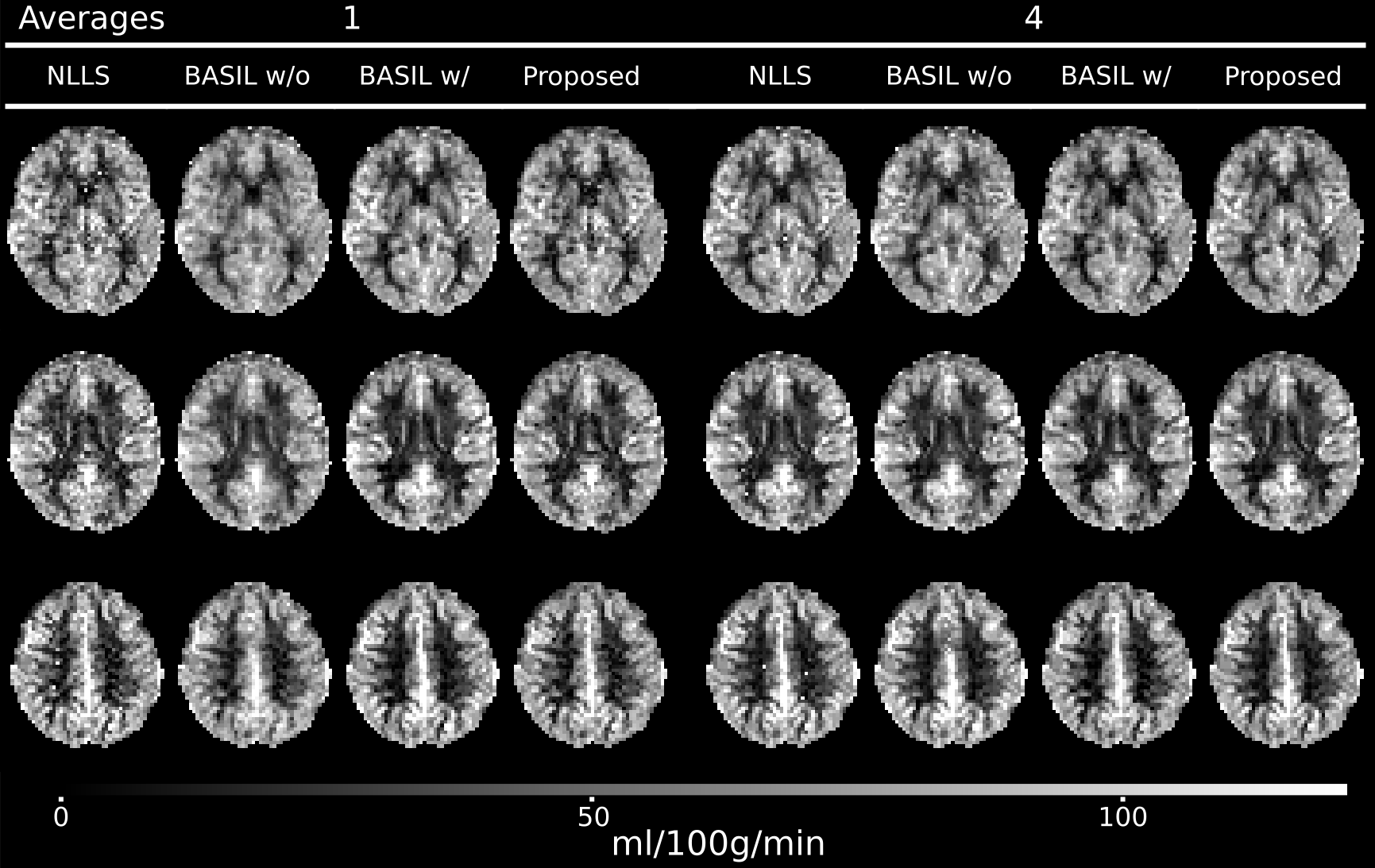}\\
 \includegraphics[width=0.95\textwidth, height=0.85\textheight, keepaspectratio]{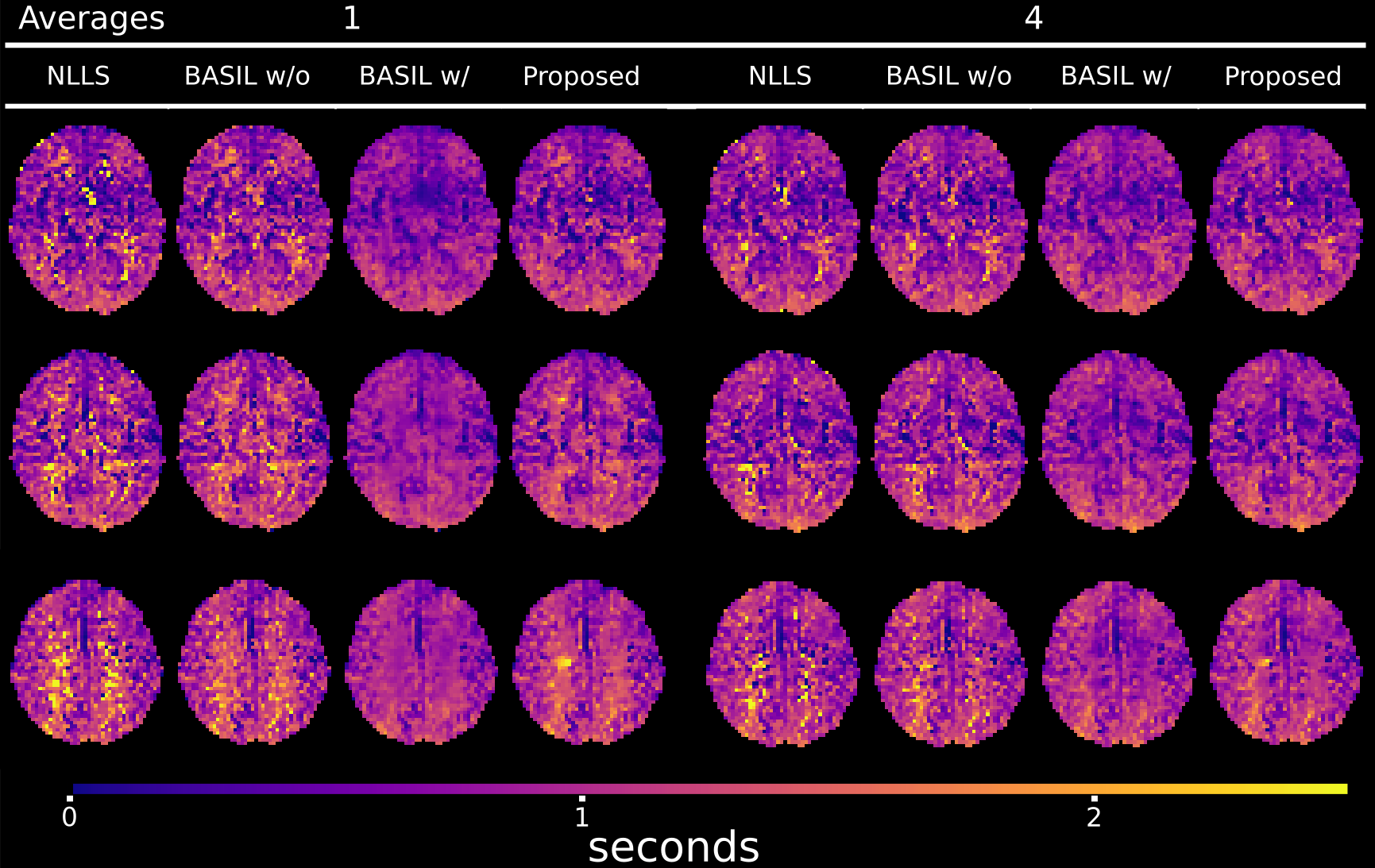}
  \caption{Three representative slices of the CBF and ATT maps of subject 5}
 \label{fig:figS1_4}
\end{figure*}

\begin{figure*}[!htbp]
\centering
 \includegraphics[width=0.95\textwidth, height=0.85\textheight, keepaspectratio]{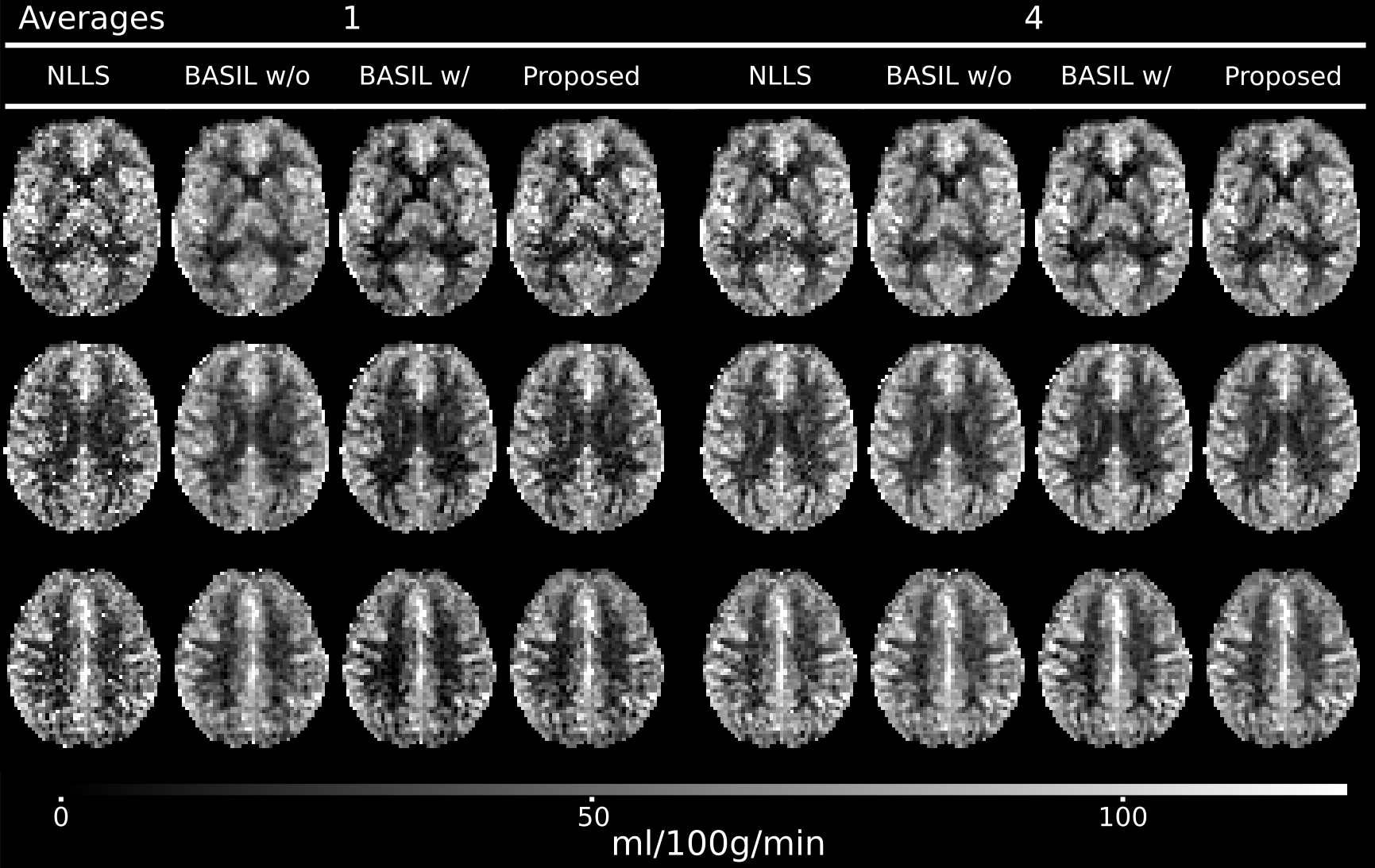}\\
 \includegraphics[width=0.95\textwidth, height=0.85\textheight, keepaspectratio]{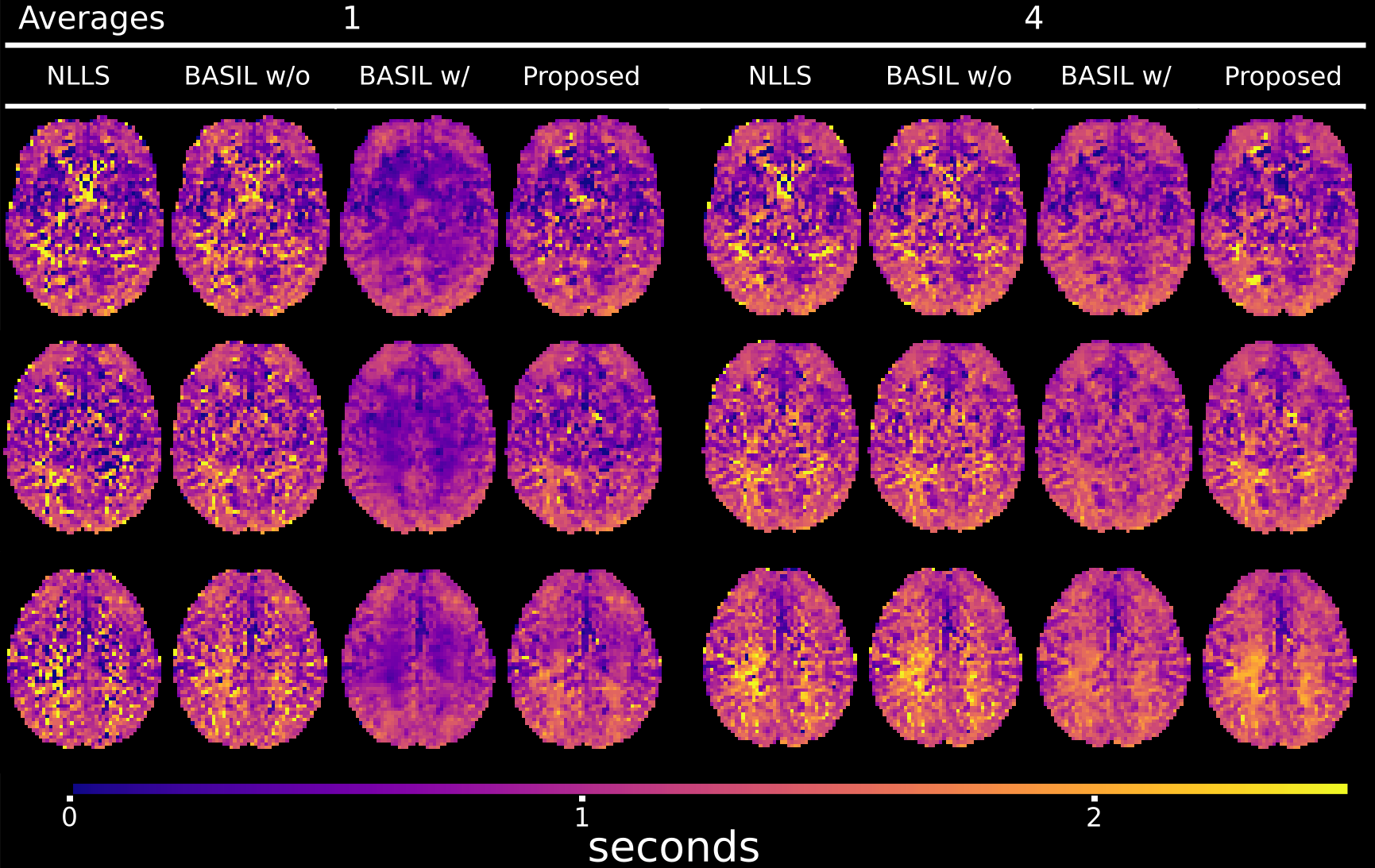}
  \caption{Three representative slices of the CBF and ATT maps of subject 6}
 \label{fig:figS1_5}
\end{figure*}
\twocolumn
\begin{figure}[!htbp]
\centering
 \includegraphics[width=0.95\columnwidth, 
height=0.9\textheight, keepaspectratio]{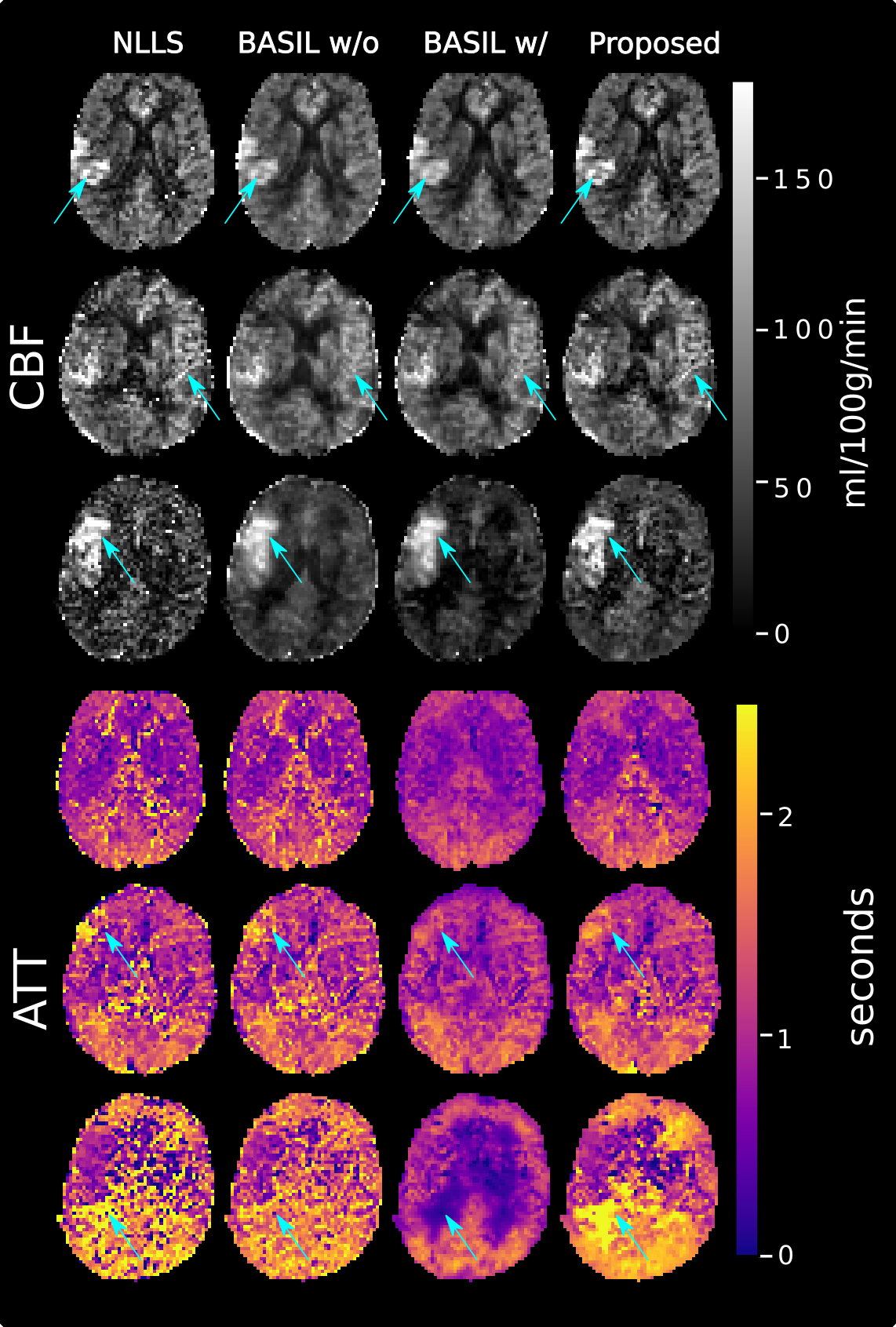}
 \responseRfour{R4}{\caption{An exemplary slice of CBF and ATT for the remaining stroke patients. 
Patients are shown in rows, different reconstruction methods in columns. 
Difference between methods are highlighted by arrows. \label{fig:figS3}}}
\end{figure}

\begin{figure}[!htbp]
\centering
 \includegraphics[width=0.95\columnwidth, 
height=0.9\textheight, keepaspectratio]{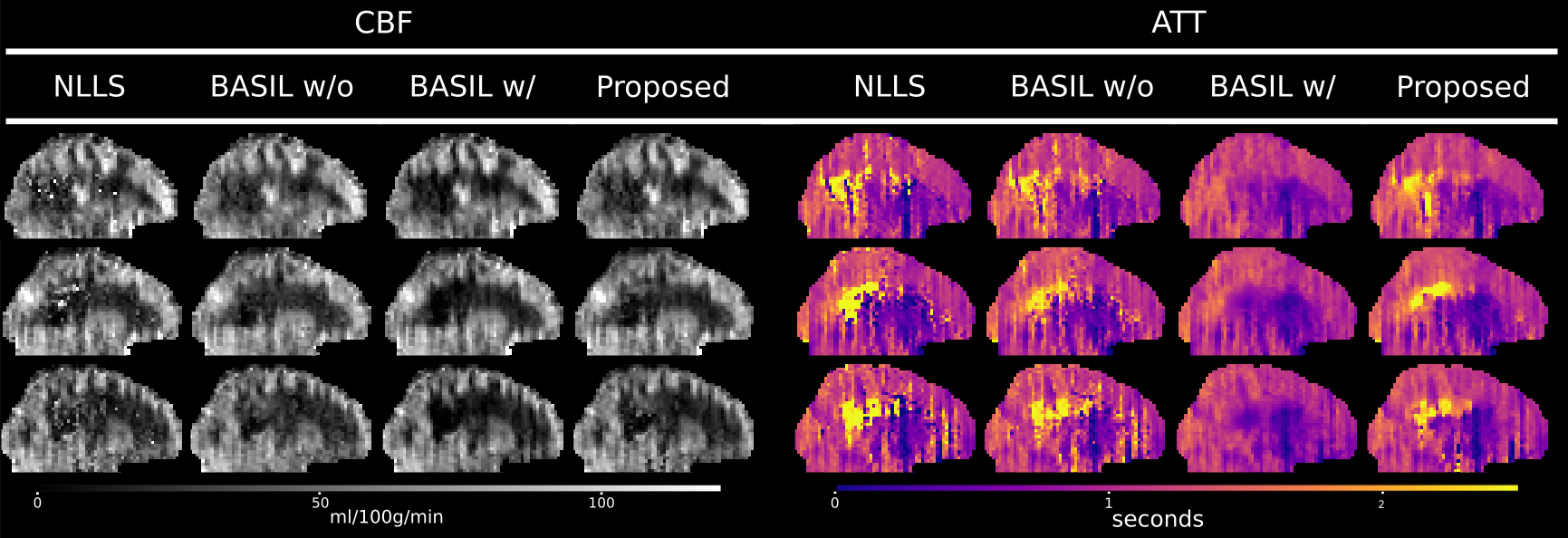}
 \responseRtwo{R2.C13}{\caption{Three exemplary sagittal views of healthy subject 3, showing the blurring induced in z-direction due to long echo train length. \label{fig:figS4}}}
\end{figure}

\begin{figure}[!htbp]
\centering
 \includegraphics[width=0.95\columnwidth, 
height=0.9\textheight, keepaspectratio]{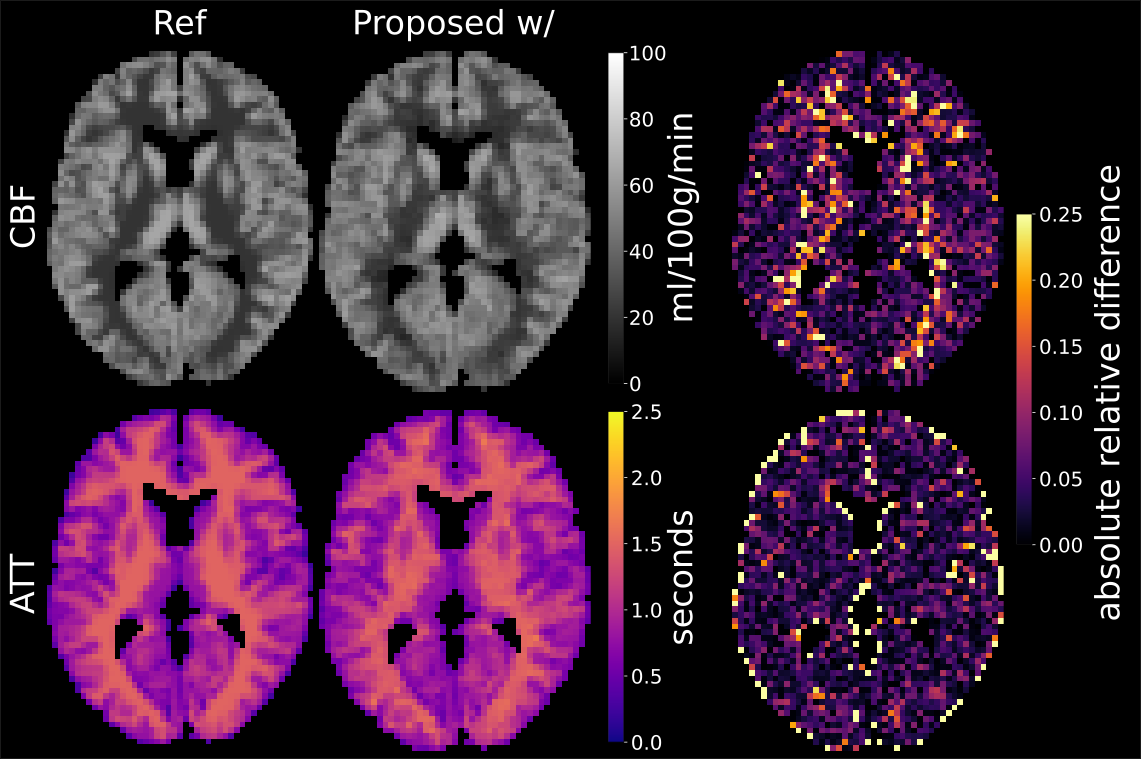}
 \responseRfour{R4.C2}{\caption{Preliminary results for a 2D slice-by-slice PASL based quantification model. Shown is an exemplary slice of Case 1. Simulated resolution amouted to 3x3x6 mm$^3$. As a 2D acquisition is assumed, a time delay between subsequent slices is introduced. \label{fig:figS5}}}
\end{figure}

%\fi

\end{document}